\documentclass[a4paper,11pt]{article}
\pdfoutput=1 % if your are submitting a pdflatex (i.e. if you have
\usepackage{jcappub} % for details on the use of the package, please
\usepackage[T1]{fontenc} % if needed

\usepackage{array,multirow,graphicx}

\title{Higgs Inflation Model with Small
\\
  Non-Minimal Coupling Constant
}

\author[a,b,1]{Alexander B. Kaganovich,\note{Corresponding author.}}
\affiliation[a]{Physics Department, Ben Gurion University of the Negev,\\Beer Sheva, Israel}
\affiliation[b]{Sami Shamoon College of Engineering, \\Beer Sheva, Israel}
\emailAdd{alexk@bgu.ac.il}
\abstract{ The Higgs sector of the Two-Measure Theory (TMT) extension of the electroweak SM (TMSM) is studied in the context of cosmology, where the only non-zero component $\varphi$ of the cosmologically averaged Higgs field plays the role of the inflaton.
The self-consistency of the system of equations obtained from the original action has the form of an algebraic constraint defining the scalar $\zeta$, which is the ratio of two volume measures, as a function of the  field $\varphi$ and its first derivatives.
The scalar $\zeta$ is present in all equations of motion and  has a significant effect on the dynamics of the fields. After the transition in the equations of motion to the Einstein frame with the spatially flat Friedmann metric, it is convenient to describe the resulting system of equations using the action $S_{eff}$ and the Lagrangian $L_{eff}$, which we call the TMT-effective action and the TMT-effective Lagrangian and from which these equations can be obtained. Due to the constraint, the original model parameters are converted in $L_{eff}$ into $\varphi$-dependent classical effective parameters. In particular, the effective  potential $U_{eff}(\varphi)$ in $L_{eff}$ has  the form 
$U_{eff}=\frac{\lambda}{4\xi^2}M_P^4\cdot F(\varphi)\cdot\tanh^4\bigl(\frac{\sqrt{\xi}\varphi}{M_P}\bigr)$, where $F(\varphi)$ is a smooth function equal to  $F(\varphi)\approx \frac{1}{2}$ for $\varphi >\sqrt{6}M_P$.
It is fundamentally important that the constant $\xi$ of non-minimal coupling to the scalar curvature can be chosen small.
 If $\xi =1/6$, then to ensure agreement  with CMB observational data,  the Higgs field self-coupling parameter $\lambda$ in the original action must be of the order of $\sim 10^{-11}$.  During cosmological evolution after the end of inflation, the decrease of $\varphi$ leads to a change in the sign of the effective Higgs mass term in $L_{eff}$. This TMSM effect provides an answer to the mystery of the Higgs potential structure and leads to spontaneous symmetry breaking. As $\varphi$ approaches VEV, the scalar function $\zeta(\varphi)$ changes in such a way that the classical TMT-effective self-coupling parameter 
$\lambda(\zeta(\varphi))$ increases by 10 orders of magnitude compared to $\lambda$, which is necessary for the implementation of the GWS theory.
 Applying the model to the very beginning of  the classical evolution of the Universe shows that under certain initial conditions, cosmological dynamics can begin with a "pathological" and even phantom regime preceding inflation.
However, if evolution begins with normal dynamics, then it proceeds only as inflation, and the problem of initial conditions for the onset of inflation does not arise.  The fermion preheating  model is described  as a preliminary study of preheatig after inflation. Mathematical and physical arguments in favor of using the TMT are presented.
}

\begin{document} 
\maketitle
\flushbottom

\section{Introduction}
\label{sec:intro}

Theoretical studies and accumulated observational data  have led to the generally accepted view that cosmological inflation was the initial stage of the classical evolution of the Universe. Despite this, one of the main problems in modern high-energy physics and cosmology remains the question of what physical object or effect drives cosmological inflation and subsequent graceful exit from it. In the  bewildering variety of existing inflation models \cite{Encyclopedia Inflationaris}, the mysterious degree of freedom responsible for inflation is ultimately described by a scalar field, usually called the inflaton. If we do not want to be satisfied with the framework of the phenomenological description, but seek to understand the nature of the inflaton based on the existing experimentally confirmed theory of particle physics, then the most suitable and, perhaps, the only candidate is the $SU(2)\times U(1)$ gauge-invariant electroweak standard model (SM). This means that the SM Higgs field must play the role of an inflaton. The corresponding scenarios are realized in  the metric-Higgs  inflation models \cite{MetricHI-1}-\cite{MetricHI-10}, as well as in the alternative approach using the Palatini formulation and developed in refs. \cite{Palatini-H-1}-\cite{Palatini-H-8}.

To compare all existing Higgs inflation models with the model proposed in this paper, we begin with a brief description of the main steps of their construction at the tree level. The electroweak SM at the tree level is defined in Minkowski space. To use
 the SM  in cosmology there is a need to reformulate  it in a general covariant form  to allow gravity to be included in the model. The next steps are to add a nonminimal coupling of the Higgs field to the scalar curvature, followed by the transition to the Einstein frame in the original action and a redefinition of the Higgs field leading to a Lagrangian with a canonically normalized field. This combination of particle physics and gravity, which we will call SM+Gravity, is the basis for how the Higgs sector of the electroweak SM is commonly used to study cosmology at energies well above the electroweak energy scale. However, the resulting models are suitable for describing inflation in accordance with the CMB data \cite{Planck}, \cite{BICEP} only if the constant of non-minimal coupling to scalar curvature is unnaturally huge: depending on the type of model, metric or Palatini, the coupling constant varies in the range $10^4\lesssim\xi \lesssim 10^8$. Nevertheless, apart from some effects due to nonminimal coupling (see, for example, review \cite{MetricHI-9}), in the SM+Gravity approach to Higgs inflation models at the tree level, the question of finding a reasonable link between low-energy particle physics and physics at the inflation energy scale remains open.
Moreover, the need to take into account quantum corrections runs into problems associated with the non-renormalizability of the resulting models, which further increases doubts about the validity of the SM+Gravity approach.

In the present paper, the Higgs inflation model is formulated in the framework of the Two-Measure Theory (TMT), which
 radically changes the approach to combining gravity with particle physics. Due to these changes, a formulation of the SM that takes into account the structure and procedure of TMT (and which we will refer to as Two-Measure Standard Model (TMSM)) can successfully overcome the above-mentioned problems of Higgs inflation.

\begin{itemize}

\item

Gravity is built into TMT in such a way that in all TMT models the equations derived from  the principle of least action describe a self-consistent system of gravity and matter, and this system of equations, represented in the Einstein frame, is valid at any energy available to classical field theory. One of the most important consequences of this feature of TMT is that, in the context of cosmology, the TMSM we study describes what the tree-level SM looks like at various stages of cosmological evolution, which is equivalent to a description at any allowed energies. The results of studying these aspects of TMSM will be presented in the next paper \cite{2-nd paper}.

\item

The radical difference between TMSM and SM+Gravity is most obvious in the way these theories describe matter in Minkowski space. In SM+Gravity, it is usually assumed that somehow the zero value of the vacuum energy is already ensured, and so the cosmological constant (CC)  in the Lagrangian is set to zero "manually". This approach means that models like Higgs+Gravity are formulated within a theory that is not self-sufficient in such a fundamental question. In TMSM, the description starts with a general coordinate-invariant initial action, including the gravitational terms, the vacuum-like terms and the SM matter. The description of matter in Minkowski space is only possible as a result of the limiting transition  in all equations of motion to the vacuum state. The latter is possible only by choosing the appropriate integration constant ${\mathcal M}$,  i.e. by fine-tuning.  Moreover, the CC problem and the method of ensuring its required value (not necessarily zero) also turns out to be built into the TMSM in a fundamentally different way compared to the SM + Gravity: if we change the integration constant ${\mathcal M}$ so that instead of zero the CC has any other value, this will lead to corresponding changes not only in the Higgs field potential (which takes place in SM+Gravity), but also 
in  {\em the values of all TMSM parameters} that are present in the equations of {\em all SM matter fields}. Therefore, despite the fact that the described fine-tuning cannot be considered a solution to the CC problem, the implementation of this fine-tuning has a significant effect on all elements of the model.

\item

{\em The main result of the present paper} is the demonstration that the Higgs+gravity sector of TMSM with  a small constant $\xi$ of non-minimal coupling to the scalar curvature allows one to construct a model of Higgs inflation that agrees well with observational CMB data. At the same time, despite the seemingly obvious expectations, there is no contradiction between this result and the fact that the TMSM also gives the correct values of the SM physical quantities at accelerator energies. Moreover, the TMSM provides a natural explanation for the mystery of the appearance of a mass term with the "wrong" sign in the Higgs potential as an effect arising during cosmological evolution due to a special property of the TMSM associated with the presence of two volume measures in the original action.

\item

{\em The second, but no less important result of this paper} concerns the problem of initial conditions for the onset of inflation and  the possible
stages of evolution preceding inflation.

\end{itemize}
Conditions that the initial kinetic and gradient energy densities of the canonically normalised scalar field  should not exceed the initial value of the potential energy density 
\begin{equation}
\rho_{kin}^{(in)}= \frac{1}{2}\dot{\varphi}_{in}^2\lesssim V(\varphi_{in}) \quad \text{and} \quad  \rho_{grad}^{(in)}=\frac{1}{2}|(\partial^k\varphi)_{in}(\partial_{k}\varphi)_{in}|\lesssim V(\varphi_{in})
\label{Cond for begin infl}
\end{equation}
are well known as the constraints neeeded for the onset of inflation. The initial understanding of this issue was developed in the first models of chaotic inflation \cite{Linde 1983}, \cite{Linde 1985}.
However, later, the need to implement flat potential models \cite{Star}, \cite{GL1}, \cite{Bardeen},   \cite{KL1}, \cite{KL2}, \cite{MetricHI-1}, in which inflation predictions agree well with the SMB observational data, turned the question of initial conditions into a problem. The seriousness and urgency of this problem became clear from the discussion in papers \cite{Stein}-\cite{singular alpha} .
A completely new approach to the problem of initial conditions for inflation based on the TMT was proposed in ref. \cite{my JCAP 2023}, where it was shown that the solution to the problem is closely related to the need to answer the question of what happened before inflation. Within the framework of the TMT model of ref. \cite{my JCAP 2023}, where the inflaton field was considered independent of the known matter fields, the problem of initial conditions for inflation turned out to be related to the Borde-Guth-Vilenkin (BGV) theorem \cite{BV1}-\cite{Vil arrows}. In the present paper, in which the inflaton field is identified with the Higgs field, we find that not only are the initial conditions (\ref{Cond for begin infl}) for inflation guaranteed to be satisfied, but also that an earlier, rather exotic stage of evolution prior  to inflation is also possible.

From the most general point of view, comparing the SM+Gravity models with TMSM, we can say that SM+Gravity belongs to the “conventional” type of models, in which the original action contains a single volume element $dV_g=\sqrt{-g}d^4x$.
In models of the Two-Measure Theory, in the  integral of the primordial action, along with terms with the standard volume measure $dV_g=\sqrt{-g}d^4x$, there are terms with an alternative, metric independent volume measure $dV_{\Upsilon}$
 constructed as the following 4-form using 4 scalar functions $\varphi_a$ ($a =1 ,..,4$)
\begin{eqnarray}
dV_{\Upsilon}=\Upsilon d^4x\equiv
\varepsilon^{\mu\nu\gamma\beta}\varepsilon_{abcd}\partial_{\mu}\varphi_{a}
\partial_{\nu}\varphi_{b}\partial_{\gamma}\varphi_{c}
\partial_{\beta}\varphi_{d}d^4x
=4!d\varphi_1\wedge d\varphi_2\wedge d\varphi_3\wedge d\varphi_4.
\label{Phi}
\end{eqnarray}
Here $\Upsilon$ is a scalar density, that is under general coordinate transformations with positive Jacobian it has the same transformation law as $\sqrt{-g}$.
In contrast to the density $\sqrt{-g}$ of the standard volume measure $dV_g$, the density $\Upsilon$ of the 4-form $dV_{\Upsilon}$ is  sign-indefinite and,
due to its continuity, can also take zero values in the general case.
A specific {\em dynamic} feature of TMT is that as a result of varying the initial action, the ratio of volume measures 
\begin{equation}
 \zeta(x)\stackrel{\mathrm{def}}{=} \frac{dV_{\Upsilon}}{dV_g}\equiv\frac{\Upsilon}{\sqrt{-g}}   
 \label{zeta}
\end{equation}
 appears in all equations of motion and it turns out  to be $\phi$-dependent. The scalar function $\zeta$ has a key role in obtaining all the new results presented in this paper.

The organization of the paper is as follows.
In section \ref{sec:TMSM},  after defining the primordial action\footnote{In "conventional" alternative gravity theories, the original action differs from the Einstein-Hilbert action only in the form of the Lagrangian, for example, by the presence of a non-minimal coupling with curvature. The transformation to the Einstein  frame, which in conventional theories is performed in the original action, simply changes the set of variables used to describe the theory, but does not change the theory itself. In TMT, the main difference is that the volume measure contains $\Upsilon$, although a modification of the Lagrangian is also possible. Therefore, the original TMT action differs from the Einstein-Hilbert action in the form of the Lagrangian {\em density}, and to bring the TMT action to the Einstein frame, the conformal transformation must exclude from the gravitational term of the action not only the non-minimal coupling, but also $\Upsilon$. In a theory obtained in this way, the constraint determining $\zeta$ cannot arise, that is, by acting in this way we would be dealing with a different theory. It is therefore incorrect to use the term "Jordan frame" in the usual sense for the set of variables in the original action of  TMT. To emphasize this distinction, we will use the terms "primordial variables", "primordial model parameters",  "primordial Lagrangian" and "primordial action" instead of "variables in the original frame",  "original model parameters", "original Lagrangian" and "original action".} and justifying the need to add a new type of vacuum-like term related to the volume measure density $\Upsilon$, we perform all the steps required by the TMT procedure. 
Having obtained the equations, we find that their consistency requires that the scalar $\zeta$ satisfy an algebraic constraint describing the dependence of $\zeta$ on the matter fields. By representing the equations in the Einstein frame and using the constraint, we arrive at the \underline{TMT-effective  equations, potential, Lagrangian, and action}\footnote{Note that this use of the term "effective" refers to the classical description and, for example, is in no way connected with the usual use of the term "effective potential" calculated taking into account quantum corrections.} 
 for the Higgs sector of the TMSM (i.e.  in the simplified TMSM, where all contributions of gauge fields and fermions are omitted). At this point, the reader wishing to understand the more fundamental reasons for the choice of the primordial action is advised to read 
 Appendix \ref{Justification}.
Analysis of the obtained (classical) TMT-effective action shows that it is applicable in the entire admissible energy spectrum: from the energies of the electroweak SM to the scale of inflation.
 In section \ref{Preliminary Higgs infl}, after choosing the model parameters that provide the required flatness of the TMT effective potential for values of the canonically normalized scalar field $\varphi>6M_P$, we study a simplified model showing its agreement  with the observational CMB data. The need to consider a simplified model (which will be justified in section \ref{impact of K-ess}) is due to the fact that the found TMT-effective action has a structure reminiscent of the K-essence models commonly used to describe the late time cosmic acceleration \cite{Kess1}-\cite{Kess4}. Important details of the shape of the TMT-effective potential and the behavior of the function $\zeta(\varphi)$ during the entire cosmological evolution are analyzed in section \ref{more about the TMT}. This allows us to understand the role of cosmological evolution in the appearance of the negative mass term in the Higgs potential.
Section \ref{orientation}, which is partly related to the paper \cite{my JCAP 2023}, focuses on the effects dictated by the need to ensure that cosmological solutions (and their initial conditions) must keep the $\zeta$ function positive.
 In section \ref{impact of K-ess} we study the influence of the K-essence type structure on the earliest stages of classical  evolution of the Universe. 
We show that the K-essence type structure may be responsible for the existence of the earliest stages of evolution, which have "pathological" dynamics.
However, if evolution starts with "normal" dynamics, then it proceeds only as inflation, and the problem of initial conditions for the onset of inflation does not arise. Finally, it is shown that the K-essence type structure does not change the results of the simplified version of the model obtained in  section \ref{Preliminary Higgs infl}, as far as the slow-roll inflation stage is concerned.
Section \ref{Further development} presents perspectives for further developments, including a summary of the main results of TMSM in ref. \cite{2-nd paper} and key findings of the preliminary preheating study.
 In  the Discussion section \ref{Discussion} we focus on   fundamentally important features of TMSM that provide the main results of the paper.
Appendix \ref{Justification} is devoted to the rationale for the TMT approach developed in this paper.
In the Appendix  \ref{append 2} it is shown that 
in the interval of $6M_P<\varphi < 14M_P$, where the TMT-effective potential  is flat  and the slow-roll approximation is valid,  the scalar $\zeta$ can vary within the range $0<\zeta \lesssim 5\cdot 10^{-3}$.

\section{Higgs sector of the Standard Model in Two-Measure Theory.
\\
 General view in the context of cosmology}
\label{sec:TMSM}

\subsection{Action}
\label{action}

Except for section \ref{Further development}, in this paper we will study  the sector of TMSM with the following primordial action
\begin{equation}
S_{primordial}=S_{gr}+ S_{H}+S_{vac},
\label{S-gr-H-VB}
\end{equation}
where
\begin{equation}
S_{gr}=\int d^4x(\sqrt{-g}+\Upsilon)\left( -\frac{M_P^2}{2}\right)\Biggl(1+\xi\frac{2|H|^2}{M_P^2}\Biggr)R(\Gamma,g)
\label{S-gr}
\end{equation} 
\begin{equation}
S_{H}=\int d^4x \left[(b_k\sqrt{-g}-\Upsilon) g^{\alpha\beta}\left(\partial_{\alpha}H\right)^{\dag}\partial_{\beta}H
 -(b_p\sqrt{-g}+\Upsilon)\lambda |H|^4-(b_p\sqrt{-g}-\Upsilon)m^2 |H|^2\right].
\label{S-H 12}
\end{equation}
\begin{equation}
S_{vac}=\int d^4x\left(-\sqrt{-g}V_1-\frac{\Upsilon^2}{\sqrt{-g}}V_2\right),
\label{S_vac}
\end{equation}
Here $M_P$  is the reduced Planck mass and $H$ is the Higgs isodoublet. We will proceed in Palatini's formulation, where $\Gamma$ denotes the affine connection, and $R(\Gamma,g)=g^{\mu\nu}R_{\mu\nu}(\Gamma)$,
$R_{\mu\nu}(\Gamma)=R^{\lambda}_{\mu\nu\lambda}(\Gamma)$ and
$R^{\lambda}_{\mu\nu\sigma}(\Gamma)\equiv \Gamma^{\lambda}_{\mu\nu
,\sigma}+ \Gamma^{\lambda}_{\gamma\sigma}\Gamma^{\gamma}_{\mu\nu}-
(\nu\leftrightarrow\sigma)$. 
Following what was said in the Introduction, all contributions from gauge fields and fermions are omitted.
In our notations, in a theory with only the volume element $\sqrt{-g}d^4x$, the model of massless scalar field non-minimally coupled  to scalar curvature
would be conformally invariant if the parameter $\xi$ were equal to $\xi=-\frac{1}{6}$.
 The model parameters $b_k>0$, $b_p>0$,
$\lambda>0$ and $m^2>0$ are chosen to be positive. If the first term in the vacuum-like contributions (\ref{S_vac}) was present in Einstein's GR,   $V_1$ would be a cosmological constant. The term with $V_2$ was first introduced in Ref. \cite{GK2}. The first reason for adding the term with $V_2$ is that the term
 $\propto \Upsilon=\zeta(x)\sqrt{-g}$, which can be expected as the contribution of quantum-gravitational effects
 to a vacuum-like action, does not contribute to the equations of motion. Therefore, the next  in powers of $\zeta$ 
vacuum-like term can be of the form $\propto \zeta^2\sqrt{-g}= \Upsilon^2/\sqrt{-g}$. 
There are also  more pragmatic  reasons. It turns out that thanks to this term,  the corresponding TMT-effective potential  and the TMT-effective action  acquire fundamentally new important properties without which the goals of this paper could not be achieved.

Given that the FLRW Universe is a model obtained by cosmological averaging, we can consider the corresponding averaging of the Higgs doublet
\begin{equation}
\langle H(x)\rangle_{cosm.av.}=\frac{1}{\sqrt{2}}\left(\begin{matrix} 0 \\ \phi(t) \end{matrix}\right), 
\label{H cosm av}
\end{equation}
and the classical homogeneous scalar field $\phi(t)$ is the only non-zero component of $\langle H(x)\rangle_{cosm.av.}$. In the model studied in this paper, $\phi(t)$ plays the role of the inflaton.

\subsection{Equations of motion in the Einstein frame}
\label{sec:Equations of motion}

Following the prescription of the TMT procedure  we start with  the equations of motion minimizing the  primordial action  (\ref{S-gr-H-VB}). 
Varying the action with respect to
 scalar functions $\varphi_{a}$ of which $\Upsilon$ is built we get 
\begin{equation}
B^{\mu}_{a}\partial_{\mu}\left[-\frac{M_P^2}{2}\Omega R(\Gamma,g)
-\frac{1}{2}g^{\alpha\beta}\phi_{,\alpha}\phi_{,\beta}-\frac{\lambda}{4}\phi^4+\frac{1}{2}m^2\phi^2
-2\zeta V_2\right]=0,
\label{varphiB}
\end{equation}
%where we used the notation
\begin{equation}
\Omega=\Omega(\phi^2)=1+\xi\frac{\phi^2}{M_P^2},
\label{Omega}
\end{equation}
\begin{equation}
B^{\mu}_{a}=\varepsilon^{\mu\nu\alpha\beta}\varepsilon_{abcd}
\partial_{\nu}\varphi_{b}\partial_{\alpha}\varphi_{c}
\partial_{\beta}\varphi_{d},
\label{B}
\end{equation}
and  the  scalar $\zeta(x)$ defined by eq.(\ref{zeta}) appears.
Since $Det (B^{\mu}_{a}) \propto\Upsilon^{3}$ it follows
that if
 \begin{equation}
\quad \textsf{everywhere} \quad \Upsilon(x)\neq 0, 
\label{Phi neq 0}
\end{equation}
the equality
\begin{equation}
-\frac{M_P^2}{2}\Omega R(\Gamma,g)-\frac{1}{2}g^{\alpha\beta}\phi_{,\alpha}\phi_{,\beta}
-\frac{\lambda}{4}\phi^4+\frac{m^2}{2}\phi^2-2\zeta V_2={\mathcal M}
\label{var varphi}
\end{equation} 
must be satisfied, where  ${\mathcal M}$ is a constant of
integration with the dimension of $(mass)^4$.

Variation with respect to $g^{\mu\nu}$ yields  the equation
\begin{eqnarray}
&&(1+\zeta)\left(-\frac{M_P^2}{2}\right)\Omega R_{\mu\nu}(\Gamma)+(b_k-\zeta)\frac{1}{2}\phi_{,\mu}\phi_{,\nu}
\nonumber
\\
&&- \frac{1}{2}g_{\mu\nu}\left[-\frac{M_P^2}{2}\Omega R(\Gamma, g) 
 +\frac{b_k}{2}g^{\alpha\beta}\phi_{,\alpha}\phi_{,\beta}-b_p\frac{\lambda}{4}\phi^4-b_p\frac{m^2}{2}\phi^2 -V_1+ \zeta^2 V_2 \right]=0
 \label{Grav.eq}
\end{eqnarray}
Using the trace of eq.(\ref{Grav.eq}) one can eliminate  the term  $-\frac{1}{2}M_P^2\Omega R(\Gamma, g)$ from eq.(\ref{var varphi}). As a result, we find that for these equations to be consistent, it is necessary that the scalar function $\zeta(x)$ satisfy the following relation
\begin{eqnarray}
&&\zeta\left[{\mathcal M}-2V_2+\frac{\lambda}{4}\phi^4-\frac{m^2}{2}\phi^2\right]
\nonumber
\\
&-&{\mathcal M}+2V_1+(2b_p-1)\frac{\lambda}{4}\phi^4+(2b_p+1)\frac{1}{2}m^2\phi^2-(1+b_k)\frac{1}{2}g^{\alpha\beta}\phi_{,\alpha}\phi_{,\beta}=0.
\label{zeta of phi and X firstly}
\end{eqnarray}
It is interesting to note that during these algebraic manipulations the term $2\zeta^2V_2$ in (\ref{zeta of phi and X firstly}) cancels out.
Following the terminology of earlier TMT models \cite{Kess4}-\cite{GK7}, we will call this a constraint.
It should be noted that in the Palatini formulation $\zeta(x)$ is not a physical degree of freedom. Therefore, when we call Eq.(\ref{zeta of phi and X firstly})  a constraint, we must keep in mind that it is different in meaning from the usual constraint in the field theory models, where it describes the relationship between dynamical degrees of freedom.

Variation of the affine connection yields the equations solution of which has the following form
\begin{equation}
\Gamma^{\lambda}_{\mu\nu}=\{^{\lambda}_{\mu\nu}\}+ (\delta^{\lambda}_{\mu}\chi,_{\nu}
+\delta^{\lambda}_{\nu}\chi,_{\mu}-
\chi,_{\beta}g_{\mu\nu}g^{\lambda\beta}),
 \label{GAM2}
\end{equation}
where $\{^{\lambda}_{\mu\nu}\}$  are the Christoffel's connection
coefficients of the metric $g_{\mu\nu}$ and
\begin{equation}
\chi(x)=\frac{1}{2}\ln\left[\bigl(1+\zeta\bigr)\Omega\right].
\label{ln in Gamma}
\end{equation}
If $\chi(x)\neq const.$ the metricity condition does not hold and consequently
geometry of the space-time with the metric $g_{\mu\nu}$ is generically non-Riemannian. In this paper, we will totally ignore a possibility to incorporate the torsion tensor, which could be an additional source for the space-time  to be different from Riemannian. 
It is easy to see that  the transformation of the metric
\begin{equation}
\tilde{g}_{\mu\nu}=\bigl(1+\zeta\bigr)\Omega g_{\mu\nu}
 \label{gmunuEin}
\end{equation}
turns  the affine connection $\Gamma^{\lambda}_{\mu\nu}$ into
the Christoffel connection coefficients of the metric
$\tilde{g}_{\mu\nu}$ and the space-time turns into (pseudo)
Riemannian. 

Gravitational equations
(\ref{Grav.eq}) expressed in terms of the metric $\tilde{g}_{\mu\nu}$ take the
canonical GR  form 
\begin{equation}
R_{\mu\nu}(\tilde{g})-\frac{1}{2}\tilde{g}_{\mu\nu}R(\tilde{g})=\frac{1}{M_P^2}T_{\mu\nu}^{(eff)}
\label{grav eq Ein}
\end{equation}
with the same Newton constant as in the original frame.
Here $R_{\mu\nu}(\tilde{g})$ and $R(\tilde{g})$ are the Ricci tensor and the scalar curvature of the metric $\tilde{g}_{\mu\nu}$, respectively.
 Therefore, gravity becomes canonical, and the set of dynamical variables using the metric $\tilde{g}_{\mu\nu}$ can be called the Einstein frame. 
$T_{\mu\nu}^{(eff)}$ on the right side of the Einstein equations (\ref{grav eq Ein})  is {\em the  TMT-effective energy-momentum tensor}
\begin{equation}
T_{\mu\nu}^{(eff)}=\frac{1}{\Omega}\Bigl(\frac{b_k-\zeta}{1+\zeta}\phi_{,\mu}\phi_{,\nu}+\tilde{g}_{\mu\nu}X_{\phi}\Bigr)
+\tilde{g}_{\mu\nu}U_{eff}^{(tree)}(\phi,\zeta(\phi);{\mathcal M}),
 \label{Tmn}
\end{equation}
where $X_{\phi}=\frac{1}{2}\tilde{g}^{\alpha\beta}\phi_{,\alpha}\phi_{,\beta}$ and
\begin{equation}
U_{eff}^{(tree)}(\phi,\zeta;{\mathcal M})=\frac{1}{\Omega^2}\left[\frac
{{\mathcal M}-V_1-V_2+(1-b_p)\frac{\lambda}{4}\phi^4-(1+b_p)\frac{m^2}{2}\phi^2}{(1+\zeta)^2}+V_2\right]
 \label{U zeta}
\end{equation}
is the $\zeta$-dependent form of the TMT-effective potential\footnote{The terms “TMT-effective energy-momentum tensor” and “TMT-effective potential” are used to denote the energy-momentum tensor and potential that appear in the  equations of motion  after  performing all the steps of the TMT procedure described above, which begins with varying the primordial TMT action and ends with the transition to the Einstein frame in the equations of motion. The superscript (tree) in $U_{eff}^{(tree)}$ is added to avoid confusion with the effective potential obtained by taking into account quantum corrections.}.
It is important to note once again that in Palatini's formalism $\zeta$ is not a dynamical variable and its value is determined by the constraint (\ref{zeta of phi and X firstly}),
 which is conveniently represented in the Einstein frame as
\begin{equation}
\zeta=\zeta(\phi, X_{\phi};\mathcal M)=\frac{\mathcal M-2V_1-(2b_p-1)\frac{\lambda}{4}\phi^4-(2b_p+1)\frac{m^2}{2}\phi^2+(1+b_k)\Omega X_{\phi}}{\mathcal M-2V_2+\frac{\lambda}{4}\phi^4-\frac{m^2}{2}\phi^2-(1+b_k)\Omega X_{\phi}}.
\label{1+ zeta no fine tun phi M}
\end{equation}
From condition (\ref{Phi neq 0}) and definition (\ref{zeta}) it follows that {\em during the entire evolution of the Universe $\zeta(x)$ cannot take a zero value and, therefore, there can only be $\zeta(x)>0$ or only $\zeta(x)<0$.}
Later on, when we use the constraint to study the dynamics of the $\phi$ field, it is fundamentally important to keep in mind that the constraint is a condition for the self-consistency of equations obtained only by variations of $\varphi_a$ and  $g_{\mu\nu}$. Therefore, {\em the origin of the restrictions imposed by the constraint on the dynamics of the $\phi$ field do not depend in any way on the dynamics of the $\phi$ field}.

To see another significant difference between similar models in TMT and in the conventional theory, it is worth writing down the result of varying the scalar field $\phi$ in the primordial action (\ref{S-gr-H-VB}):
\begin{equation}
\frac{1}{\sqrt{-g}}\partial_{\mu}\left((b_k-\zeta)\sqrt{-g}g^{\mu\nu}\partial_{\nu}\phi\right)
+(1+\zeta)\xi R(\Gamma,g)\phi+(b_p+\zeta)\lambda\phi^3+(b_p-\zeta)m^2\phi=0.
\label{phi original}
\end{equation}
In the Higgs inflation models {\em in the conventional theory (both in the metric and in the Palatini formulation), the transition to the Einstein frame performed directly in the original action} completely eliminates the non-minimal coupling of the scalar field to the scalar curvature.  {\em In the TMSM under study, as in all TMT  models, the transition to the Einstein frame (\ref{gmunuEin}) is carried out in the equations of motion} and we notice that  in the $\phi$-field equation (\ref{phi original}) presented in the Einstein frame using (\ref{gmunuEin}), the term of the non-minimal coupling $\propto \xi R(\tilde{g})\phi$ is preserved. After passing to the Einstein frame in eq.(\ref{phi original}),   the $\phi$-field equation  takes the following form
\begin{equation}
\frac{1}{\sqrt{-\tilde{g}}}\partial_{\mu}\Bigg(\frac{b_k-\zeta}{(1+\zeta)\Omega}\sqrt{-\tilde{g}}\tilde{g}^{\mu\nu}\partial_{\nu}\phi\Bigg)
+\frac{\xi}{\Omega^2}R(\tilde{g})\phi+\frac{b_p+\zeta}{(1+\zeta)^2\Omega^2}\lambda\phi^3+\frac{b_p-\zeta}{(1+\zeta)^2\Omega^2}m^2\phi
=0,
 \label{phi eq Ein frame}
\end{equation}
where $R(\tilde{g})$ is determined from the Einstein equations (\ref{grav eq Ein}) with the TMT-effective energy-momentum tensor (\ref{Tmn}), and $\zeta$ is given by the constraint (\ref{1+ zeta no fine tun phi M}). The procedure performed means that {\bf we are dealing with an explicitly formulated self-consistent system of the Higgs field and gravity.}

\subsection {Vacuum of the Higgs field in the context of cosmology}
\label{sec:Higgs vacuum}

The dependence of $U_{eff}^{(tree)}(\phi,\zeta;{\mathcal M})$, eq.(\ref{U zeta}), on $\zeta$, which, in turn, changes during the entire cosmological evolution (from inflation to transition to vacuum), plays a decisive role in all the physical results of the theory. Below, the TMT-effective potential will be  obtained using eqs.(\ref{U zeta}) and (\ref{1+ zeta no fine tun phi M}).
From now on we will assume that $V_1 < 0$, $V_2 < 0$ and $(electroweak \, scale)^4\lll|V_1|\sim |V_2|< M_P^4$ and we will see that
by choosing the integration constant $\mathcal{M}$ one can ensure that the TMT-effective potential has a minimum at some value 
$\phi |_{min} = \sigma\lll M_P$.
 One can treat $\sigma$ as  a vacuum expectation value (VEV) of $\phi$ obtained in the context of the studied cosmological model. The corresponding value $\zeta_v$ in the vacuum is obtained by
making use of the constraint (\ref{1+ zeta no fine tun phi M})
\begin{equation}
\zeta_v=\frac{\mathcal M+2|V_1|-(2b_p-1)\frac{\lambda}{4}\sigma^4-(2b_p+1)\frac{m^2}{2}\sigma^2}
{\mathcal{M}+2|V_2|+\frac{\lambda}{4}\sigma^4-\frac{m^2}{2}\sigma^2}
\label{zeta v before choice M}
\end{equation}
Then eq.(\ref{phi eq Ein frame}) reduces to an equation with a canonically normalized scalar field
\begin{equation}
\frac{1}{\sqrt{-\tilde{g}}}\partial_{\mu}\bigl(\sqrt{-\tilde{g}}\tilde{g}^{\mu\nu}\partial_{\nu}\phi\bigr)+V'_{eff}(\phi)|_{near\, vac}=0,
\label{phi eq via Vprime}
\end{equation}
where  the derivative $V'_{eff}(\phi)|_{near\, vac}$  in the vicinty of the  vacuum has the following form
\begin{eqnarray}
V'_{eff}(\phi)|_{near\, vac}&=&\frac{1}{(1+\zeta_v)(b_k-\zeta_v)}\bigl[(b_p+\zeta_v)\lambda\phi^2
+(b_p-\zeta_v)m^2\bigr]\phi
\nonumber
\\
&+&\frac{2\xi}{(b_k-\zeta_v)M_P^2}\Bigl[-\mathcal{M}+2\zeta_v|V_2|-\frac{\lambda}{4}\sigma^4+\frac{m^2}{2}\sigma^2\Bigr]\phi.
\label{Vprime near vac}
\end{eqnarray}
Here we neglected $\frac{\xi\sigma^2}{M_P^2}\lll 1$ in $\Omega$ and used eq.(\ref{var varphi}) rewritten in the Einstein frame. One should pay special attention that in order to obtain equation (\ref{phi eq via Vprime}) in canonical form, it was necessary to devide eq.(\ref{phi eq Ein frame}), considered near the vacuum, by a constant $\frac{b_k-\zeta_v}{1+\zeta_v}$.
In secs.\ref{sub general view} and \ref{pathol and normal} we will return to the discussion of the source of this question in more general case and will see that it is directly related to the K-essence type structure of the theory, which is responsible for the important results of the paper.

The value of the TMT effective potential in the  vacuum, obtained using eqs.(\ref{U zeta}) and (\ref{zeta v before choice M}), depends on the  integration constant ${\mathcal M}$
\begin{equation}
U_{eff}^{(tree)}|_{(vac)}=
\frac{{\mathcal M}^2-4V_1V_2+2(2b_p|V_2|+{\mathcal M})\frac{\lambda}{4}\sigma^4+2({\mathcal M}-2b_p|V_2|)\frac{m^2}{2}\sigma^2
+\mathcal{O}(\lambda^2\sigma^8)}
{4\left[{|V_1|+|V_2|+\mathcal M}+\mathcal{O}(\lambda \sigma^4)\right]}=\Lambda({\mathcal M}),
 \label{U vac}
\end{equation}
where we introduced the notation $\Lambda({\mathcal M})$ to emphasize that with an appropriate choice of ${\mathcal M}$, models with a very wide range of values of the cosmological constant can be realized. The natural constraint on $\Lambda$ is $0\leq\Lambda\lesssim (electroweak \, scale)^4$ and it  is achieved if ${\mathcal M}^2\approx 4V_1V_2$. 
With the choice of the integration constant\footnote{By choosing $\delta$ appropriately, one can ensure that $\Lambda =0$. This will exploit in section \ref{Preheating}. However, this is nothing more than fine tuning, that is, in this way we do not solve the CC problem.} 
\begin{equation}
 {\mathcal M}=2\sqrt{V_1V_2}(1+\delta), \qquad {\text where} \qquad \delta\sim \frac{\lambda \sigma^4}{|V_2|}\lll 1,
\label{M 2sqrt}
\end{equation}
 which will be used in what follows,  eq. (\ref{zeta v before choice M})
gives
\begin{equation}
\zeta_v=\sqrt{\frac{V_1}{V_2}}+ \mathcal{O}\left(\frac{\lambda \sigma^4}{|V_2|}\right).
\label{zeta v}
\end{equation}
We then find that near vacuum the second term in eq.(\ref{Vprime near vac}) is negligible compared to the first term:
\begin{eqnarray}
&&\Bigg|\frac{2\xi}{(b_k-\zeta_v)M_P^2}\Bigl[-\mathcal{M}+2\zeta_v|V_2|-\frac{\lambda}{4}\sigma^4+\frac{m^2}{2}\sigma^2\Bigr]\Bigg|\phi
\label{small corr to v2}
\\
&=&\frac{2\xi}{b_k-\zeta_v}\frac{\sigma^2}{M_P^2}\Bigg|-\frac{\lambda}{4}\sigma^2+\frac{m^2}{2}\Bigg|\phi \lll \frac{(b_p+\zeta_v)\lambda\phi^2+(b_p-\zeta_v)m^2}{(1+\zeta_v)(b_k-\zeta_v)}\phi.
\nonumber
\\
\end{eqnarray}
In this estimate, we took into account that the implementation of Higgs inflation studied in the next section admits a large region in the parameter space that does not contradict this estimate. Then   from eq.(\ref{Vprime near vac}) it follows that if $\zeta_v>b_p$, spontaneous symmetry breaking (SSB) occurs with
 the following expression
 for the VEV of the classical scalar field $\phi$
\begin{equation}
\sigma^2=\frac{\zeta_v-b_p}{b_p+\zeta_v}\cdot\frac{m^2}{\lambda}\left[1+ \mathcal{O}\left(\frac{\lambda\sigma^4}{|V_2|}\right)\right].
\label{v2}
\end{equation}
Eqs. (\ref{phi eq via Vprime}) and (\ref{Vprime near vac}) near vacuum obviously reduce to the form of the SM equation of the Higgs field
 \begin{equation}
\frac{1}{\sqrt{-\tilde{g}}}\partial_{\mu}\bigl(\sqrt{-\tilde{g}}\tilde{g}^{\mu\nu}\partial_{\nu}\phi\bigr)+\lambda_{\sigma}\phi^3-m_{\sigma}^2\phi=0
\label{h eq}
\end{equation}
and {\em the notations}\footnote{We use the subscript $\sigma$ here to explicitly indicate that we are dealing with vacuum values of the TMT-effective 
parameters. Indeed, one might expect that, as follows from (\ref{phi eq Ein frame}) and the constraint (\ref{1+ zeta no fine tun phi M}), 
$\phi$-dependent TMT-effective parameters $\lambda_{eff}$ and $m^2_{eff}$ appear as $\phi$ moves away from the vacuum. However, this is only partly true, namely for $\sigma\ll\phi\lesssim  M_P$. As we will see in sections \ref{sub general view} and \ref{more about the TMT}, substituting $\zeta$ from the constraint
(\ref{1+ zeta no fine tun phi M}) into eq.(\ref{U zeta}) makes the TMT-effective potential essentially non-polynomial for $\phi \gtrsim M_P$, which does not allow us to use the terms TMT-effective parameters $\lambda_{eff}$ and $m^2_{eff}$ in the general case.}
\begin{equation}
\lambda_{\sigma}=\frac{b_p+\zeta_v}{(1+\zeta_v)(b_k-\zeta_v)}\lambda,    \qquad   m_{\sigma}^2=\frac{\zeta_v-b_p}{(1+\zeta_v)(b_k-\zeta_v)}m^2
\label{lambda and m2 SM}
\end{equation}
are introduced for   the values of quartic selfinteraction and squared mass parameters near the vacuum $\sigma$.
Note that up to corrections of the order $\sim\mathcal{O}\left(\frac{\lambda\sigma^4}{|V_2|}\right)$
\begin{equation}
\sigma^2=\frac{m_{\sigma}^2}{\lambda_{\sigma}}.
\label{v2 m mv}
\end{equation}

If we restrict ourselves to studying the TMSM near vacuum, then the parameter $m$ will remain undefined, as well as the mass parameter in the GWS theory. The presence in the relations (\ref{lambda and m2 SM}) of factors depending on the model parameters $\zeta_v=\sqrt{\frac{V_1}{V_2}}$, $b_k$ and $b_p$ means that there is a very wide range of these parameters in which the required VEV SM can be ensured\footnote{However, in the model studied in this paper, we completely excluded gauge bosons and fermions and, accordingly, completely ignored the need to ensure gauge invariance. A suitably extended model that modifies the above results and allows adequate comparison with the GWS theory is studied in ref. \cite{2-nd paper}, where it is shown that the VEV of the Higgs field, which should be compared with the VEV in the GWS theory, is not the value of $\sigma$ given by
eq.(\ref{v2 m mv}), but the value of $\sqrt{\frac{b_k-\zeta_v}{1+\zeta_v}}\sigma$ (see also section \ref{concise summary}). As will be shown in ref.\cite{2-nd paper}, to demonstrate this it is necessary to take into account the differences in the description   of TMSM in the so-called cosmological  frame (CF) and in the local particle physics  frame (LPPF). From this point of view, in studying cosmology in this paper, we limit ourselves to work in the CF, where the kinetic term in the TMT-effective Lagrangian is non-canonical, see below, eq.(\ref{Leff})}. The fundamentally new effect of the model proposed in this paper is that, thanks to these three model parameters,
 {\em there is also freedom in the choice of the parameter $\lambda$.}
As we will see in section \ref{Preliminary Higgs infl}, this allows the implementation of the Higgs inflation model with a small constant $\xi$ of the non-minimal Higgs coupling to the scalar curvature.

\subsection{  General view of the TMSM
\\
from the electroweak energy scale to the inflation scale}
\label{sub general view}

In section \ref{sec:Equations of motion}, TMT-effective equations in the Einstein frame were obtained. To understand
the main features of TMSM at energies above the electroweak scale, it is necessary to return
to these equations, taking into account the results of section \ref{sec:Higgs vacuum} relating to what
happens near the Higgs vacuum in the CF (see footnote 6).
From the requirement that the TMT-effective cosmological constant be negligibly
small, we obtained that, neglecting insignificant corrections, the integration constant is
$\mathcal{M}=2\sqrt{V_1V_2}$, eq.(\ref{M 2sqrt}). 
From now on, we will use the following parametrization for   $V_2$
\begin{equation}
|V_2|=(qM_P)^4, \quad q\leq 1.
\label{V2 via q4}
\end{equation}
Then the constraint (\ref{1+ zeta no fine tun phi M}) determines $\zeta=\zeta(\phi, X_{\phi})$ as follows
\begin{equation}
\zeta=\frac{2q^4\zeta_v(1+\zeta_v)-(2b_p-1)\frac{\lambda}{4}\frac{\phi^4}{M_P^4}-
(2b_p+1)\frac{m^2}{2M_P^2}\frac{\phi^2}{M_P^2}+(1+b_k)\Omega \frac{X_{\phi}}{M_P^4}}{2q^4(1+\zeta_v)+
\frac{\lambda}{4}\frac{\phi^4}{M_P^4}-\frac{m^2}{2M_P^2}\frac{\phi^2}{M_P^2}-(1+b_k)\Omega \frac{X_{\phi}}{M_P^4}}.
\label{constraint using M and zetav}
\end{equation}
Substituting $\zeta(\phi, X_{\phi})$ into the $\zeta$-dependent form of $T_{\mu\nu}^{(eff)}$ in eqs.(\ref{Tmn}) and (\ref{U zeta}) with
subsequent simple but rather lengthy algebraic calculations,  the expression for the TMT-effective energy-momentum tensor reduces to 
\begin{equation}
T_{\mu\nu}^{(eff)}=\frac{K_1(\phi)}{\Omega}\cdot\big(\phi_{,\mu}\phi_{,\nu}
-\tilde{g}_{\mu\nu}X_{\phi}\big)-K_2(\phi)\cdot \frac{X_{\phi}}{M_P^4}\Big(\phi_{,\mu}\phi_{,\nu}
-\frac{1}{2}\tilde{g}_{\mu\nu}X_{\phi}\Big)
+\tilde{g}_{\mu\nu}U_{eff}^{(tree)}(\phi),
 \label{Tmn 0 phi K1 K2}
\end{equation}
where the TMT effective potential $U_{eff}^{(tree)}(\phi)$ and
 functions $K_1(\phi)$ and $K_2(\phi)$
are defined by the following expressions:
\begin{equation}
U_{eff}^{(tree)}(\phi)=\frac{q^4\Bigl[(\zeta_v+b_p)\lambda\phi^4-2(\zeta_v-b_p)m^2\phi^2 \Bigr]+\frac{1}{M_P^4}\bigl(\frac{\lambda}{4}\phi^4-\frac{m^2}{2}\phi^2\bigr)^2}{4\Omega^2\left[q^4(1+\zeta_v)^2+(1-b_p)\frac{\lambda}{4}\frac{\phi^4}{M_P^4}-(1+b_p)\frac{m^2\phi^2}{2M_P^4}\right]},
\label{V eff}
\end{equation}

\begin{equation}
K_1(\phi)=\frac{2q^4(1+\zeta_v)(b_k-\zeta_v)+(2b_p+b_k-1)\frac{\lambda}{4}\frac{\phi^4}{M_P^4}+(2b_p+1-b_k)\frac{m^2\phi^2}{2M_P^4}}
{2\left[q^4(1+\zeta_v)^2+(1-b_p)\frac{\lambda}{4}\frac{\phi^4}{M_P^4}-(1+b_p)\frac{m^2\phi^2}{2M_P^4}\right]},
\label{K1 phi}
\end{equation}

\begin{equation}
K_2(\phi)=\frac{(1+b_k)^2}{2\left[q^4(1+\zeta_v)^2+(1-b_p)\frac{\lambda}{4}\frac{\phi^4}{M_P^4}-(1+b_p)\frac{m^2\phi^2}{2M_P^4}\right]}.
\label{K2 phi}
\end{equation}

Instead of the system of the Einstein equation (\ref{grav eq Ein}) with the energy-momentum tensor
(\ref{Tmn 0 phi K1 K2})-(\ref{K2 phi})   and the field $\phi$-equation obtained by substituting $\zeta$ from the constraint (\ref{constraint using M and zetav}) into eq.(\ref{phi eq Ein frame}), it is much more convenient to work with the
TMT-effective action, the variation of which gives these equations. As usual, if $\tilde{g}^{\alpha\beta}\phi_{,\alpha}\phi_{,\beta}>0$,  the energy-momentum tensor $T_{\mu\nu}^{(eff)}$ can be rewritten in the form of a perfect fluid. Then the pressure density plays the role of the matter Lagrangian  in the effective action, and we arrive
at the {\em tree level TMT-effective action}
\begin{equation}
S_{eff}^{(TMT)}=\int \left(-\frac{M_P^2}{2}R(\tilde{g})+L_{eff}^{(tree)}(\phi, X_{\phi}))\right)\sqrt{-\tilde{g}}d^4x,
\label{Seff}
\end{equation}
with the following tree level TMT-effective Lagrangian
\begin{equation}
L_{eff}^{(tree)}(\phi, X_{\phi})=K_1(\phi)\frac{X_{\phi}}{\Omega}-\frac{1}{2}K_2(\phi)\frac{X_{\phi}^2}{M_P^4}-U_{eff}^{(tree)}(\phi).
\label{Leff}
\end{equation}
Note that $L_{eff}^{(tree)}(\phi, X_{\phi}))$ has a form typical for K-essence models \cite{Kess1}-\cite{Kess4}, and a manifestation of this is also the non-canonical form of the kinetic term in the $\phi$ equation (\ref{phi eq Ein frame}).
 To extract the canonical kinetic term the equation must be divided by $\frac{b_k-\zeta}{(1+\zeta)\Omega}$.
When studying the Higgs field equation in the vicinity of a vacuum in section \ref{sec:Higgs vacuum}, in the paragraph after eq.(\ref{Vprime near vac}), we already encountered this feature, where, however, we had  $\zeta=\zeta_v=const$ and $\Omega=1$.

In conclusion of this section, it can be stated that in the model under consideration, {\em the equations and the TMT-effective (classical) action  are applicable in the energy range from the electroweak energy scale (already considered in section \ref{sec:Higgs vacuum}) to the inflationary one}, which we study in the next section.

\section{ The Two-Measure SM  with small $\xi$ and Higgs inflation.
\\
 Preliminary consideration}
\label{Preliminary Higgs infl}

\subsection{About the shape of the TMT-effective potential and other features of the model}
\label{about the shape}

The model contains 6 parameters $V_1$, $V_2$, $b_p$, $b_k$, $\lambda$ and $m$. Instead of $V_2<0$ it is convenient
to use parameter $q^4\leq 1$ defined by the relation (\ref{V2 via q4}). Instead of $V_1<0$ we use $\zeta_v$ defined by eq.(\ref{zeta v}).
It is natural to assume that $V_1$ and $V_2$ have the same order.  Further in the paper, for numerical estimates and plotting graphs, we will need to select the values of $V_1$ and $V_2$. For simplicity, we choose 
\begin{equation}
V_1=V_2 \quad \Longleftrightarrow \quad \zeta_v\equiv\sqrt{\frac{V_1}{V_2}}=1.
 \label{V1=V2}
\end{equation}
Bearing in mind that the factor $\frac{V_1}{V_2}\sim\mathcal{O}(1)$ can lead to minor changes in the values of the calculated parameters, we can state that this choice is made without loss of generality.  As a
specific example of small $\xi$ from now on, we will always use $\xi=\frac{1}{6}$. We also choose
\begin{equation}
b_p=\frac{1}{2}(1+\delta_p) \quad 0<\delta_p\ll 1
 \label{bp}
\end{equation}
The impact of the choice $b_p\approx\frac{1}{2}$ and the value of $\delta_p$ will become clear in sections \ref{more about the TMT} and \ref{sub initial}.

To apply the model to the study of the inflationary epoch, we would like to change
field $\phi$ to the canonically normalized scalar field. For this to be done it is usually enough
to redifine $\phi$ by means of the relation
\begin{equation}
\phi=\frac{M_P}{\sqrt{\xi}}{\sinh}\left(\sqrt{\xi}\frac{\varphi}{M_P}\right).
\label{phi via canonical varphi}
\end{equation}
However, as can be seen from eq.(\ref{Leff}), the redefinition (\ref{phi via canonical varphi}) absorbs only the factor $\Omega^{-1} $ in front of the first term, but does not provide canonical normalization of the field $\varphi$.
The reason is that the TMT-effective action (\ref{Seff}), (\ref{Leff}) has the structure of K-essence type models, 
which, in turn, is due to the presence in the primordial TMSM action   of two volume elements $dV_g$ and $dV_{\Upsilon}$ .
  Thus, we are limited to only the redefinition (\ref{phi via canonical varphi}). The TMT-effective energy-momentum tensor given by eq.(\ref{Tmn 0 phi K1 K2}) now takes the following form
\begin{equation}
T_{\mu\nu}^{(eff)}=K_1(\varphi)\cdot\big(\varphi_{,\mu}\varphi_{,\nu}
-\tilde{g}_{\mu\nu}X_{\varphi}\big)-{\tilde K}_2(\varphi)\cdot \frac{X_{\varphi}}{M_P^4}\Big(\varphi_{,\mu}\varphi_{,\nu}
-\frac{1}{2}\tilde{g}_{\mu\nu}X_{\varphi}\Big)
+\tilde{g}_{\mu\nu}U_{eff}^{(tree)}(\varphi)
 \label{Tmn 0 phi K1 K2 varphi}
\end{equation}
and we obtain the TMT-effective action in terms of $\varphi$
\begin{equation}
S_{eff}^{(TMT)}=\int \left(-\frac{M_P^2}{2}R(\tilde{g})+L_{eff}^{(tree)}(\varphi,X_{\varphi})\right)\sqrt{-\tilde{g}}d^4x
\label{Seff varphi}
\end{equation}
with the following tree-level TMT-effective Lagrangian rewritten in terms of  $\varphi$ and $X_{\varphi}$
\begin{equation}
L_{eff}^{(tree)}(\varphi,X_{\varphi})=K_1(\varphi)X_{\varphi}-\frac{1}{2}{\tilde K}_2(\varphi)\frac{X_{\varphi}^2}{M_P^4}
-U_{eff}^{(tree)}(\varphi),
\label{Leff varphi}
\end{equation}
where
\begin{equation}
U_{eff}^{(tree)}(\varphi) =\frac{\lambda M_P^4}{4\xi^2}{\tanh}^4z\cdot F(z),
\label{Veff varphi tanh no fine tun 1}
\end{equation}
\begin{equation}
F(\varphi)=\frac{(\zeta_v+b_p)q^4+
\frac{\lambda}{16\xi^2}\sinh^4z-\frac{\lambda m^2}{4\xi M_P^2}\sinh^2z+\frac{m^4}{4\lambda M_P^4}
-2 (\zeta_v-b_p)q^4\frac{\xi m^2}{\lambda M_P^2}\cdot \sinh^{-2}z }
{(1+\zeta_v)^2q^4+(1-b_p)\frac{\lambda}{4\xi^2}{\sinh}^4z -(1+b_p)\frac{m^2}{2 M_P^2\xi}\sinh^2z}
\label{Veff varphi tanh no fine tun 2}
\end{equation}
\begin{equation}
z=\sqrt{\xi}\frac{\varphi}{M_P}, \qquad X_{\varphi}=\frac{1}{2}\tilde{g}^{\alpha\beta}\varphi_{\alpha}\varphi_{\beta},
\label{z X}
\end{equation}
\begin{equation}
K_1(\varphi)=\frac{2q^4(1+\zeta_v)(b_k-\zeta_v)+(2b_p+b_k-1)\frac{\lambda}{4\xi^2}\sinh^4z
+(2b_p+1-b_k)\frac{m^2}{2 M_P^2\xi} \sinh^{2}z}
{2\Bigl[q^4(1+\zeta_v)^2+(1-b_p)\frac{\lambda}{4\xi^2}\sinh^4z-(1+b_p)\frac{m^2}{2M_P^2\xi} \sinh^{2}z\Bigr]},
\label{K1 varphi}
\end{equation}
\begin{equation}
{\tilde K}_2(\varphi)=\frac{(1+b_k)^2\cosh^4z}{2\Bigl[q^4(1+\zeta_v)^2+(1-b_p)\frac{\lambda}{4\xi^2}\sinh^4z-(1+b_p)\frac{m^2}{2M_P^2\xi} \sinh^{2}z\Bigr]}.
\label{K2 varphi}
\end{equation}
The last expression for ${\tilde K}_2(\varphi)$ takes into account that $X_{\phi}=\cosh^2z\cdot X_{\varphi}$.

The constraint (\ref{constraint using M and zetav}) represented through $\varphi$ becomes
\begin{eqnarray}
&&\zeta(\varphi, X_{\varphi})
\label{zeta via varphi}
\\
&=&\frac{2q^4\zeta_v(1+\zeta_v)-(2b_p-1)\frac{\lambda}{4\xi^2}\sinh^4z-
(2b_p+1)\frac{m^2}{2\xi M_P^2}\sinh^2z+(1+b_k)\cosh^4z \cdot\frac{X_{\varphi}}{M_P^4}}
{2q^4(1+\zeta_v)+\frac{\lambda}{4\xi^2}\sinh^4z-\frac{m^2}{2\xi M_P^2}\sinh^2z-(1+b_k)\cosh^4z \cdot\frac{X_{\varphi}}{M_P^4}},
\nonumber
\end{eqnarray}
Subsequently, it will be shown that in the process of inflation, which occurs in the slow-roll
regime, the difference from the case of a canonically normalized scalar field turns out to be
insignificant.

Despite the fact that in the model under study the TMT-effective potential $U_{eff}^{(tree)}(\varphi)$ has a
plateau-like character, there are two significant circumstances, without a detailed analysis
of which it is impossible to obtain sufficiently definite quantitative predictions for comparing
them with observational cosmological data. The first of them is the difference in the form
of the function $U_{eff}^{(tree)}(\varphi)$  from those known in the literature. The second is due to the fact
that TMT-effective action has a K-essence type structure. Without an adequate response
to these problems, it would be impossible, for example, to ensure that inflation proceeds in
a slow-roll regime. As will be shown in the following sections, there is a fairly wide range
of model parameters in which the influence of the K-essence structure on the predicted
inflation parameters turns out to be extremely insignificant. A noticeable difference from
the predictions of a model with the same potential, but with a standard, non-K-essence
structure can only exist  at the end of inflation, which will be discussed in section \ref{K slow roll}. Below,
in this section, we will completely ignore the K-essence structure, and therefore the results
obtained should be considered as preliminary.

\begin{figure}
\includegraphics[width=13.0cm,height=8cm]{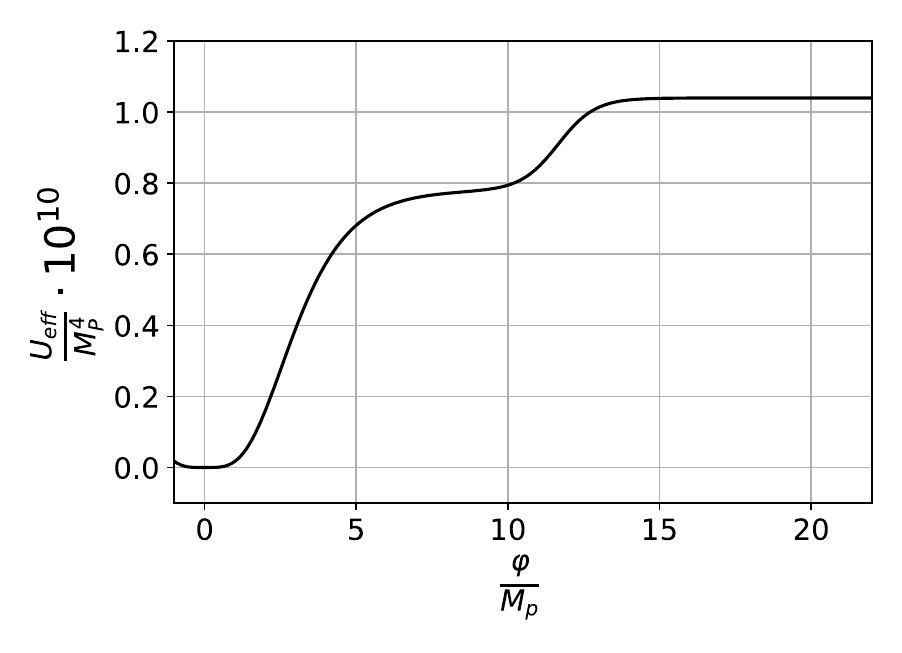}
%\end{capture}
\caption{The graph of the TMT-effective  potential $U_{eff}^{(tree)}(\varphi)$, defined by eqs.(\ref{Veff varphi tanh no fine tun 1}) and (\ref{Veff varphi tanh no fine tun 2}) for the model parameters $\lambda = 2.3\cdot 10^{-11}$, \,
 $\xi=\frac{1}{6}$, \,  $m=0.7GeV$, \, $b_p\approx 0.5$ and $q^4=3\cdot 10^{-4}$. This choice of $q^4$ means that  $|V_1|= |V_2|\approx(10^{16}GeV)^4$, and this leads to a potential with two clearly defined plateaus   (see also eqs.(\ref{1-st plateau}) and (\ref{2-nd plateau}))}
\label{fig1}
\end{figure}

To get an idea of the shape of the TMT-effective potential $U_{eff}^{(tree)}(\varphi)$, we note that its formula
differs by the factor $F(\varphi)$, eq.(\ref{Veff varphi tanh no fine tun 2}), from the corresponding formulas for the potential
in the T-model with n = 2 in the
notations of ref. \cite{KL1} and for the form of the Eistein-frame potential of the Palatini inflation models \cite{Palatini-H-1} (see also reviews
\cite{Palatini-H-7}, \cite{Palatini-H-8} and references therein). It is well
known that to ensure agreement with Planck's data, the height of the potential plateau
should be of the order $\sim 10^{-10}M_P^4$.
Therefore, depending on the relationship between
parameters $q^4$ and $\frac{\lambda}{4\xi^2}$ two types of TMT potential shape are possible: with one plateau
and with two plateaus.
An intermediate form is also possible, when the presence of a second
plateau is weakly expressed, but we will consider only cases with clearly defined one or two
plateaus.
When $\varphi>\frac{M_P}{\sqrt{\xi}}$ is such that numerator and denominator in $F(\varphi)$ are dominated by the terms $\propto \sinh^4z$,
then $F(\varphi)\approx \frac{1}{4(1-b_p)}$ and the TMT-effective potential $U_{eff}^{(tree)}(\varphi)$ has one plateau
\begin{equation}
U_{eff}^{(tree)}(\varphi)\Big|_{\varphi\gg\frac{M_P}{\sqrt{\xi}}} =
\frac{\lambda M_P^4}{16(1-b_p)\xi^2}{\tanh}^4\Bigl(\sqrt{\xi}\frac{\varphi}{M_P}\Bigr)
\label{1-st plateau}
\end{equation}
However, generically it is possible that for lower values of $\varphi>\frac{M_P}{\sqrt{\xi}}$, such that 
$(\zeta_v+b_p)q^4>\frac{\lambda}{16\xi^2}\sinh^4\Bigl(\sqrt{\xi}\frac{\varphi}{M_P}\Bigr)$ and $(1+\zeta_v)^2q^4>(1-b_p)$,
 there is a second plateau\footnote{A similar situation can also be seen in ref. \cite{my JCAP 2023}} described
mainly by
\begin{equation}
U_{eff}^{(tree)}(\varphi)\Big|_{\varphi >\frac{M_P}{\sqrt{\xi}}} =\frac{\lambda(\zeta_v+b_p) M_P^4}{4(1+\zeta_v)^2\xi^2}{\tanh}^4\Bigl(\sqrt{\xi}\frac{\varphi}{M_P}\Bigr).
\label{2-nd plateau}
\end{equation}
 A typical shape of the TMT-effective potential $U_{eff}^{(tree)}(\varphi)$ with two plateaus is shown in figure \ref{fig1}.
 Apparently the
second plateau must be responsible for the last stage of inflation, and the potential with two plateaus may be of interest as another type of inflation pattern. However,  in this paper we will not consider this possibility in detail, but will focus on studying the model in the parameter range, where the effect of the existence of a second potential plateau does not manifest itself explicitly.

\begin{figure}
\includegraphics[width=13.0cm,height=8cm]{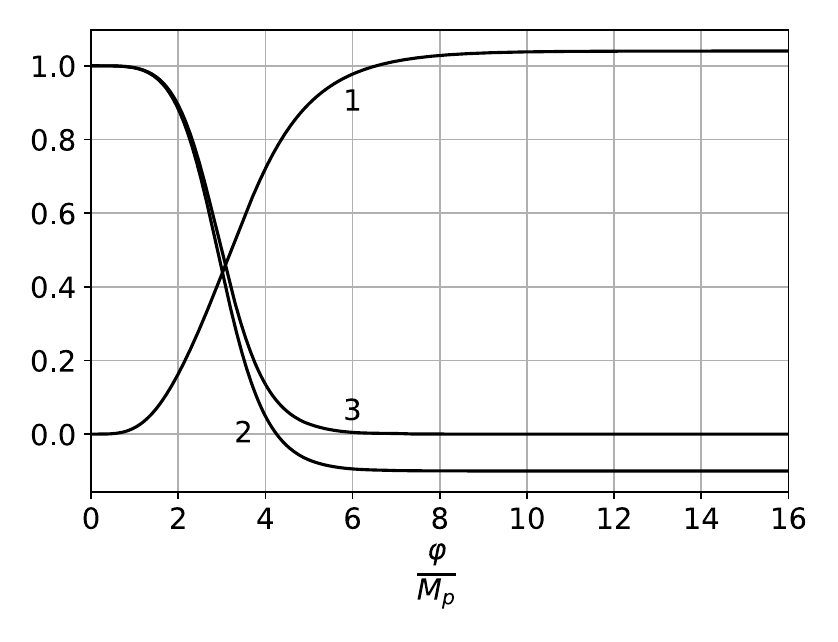}
%\end{capture}
\caption{
The curve labeled 1 is a plot of the TMT-effective potential $\frac{U_{eff}^{(tree)}(\varphi)}{M_P^4}\cdot 10^{10}$, defined by eqs.(\ref{Veff varphi tanh no fine tun 1})
 and (\ref{Veff varphi tanh no fine tun 2}). It has one plateau due to the choice of $q^4=3\cdot 10^{-10}$, i.e. $|V_1|=|V_2|\approx (10^{16}GeV)^4$, while other model parameters are the same as in Fig. 1.
The curves labeled 2 and 3 are plots of the scalar function $\zeta(\varphi)$ defined by the constraint (\ref{zeta via varphi}), where the terms $\propto \frac{X_{\varphi}}{M_P^4}$ are assumed to be negligible compared to the other terms. Curve 2 is the plot of $\zeta(\varphi)$ if the model parameter $b_p=0.55$ is chosen, and the plot intersects $\zeta=0$ at $\varphi=4.25M_P$. Curve 3 is the plot of $\zeta(\varphi)$ if the model parameter $b_p=0.5(1+10^{-8})$ is chosen. In this case, curve 3 intersects $\zeta=0$ at $\varphi\approx 14M_P$.}
\label{fig2}
\end{figure}

\begin{itemize}

\item{\em The TMT-effective potential with one plateau}

Let us now analyze in more detail the TMT-effective potential $U_{eff}^{(tree)}(\varphi)$ with one plateau. 
With the choice described by eqs.(\ref{V1=V2}) and (\ref{bp}), and using the parameter values $\lambda = 2.3\cdot 10^{-11}$, \,
 $\xi=\frac{1}{6}$ tested in sec.\ref{sub Higgs slow preliminary}, and
 $m=0.7GeV$, one can see that the TMT-effective potential  with one plateau is reached\footnote{For $\varphi>\frac{M_P}{\sqrt{\xi}}=\sqrt{6}M_P$, we find that  $F(\varphi)\approx \frac{1}{2}$ with a relative error $\frac{|F(\varphi)-\frac{1}{2}|}{1/2}<2.2\cdot 10^{-3}$.}  if $q^4\lesssim 10^{-7}$.
 For further analysis of the model, we choose
\begin{equation} 
q^4=3\cdot 10^{-10}, \quad \text{that is} \quad V_1=V_2\approx -(10^{16}GeV)^4.
\label{q4 10-10}
\end{equation}
 The graph of the TMT-effective potential $U_{eff}^{(tree)}(\varphi)$, given by eqs.(\ref{Veff varphi tanh no fine tun 1}) and 
(\ref{Veff varphi tanh no fine tun 2}), is shown in figure \ref{fig2}.
 Substituting $\xi=\frac{1}{6}$ and $b_p=\frac{1}{2}$ into (\ref{1-st plateau}), we find that for $\varphi>6M_P$ the TMT-effective potential  is described with high accuracy by the formula
\begin{equation}
U_{eff}^{(tree)}(\varphi)|_{\varphi>6M_P}=\frac{\lambda M_P^4}{8\xi^2}{\tanh}^4\Bigl(\frac{\varphi}{\sqrt{6}M_P}\Bigr)\approx
 \frac{\lambda M_P^4}{8\xi^2}\Bigl(1-8e^{-\sqrt{\frac{2}{3}}\frac{\varphi}{M_P}} \Bigr).
\label{1-st plateau 1}
\end{equation}
 Note that due to $F(\varphi)\approx \frac{1}{2}$ at $\varphi>6M_P$, the formula for the TMT-effective potential (\ref{1-st plateau 1}) differs from  the    form of the potential in the Einstein frame in the Palatini-Higgs models \cite{Palatini-H-1}, \cite{Palatini-H-7}, \cite{Palatini-H-8} by a factor $\frac{1}{2}$.
But the fundamental difference from the Palatini-Higgs models is that here $\xi=\frac{1}{6}$. In this respect, the potential of the T-model \cite{KL1} is more suitable for comparison: the TMT-effective potential $U_{eff}^{(tree)}(\varphi)|_{\varphi>6M_P}$ coincides with the T-model potential in the region 
$\varphi>6M_P$ for the case n=2 in the notation of the paper \cite{KL1}.

\end{itemize}

\subsection{Higgs inflation. Preliminary consideration of the slow-roll approximation}
\label{sub Higgs slow preliminary}
 
Continuing our preliminary consideration, let us start with the $\varphi$ field equation and the
Friedmann equation in the spatially flat FLRW Universe completely ignoring the K-essence type structure of the TMT-effective action
(\ref{Seff varphi}), (\ref{Leff varphi})
\begin{equation}
\ddot{\varphi}+3\frac{\dot{a}}{a}\dot{\varphi}+U_{eff}^{(tree)\, \prime}=0,
\label{FLRW varphi}
\end{equation} 
\begin{equation}
\frac{\dot{a}^2}{a^2}=\frac{1}{3M_P^2}\Bigl(\frac{1}{2}\dot{\varphi}^2+ U_{eff}^{(tree)}(\varphi)\Bigr)
\label{Frid prelim}
\end{equation} 
An overdot denotes $d/dt$ and a prime denotes $d/d\varphi$. In slow-roll approximation
$\frac{1}{2}\dot{\varphi}^2\ll U_{eff}$ and $|\ddot{\varphi}|\ll |U_{eff}^{(tree)\, \prime}|\approx 3H|\dot{\varphi}|$
these equations reduce to
\begin{equation}
3\frac{\dot{a}^2}{a^2}\dot{\varphi}\simeq U_{eff}^{(tree)\, \prime}(\varphi),
\label{FLRW varphi slow}
\end{equation} 
\begin{equation}
\frac{\dot{a}^2}{a^2}\simeq\frac{1}{3M_P^2} U_{eff}^{(tree)}(\varphi),
\label{Frid prelim slow}
\end{equation} 
from where it follows that the flatness conditions imposed on $U_{eff}^{(tree)}(\varphi)$
\begin{equation}
\epsilon=\frac{M_P^2}{2}\Biggl(\frac{U_{eff}^{(tree)\, \prime}}{U_{eff}^{(tree)}}\Biggr)^2\ll 1, \qquad |\eta|=M_P^2\Biggl|\frac{U_{eff}^{(tree)\, \prime\prime}}{U_{eff}^{(tree)}} \Biggr|\ll 1
\label{epsilon eta prelim}
\end{equation} 
must be satisfied. In sec.\ref{impact of K-ess} we will study in detail the modification of the flatness conditions
if the K-essence type structure is taken into account.

 At this stage of the study,  our main goal is to understand how  Higgs inflation with a small model parameter $\xi$ of nonminimal coupling   becomes possible in  the  TMSM. First of all, choosing the value of $\xi=\frac{1}{6}$ we must find the value of the primordial parameter $\lambda$ in the TMT-effective potential (\ref{1-st plateau 1}) in such a way that the model predictions agree with the CMB data. In light of what was said at the very end of the previous subsection, the predictions for the values of all perturbation parameters in the model under study must coincide with the corresponding predictions of the T-model. Therefore, to find $\lambda$, it is sufficient, for example, to consider the prediction for scalar perturbations $\mathcal{A}_s$.
Using the standard  method for estimating the duration of inflation, measured by the number of e-folds $N$, we obtain
\begin{equation}
\epsilon\approx\frac{1}{8\xi N^2} \qquad \text{and} \qquad \eta\approx -\frac{1}{N}.
\label{epsilon via N}
\end{equation}
The power spectrum of scalar perturbations $\mathcal{A}_s$, measured in the CMB at large scales, can be expressed  \cite{Lyth Riotto}   in terms of
 the slow-roll parameter $\epsilon$   as  $\mathcal{A}_s\approx\frac{1}{24\pi^2\epsilon}\frac{U_{eff}}{M_P^4}$,
and for large $N$ we get
\begin{equation}
\mathcal{A}_s\approx\frac{\lambda N^2}{24\pi^2\xi}.
\label{As via xi and lambda}
\end{equation}
With  the observational  data bound \cite{Planck} $\mathcal{A}_s\approx 2.1\cdot 10^{-9}$  we get the relation
\begin{equation}
\lambda\approx 5\cdot 10^{-7}\frac{\xi}{N^2}.
\label{lambda via xi}
\end{equation}
For $N=60$, with our choice of parameter $\xi=\frac{1}{6}$, we obtain
\begin{equation}
\lambda\approx 2.3\cdot 10^{-11}.
\label{lambda 10 -11}
\end{equation}

With $\xi=\frac{1}{6}$, for the scalar spectral index $n_s=1-6\epsilon +2\eta$ and the tensor-to-scalar ratio $r=16\epsilon$ we find for $N=60$
\begin{equation}
n_s=1-\frac{2}{N}-\frac{3}{4\xi N^2}\approx 0.965 \qquad \text{and} \qquad r=\frac{2}{\xi N^2}\approx 0.003,
\label{ns r}
\end{equation}
which are in good agreement with the observational results of the Planck 2018 survey \cite{Planck} and the BICEP/Keck measurement \cite{BICEP}.

Now, staying within the framework of a preliminary study, we will take the final step to demonstrate how the TMSM model proposed in this paper allows us to implement the Higgs inflation model with $\xi=\frac{1}{6}$ while satisfying the observed cosmological data due to $\lambda\sim 10^{-11}$, on the one hand, and to satisfy the requirement of $\lambda\sim 0.1$ near vacuum imposed by the SM, on the other hand.
This is possible due to the arbitrariness of the primordial parameter $\lambda$ in TMSM, as discussed at the end of section \ref{sec:Higgs vacuum}. Indeed, as can be seen from eqs.(\ref{lambda and m2 SM}) and (\ref{v2 m mv}), to obtain the desired VEV  it suffices to choose  the model parameter $b_k$ such that the smallness of $b_k-\zeta_v=b_k-1$ compensates for the smallness of $\lambda$.  But as already noted in footnote 5, in ref. \cite{2-nd paper} it will be shown that the VEV of the Higgs field, which should be compared with the VEV in the SM, differs from the value of $\sigma$ given by eq.(\ref{v2 m mv}).  However, in this paper, the exact value of $b_k-1$ does not play a significant role. Therefore, here it is sufficient to restrict ourselves to the order of magnitude of the result obtained in ref. \cite{2-nd paper}:  $b_k-\zeta_v\sim  10^{-5}$.
Thus, our goal is achieved by simply choosing a value for the parameter $b_k$ very close to
$\zeta_v=1$. Turning to how the primordial action (\ref{S-gr-H-VB}), (\ref{S-H 12}) is structured, we understand
that such a choice seems quite natural. 
Indeed, if we recall that the volume measure with which the kinetic term enters the  primordiial action  (\ref{S-H 12}), $(b_k\sqrt{-g}-\Upsilon)d^4x=
\sqrt{-g}(b_k-\zeta)d^4x$, the extremely small deviation of $b_k$ from the vacuum value of $\zeta$, i.e. $\zeta_v$, can be interpreted as the effect of the violation of some unknown symmetry of a more fundamental theory.

\section{More about the $\zeta(\phi)$ function and the TMT-effective potential.
\\
TMSM's answer to the mystery of the Higgs potential structure}
\label{more about the TMT}

With the obtained value of the parameter $\lambda$, eq.(\ref{lambda 10 -11}), we return to the discussion of the shape of the TMT-effective potential $U_{eff}^{(tree)}(\varphi)$, eqs.(\ref{Veff varphi tanh no fine tun 1}) and (\ref{Veff varphi tanh no fine tun 2}), the graph of which in the entire region $\varphi\geq 0$ is shown\footnote{Remember that we chose $q^4=3\cdot 10^{-10}$ which ensures the TMT potential with one plateau.} in figure \ref{fig2}. 
Two graphs of $\zeta(\varphi)$ for the case of $X_{\varphi}=0$ are also presented in figure \ref{fig2}; for these two graphs, in addition to the parameters listed above, two corresponding values of $\delta_p=0.1$ and $\delta_p= 10^{-8}$ are chosen in the definition $b_p=0.5(1+\delta_p)$ given by eq.(\ref{bp}).   As will be explained in section \ref{orientation}, for any admissible values of $\varphi$ the function $\zeta$ must be positive. Therefore, from the shape of curve 2 in figure \ref{fig2} it follows that the choice $\delta_p=0.1$ is unsatisfactory. It should be noted that the graph of $\zeta(\varphi)$ for $\delta_p= 10^{-8}$, represented by curve 3 in figure \ref{fig2}, is to some extent illustrative, since the case $X_{\varphi}=0$ is unrealistic in general.
 However, as shown in appendix \ref{append 2}, in the regime of slow-roll inflation, which occurs on the plateau part of $U_{eff}^{(tree)}(\varphi)$, i.e., for $\varphi > 6M_P$, the contribution of $X_{\varphi}$ to the constraint (\ref{zeta via varphi}) is bounded from above by the following estimate
$\frac{X_{\varphi}}{M_P^4}\approx 128\lambda \cdot e^{-2\sqrt{\frac{8}{3}}\frac{\varphi}{M_P}} <9\cdot 10^{-18}$. Therefore, for $\varphi > 6M_P$, the curve 3  in figure \ref{fig2}  can be considered as quite reliable graph for  $\zeta$. In the opposite region, where $\varphi \lll M_P$, one can expect that $X_{\varphi}\lll M_P^4$ and therefore the $\zeta(\varphi)$ graph is also reliable. After inflation ends, the contribution of $X_{\varphi}$ to the constraint (\ref{zeta via varphi}) may increase, but we will assume that it does not lead to qualitative changes\footnote{The contribution of $X_{\phi}/M_P^4$ to the constraint (\ref{constraint using M and zetav}) is negligible if $\frac{\dot{\phi}^2}{M_P^4}\ll 3.6\cdot 10^{-9}$}. From the graph with 
$\delta_p= 10^{-8}$ it follows that
\begin{itemize}
\item
 inflation can start from sufficiently large initial values of the scalar field $\varphi \approx 14M_P$;
\item
 the decrease in $\varphi$ at the inflationary stage is accompanied by a monotonous increase in
 $\zeta(\varphi)$ from positive values very close to zero\footnote{The fundamental significance of the condition $\zeta>0$ will be the subject of study in section \ref{orientation}.}.
\end{itemize}
 For 
$\varphi \ll M_P$, with a decrease in $\varphi$, $\zeta(\varphi)$ also increases monotonically, asymptotically approaching the value $\zeta=\zeta_v=1$.

The representation of $U_{eff}^{(tree)}(\varphi)$ in figure \ref{fig2} does not allow to see important details of the shape, which  differ greatly  in  the regions 
$\varphi> M_P$ and $\varphi\ll M_P$. This is due to the greatly  different behaviour of the factor $F(\varphi)$, eq.(\ref{Veff varphi tanh no fine tun 2}), in  the regions $\varphi> M_P$ and $\varphi\ll M_P$. The analysis simplified if we use the value of the primordial model parameter $m\approx 0.7 GeV$ obtained in ref. \cite{2-nd paper}, that is, $\frac{m^2}{M_P^2}\approx 8\cdot 10^{-38}$. 
A direct evaluation shows that the entire region $\varphi\geq 0$ can be divided into two opposite regions with the following asymptotic representations for the TMT-effective  potential
\begin{equation}
 U_{eff}^{(tree)}(\varphi) = \begin{cases}
\frac{\lambda M_P^4}{8\xi^2}{\tanh}^4z\cdot \frac{3q^4+\frac{\lambda}{8\xi^2}\sinh^4z}{4q^4+\frac{\lambda}{8\xi^2}\sinh^4z}& \text{for} \quad \varphi> M_P\\
\frac{\zeta_v+b_p}{(1+\zeta_v)^2}\frac{\lambda}{4}\phi^4-\frac{\zeta_v-b_p}{(1+\zeta_v)^2}\frac{m^2}{2}\phi^2 & \text{for}\quad \phi\ll 10^{-4}M_P,
\end{cases} 
\label{Ueff in all}
\end{equation}
where we also took into account that $\varphi\approx\phi$ when $\varphi\ll M_P$.

\begin{figure}
\includegraphics[width=10.0cm,height=6cm]{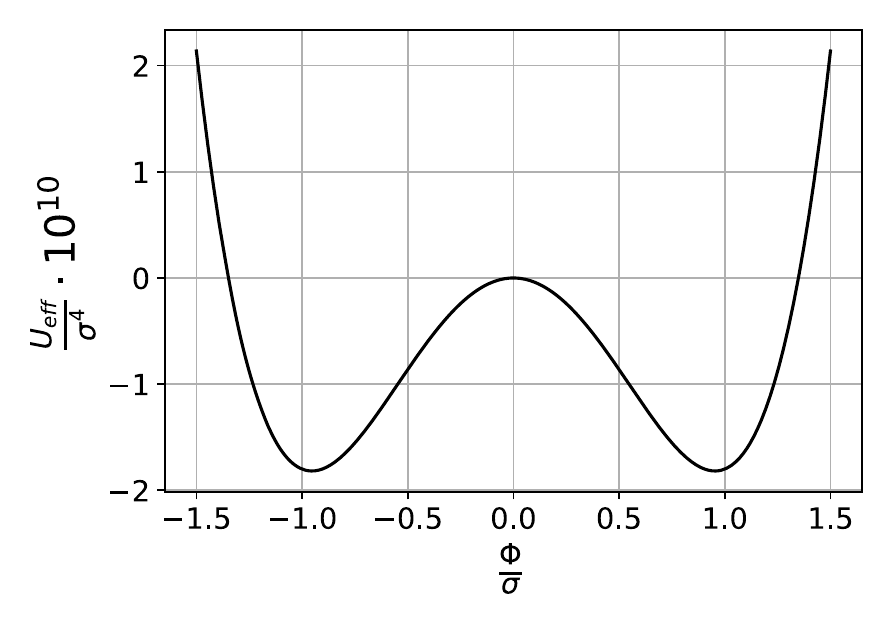}
%\end{capture}
\caption{The plot of the TMT-effective potential $U_{eff}^{(tree)}(\phi)$ for $\phi\ll 10^{-4}M_P$, defined in eq.(\ref{Ueff in all}),
has the well-known form of the Higgs potential. In the $\phi$-equation of motion with the canonical kinetic term, the derivative of this potential 
turns out to be  divided by a constant factor $\frac{b_k-\zeta_v}{1+\zeta_v}$, where $b_k-\zeta_v=b_k-1\approx 10^{-5}$.}
\label{fig3}
\end{figure}

The result in region $\phi\ll 10^{-4}M_P$, presented in figure \ref{fig3}, will be more understandable if we consider the expression for the TMT-effective Lagrangian 
$L_{eff}^{(tree)}(\varphi,X_{\varphi})$ given by eqs.(\ref{Leff varphi})-(\ref{K2 varphi}). An important point is that the kinetic term in $L_{eff}^{(tree)}(\varphi,X_{\varphi})$ is non-canonical. Considering that, as follows from eqs.(\ref{K1 varphi}) and (\ref{K2 varphi})
\begin{equation}
K_1|_{\phi\ll 10^{-4}M_P}=\frac{b_k-\zeta_v}{1+\zeta_v}, \quad {\tilde K}_2|_{\phi\ll 10^{-4}M_P}=\frac{(1+b_k)^2}{2q^4(1+\zeta_v)^2},
\label{K1 phi ll MP }
\end{equation}
we arrive at the following expression
\begin{equation}
L_{eff}^{(tree)}|_{\phi\ll 10^{-4}M_P}=\frac{b_k-\zeta_v}{1+\zeta_v}X_{\phi}-\frac{(1+b_k)^2}{4q^4(1+\zeta_v)^2}\frac{X_{\phi}}{M_P^4}X_{\phi}
-U_{eff}^{(tree)}(\phi)|_{\phi\ll 10^{-4}M_P}.
\label{Leff phi ll MP}
\end{equation}
In the region $\phi\ll 10^{-4}M_P$ it is natural to assume that 
$|X_{\phi}|\lesssim |U_{eff}(\phi)|_{\phi\ll 10^{-4}M_P}|\ll q^4M_P^4\sim (10^{16}GeV)^4$. Then it is easy to see that the second term in (\ref{Leff phi ll MP}) is many orders of magnitude smaller than the first. In this approximation, the field $\phi$ equation of motion, obtained using the Lagrangian
$L_{eff}^{(tree)}|_{\phi\ll 10^{-4}M_P}$ and then dividing by a constant factor $\frac{b_k-\zeta_v}{1+\zeta_v}$, coincides with the $\phi$-equation near the vacuum, described by eqs.(\ref{h eq}) and (\ref{lambda and m2 SM}), found in section \ref{sec:Higgs vacuum}.

\begin{itemize}
\item
We can now understand how in the TMSM, despite the fact that the parameter $m^2$ in the primordial action is positive, the mass term near the vacuum turns out to have the "wrong" sign, which provides a spontaneous symmetry breaking. This {\em mystery of the Higgs potential is explained in the TMSM} as a consequence of the behavior of the function $\zeta(\varphi)$ during cosmological evolution, which was discussed at the beginning of this section.  This can be best seen by analyzing the $\phi$-equation in the $\zeta$-dependent form (\ref{phi eq Ein frame}) (which  also has a non-kanonical kinetic term). In the process of the increase of the function $\zeta(\varphi)$ described above, at a certain moment of cosmological evolution $\zeta(\varphi)$ crosses the value $b_p\approx\frac{1}{2}$; as a result, the mass term proportional to
$(b_p-\zeta)m^2\phi$ in eq.(\ref{phi eq Ein frame}) changes sign to the "wrong" one.

\end{itemize}

\section{Constraint, orientability  of the space-time manifold
\\
 and their impact on the initial coditions}
\label{orientation}

The question of why the initial conditions for the solution of cosmological equations ensure the occurrence of inflation is of fundamental theoretical importance and is known in the literature \cite{Linde 1983}, \cite{Linde 1985}, \cite{Stein}-\cite{singular alpha}  as {\em the problem of the initial conditions for inflation}. In this section we will show that in the model under study the conditions described by inequalities (\ref{Cond for begin infl}) are guaranteed to be met. However, in section \ref{impact of K-ess} we will discover that the effects of the TMT-effective  K-essence type structure  lead to the need to take into account that, instead of inflation, cosmological evolution may begin with some exotic scenario obeying completely different laws of physics. For this reason, we are forced in this section to formulate the problem and the method existing in TMSM for solving it in a broader context. Namely, if the solutions of cosmological equations describe the evolution of the universe (in classical space-time), then: 1)  the basic structure of the TMT, which also underlies the TMSM, imposes dynamics-independent constraints on the admissible initial conditions of solutions; 2) in the following section \ref{impact of K-ess}  we will show that the dynamics features in the model under study impose stronger constraints on the admissible initial conditions than those mentioned in item 1).

\subsection{Brief summary of ref. \cite{my JCAP 2023}:  guaranteed initial conditions for inflation
\\
 and the Borde-Guth-Vilenkin theorem}
\label{brief summary}

A solution (\ref{var varphi}) of eq.(\ref{varphiB}),  the general form of which is the same as in ref. \cite{my JCAP 2023}, exists under the condition $\Upsilon(x)\neq 0$, Eq.(\ref{Phi neq 0}), i.e, only if
$\Upsilon(x)>0$ or only if $\Upsilon(x)<0$. Therefore, {\em only those solutions of the system of equations are
valid for which the corresponding $\Upsilon(x)$ is sign-definite}, since Eq.(\ref{varphiB}) is one of the equations
resulting from the principle of least action. It should be noted here that, as usual, by
default we assume that the original metric $g_{\mu\nu}$ in the primordial action is regular, that is
$g=\det(g_{\mu\nu})<0$. Therefore, the validity of the solutions regarding the fulfillment of the
condition on the sign of $\Upsilon$ can be controlled by checking the sign of the scalar $\zeta=\frac{\Upsilon}{\sqrt{-g}}$.
In particular, in the model under consideration, with our choice of parameters and constant
of integration, the value of $\zeta$ in vacuum, eq.(\ref{zeta v}), is positive: $\zeta_v>0$. Thus, only those
cosmological solutions (together with their initial conditions) are valid for which $\zeta(x)$ is
positive throughout the evolution of the universe. In other words, these solutions lose their
validity when we try to extend them to the region where $\zeta$ crosses zero and becomes negative.
The condition $\Upsilon(x)>0$ is equivalent to the sign-definitness of the volume 4-form $\Upsilon(x)d^4x$
on the 4D-space-time manifold. From the geometrical point of view this means that the
differential manifold of our Universe is orientable\footnote{For more mathematical aspects of this issue see, e.g., books \cite{Lee}, \cite{Hawking}
 and section 4 in ref. \cite{my JCAP 2023}.}.

If $\zeta(x)$ vanishes for a certain field configuration, then this situation is of particular
interest and was studied in detail in paper \cite{my JCAP 2023} within a fairly simple model. As a result
of the condition $\Upsilon(x)>0$, or equivalently $\zeta(x)>0$, we discovered that, if $X_{\varphi}>0$, there
exists an upper bound $\varphi_0$ on admissible values of the inflaton field $\varphi$ for which
\begin{equation}
\Upsilon\equiv\zeta(\varphi,X_{\varphi})\cdot\sqrt{-g}\rightarrow 0^{+} \quad {\text and} \quad X_{\varphi}\rightarrow 0^{+} 
\quad {\text as} \quad  \varphi\rightarrow \varphi_0^{\,\,\,-}.
\label{results of JCAP}
\end{equation}
A detailed study has established that if $X_{\varphi}^{(in)}>0$,  that is the initial kinetic energy density
$\rho_{kin}^{(in)}$ prevails over initial gradient energy density $\rho_{grad}^{(in)}$, then there is an interval of initial
values $\varphi_{in}^{(min)}\leq\varphi_{in}<\varphi_0$, where $\rho_{kin}^{(in)}$ and $\rho_{grad}^{(in)}$
  cannot exceed the potential energy density
and hence the initial conditions necessary for the onset of infation are satisfied. In addition,
it was shown that $\Upsilon(x)=0$ is a spacelike hypersurface such that in the inflating spacetime of
the cosmological model studied in ref. \cite{my JCAP 2023}, timelike geodesics cannot be extended to the
past beyond $\Upsilon(x)=0$. This conclusion is consistent with the statement of the BGV theorem \cite{BGV},
and we had reason to assert that the hypersurface $\Upsilon(x)=0$ is the boundary $\mathcal{B}$, which appears
in the formulation of the theorem. The analyze made in ref. \cite{my JCAP 2023} shows that $\mathcal{B}$ is the
boundary of the space-time manifold of our Universe. Here we will limit ourselves to this
brief summary of the essence of the discovered effect and move on to the analysis of the
constraint, its features and consequences for the Higgs inflation model studied in this paper.

\subsection{Constraint,  initial conditions 
\\
and  orientability of the space-time manifold
 }
\label{sub initial}

The constraint (\ref{zeta via varphi}) has a fundamental role throughout the entire process of cosmological
evolution, but we will consider here only the case of $\sqrt{\xi}\varphi>M_P$ applicable to the inflationary stage.
We are going to show that {\em the condition $\zeta>0$ imposes restrictions on the possible
values of $\varphi$ and $X_{\varphi}$  of the solutions to the cosmological equations.} In the context of this
paper, the most important question for us is to find out  whether the application of these
restrictions to the initial values $\varphi_{in}$ and $X_{\varphi}^{(in)}$
ensures the onset of inflation. However, the need to take into account the influence  of the K-essence type structure forces us to limit ourselves here to finding the quantitative restrictions imposed on the initial values $\varphi_{in}$ and $X_{\varphi}^{(in)}$ by the condition $\zeta>0$, and to postpone the study of their role in the onset of inflation until section \ref{pathol and normal}.

It turns out
that the signs with which the terms with $X_{\varphi}$ enter into the constraint (\ref{zeta via varphi}) are opposite
to the signs of the corresponding terms in the model of ref. \cite{my JCAP 2023}. This difference leads to
restrictions on  the possible values of $\varphi$ and $X_{\varphi}$ of  {\em the solutions} to the cosmological equations,
 which differ significantly from the model results of ref. \cite{my JCAP 2023}. Therefore, the restrictions on {\em the  initial values} $\varphi_{in}$ and $X_{\varphi}^{(in)}$ {\em of these solutions} also differ significantly from the model results of ref. \cite{my JCAP 2023}. 
For $\sqrt{\xi}\varphi >M_P$
 the constraint (\ref{zeta via varphi})) can be conveniently rewritten by
omitting terms $\propto\frac{m^2}{M_P^2}$
\begin{equation}
\zeta(\varphi,X_{\varphi})=\frac{32q^4e^{-\sqrt{\frac{8}{3}}\frac{\varphi}{M_P}}-\frac{\lambda}{8\xi^2}\delta_p
+\frac{X_{\varphi}}{M_P^4}}
{32q^4e^{-\sqrt{\frac{8}{3}}\frac{\varphi}{M_P}}+\frac{\lambda}{8\xi^2}-\frac{X_{\varphi}}{M_P^4}},
\label{constr for initial}
\end{equation}
where we took $\xi=\frac{1}{6}$, $\zeta_v=1$, $b_k=1$ and the presentation (\ref{bp}) of $b_p$ via $\delta_p$ was used. It is
obvious that $\zeta(\varphi,X_{\varphi})>0$ if and only if the values of $\varphi$ and $X_{\varphi}$ satisfy the following double
inequality
\begin{equation}
\frac{\lambda}{8\xi^2}\delta_p-32q^4e^{-\sqrt{\frac{8}{3}}\frac{\varphi}{M_P}}<\frac{X_{\varphi}}{M_P^4}
<\frac{\lambda}{8\xi^2}+32q^4e^{-\sqrt{\frac{8}{3}}\frac{\varphi}{M_P}}.
\label{double inequal}
\end{equation}
It is important to note that the way this double inequality is derived has nothing to do with the K-essence type structure.
Both the solutions to the cosmological equations obtained from the TMT-effective action (\ref{Seff varphi})-(\ref{K2 varphi}) and the initial conditions of these solutions must satisfy the requirements imposed by this double inequality.
Therefore,  they must be fulfilled throughout the entire process of cosmological evolution, driven by the classical scalar field $\varphi$ (under condition 
$\sqrt{\xi}\varphi>M_P$), including the very beginning of evolution. Consequently, by applying the inequalities for the initial values\footnote{It should be kept in mind that $\varphi_{in}$ and $X_{\varphi}^{(in)}$ are treated as independent of each other.}
 $\varphi_{in}$ and $X_{\varphi}^{(in)}$, we will be able to see what restrictions these inequalities impose on the initial conditions, which is of particular interest in the light of clarifying  the issue of the occurrence of inflation.

To interpret the results, one must keep in mind that $X_{\varphi}$ equals to the difference of the canonical\footnote{It should be noted that only the regular form of kinetic and gradient energy densities is used here, and the kinetic and gradient energy densities modified by the K-essence type structure will be considered in section \ref{impact of K-ess}.}
kinetic and gradient energy densities
\begin{equation}
X_{\varphi}=\frac{1}{2}\tilde{g}^{\alpha\beta}\varphi_{,\alpha}\varphi_{,\beta}=\frac{1}{2}\left(\dot{\varphi}^2-
\frac{1}{a^2}(\nabla\varphi)^2\right)
=\rho_{kin}-\rho_{grad},
\label{X as difference}
\end{equation}
Therefore, the initial 
$X_{\varphi}^{(in)}=\rho_{kin}^{(in)}-\rho_{grad}^{(in)}$ can be positive or negative depending on how inhomogeneous and anisotropic at the beginning of classical evolution was the space domain  whose expansion generates our Universe.

Let's look at the implications that can be drawn from the inequalities (\ref{double inequal}) with the model parameters used in figure \ref{fig2}, where we chosed $q^4=3\cdot 10^{-10}$. The shape of the potential $U_{eff}^{(tree)}(\varphi)$ shows that, with the exception of the very end, inflation can occur at $\varphi\gtrapprox 6M_P$; if evolution starts from other scenario, possibility of which will be found in sec.\ref{impact of K-ess}, it is natural to assume that this also occurs at $\varphi\gtrapprox 6M_P$.
 The height of the potential plateau is between $10^{-10}M_P^4\lesssim U_{eff}(\varphi)\lesssim 1.04\cdot 10^{-10}M_P^4$.
The latter will be used in the numerical estimates made below.
The remaining free parameter is $0<\delta_p<1$ and it turns out that changing its value can lead to qualitatively different results. 
As numerical examples, consider two cases $\delta_p=0.1$  and $\delta_p=10^{-8}$, corresponding to two graphs
$\zeta(\varphi)$ for $X_{\varphi}=0$ in figure \ref{fig2}.

\begin{itemize} 

\item 
In the case $\delta_p=0.1$, that is $b_p=0.55$, the inequalities read
\begin{equation}
10^{-11}\lesssim\frac{X_{\varphi}}{M_P^4}\lesssim \frac{\lambda}{8\xi^2}\approx \frac{U_{eff}^{(tree)}(\varphi)}{M_P^4}\approx 10^{-10}.
\label{double inequal 01}
\end{equation}
In the case of a homogeneous and isotropic universe, where $X_{\varphi}=\frac{1}{2}\dot{\varphi}^2$, we see that
   $\dot{\varphi}^2$ is limited to positive values. There are at least two reasons why this result is unacceptable. First, the existence of this lower positive limit for $X_{\varphi}$ would lead to the completely unnatural conclusion that at the very beginning of cosmological evolution in classical space-time the existence of regions in which $\frac{1}{a^2}(\nabla\varphi)^2>\frac{1}{2}\dot{\varphi}^2$  is impossible. Second, note that,  as follows from the plots of $\zeta(\varphi)$ for $X_{\varphi}=0$  in Fig.\ref{fig2}, in the case of $\delta_p=0.1$, values of $X_{\varphi}>0$ close to zero are prohibited for $\varphi> 4.25. M_P$, since in this case $\zeta<0$. All of the above makes the choice of the parameter $b_p=0.55$ undesirable.

\item

In the case
\begin{equation}
  b_p=0.5(1+10^{-8}),
\label{bp 6 and 8}
\end{equation}
for $\varphi> 6 M_P$, the inequalities (\ref{double inequal})  are reduced to
\begin{equation}
-5.3\cdot 10^{-13}\lesssim\frac{X_{\varphi}}{M_P^4}\lesssim \frac{\lambda}{8\xi^2}\approx \frac{U_{eff}^{(tree)}(\varphi)}{M_P^4}\approx  10^{-10}
\label{double inequal 6 and 8}
\end{equation}
with very high accuracy.
Obviously, in this example, at the very beginning of cosmological evolution in classical space-time, the existence of regions in which $\frac{1}{a^2}(\nabla\varphi)^2>\frac{1}{2}\dot{\varphi}^2$ is not forbidden.
For  initial values of $\varphi_{in}$ and $(\partial_{\alpha}\varphi)_{in}$ we obtain
\begin{equation}
-5.3\cdot 10^{-13}M_P^4\lesssim\rho_{kin}^{(in)}-\rho_{grad}^{(in)}\lesssim U_{eff}^{(tree)}(\varphi_{(in)})\approx 10^{-10}M_P^4
\label{double inequal 6 and 8 in}
\end{equation}

Discussing the possible sign of $X_{\varphi}^{(in)}=\rho_{kin}^{(in)}-\rho_{grad}^{(in)}$, it cannot be ruled out that 
$\rho_{kin }^{(in)}>U_{eff}^{(tree)}(\varphi_{in})$ and $\rho_{grad}^{(in)}>U_{eff}^{(tree)}(\varphi_{in})$ , but $X_{\varphi}^{(in)}< U_{eff}^{(tree)}(\varphi_{in})$. However, this possibility seems unlikely, and we will not consider it. Taking into account this remark, in the case where $\rho_{kin}^{(in)}>\rho_{grad}^{(in)}$, the inequality (\ref{double inequal 6 and 8 in}) reduces to
\begin{equation}
\frac{1}{2}\dot{\varphi}_{in}^2\lesssim U_{eff}^{(tree)}(\varphi_{(in)})\approx  10^{-10}M_P^4;
\label{kin bigger grad 6 and 8 in}
\end{equation}
and if $\rho_{grad}^{(in)}>\rho_{kin}^{(in)}$ then the inequality 
(\ref{double inequal 6 and 8 in}) reduces to  
\begin{equation}
\frac{1}{a^2}(\nabla\varphi)_{in}^2\lesssim 5.3\cdot\cdot 10^{-13}M_P^4\ll U_{eff}^{(tree)}(\varphi_{in}).
\label{grad  bigger  kin 6 and 8 in}
\end{equation}

\end{itemize}
 Thus, we come to the conclusion that there is a quite wide range of model parameters in which conditions  (\ref{Cond for begin infl}) for the onset of inflation  are met .   At this stage,  it is worth noting once again that {\em the described  results follow only from the condition $\zeta>0$, which, as discussed above, means that the space-time manifold of our Universe is orientable.}
Recall also that, as explained in the paragraph after eq.(\ref{1+ zeta no fine tun phi M}), the origin of the restrictions imposed by the constraint on the dynamics of the field $\phi$ are in no way dependent on the dynamics of the field $\phi$;
this is true in general and also in this particular case with respect to the condition $\zeta>0$.
However, as noted at the beginning of this section, in order to assert that the conditions necessary for the onset of inflation are met, we must take into account the influence of the K-essence type structure.

\section{Impact of the K-essence type structure 
\\
 on the first stages of classical cosmological evolution}
\label{impact of K-ess}

\subsection{General overview and  possible scenarios 
\\
for the beginning of cosmological evolution }
\label{pathol and normal}

 Using the TMT-effective potential (\ref{Tmn 0 phi K1 K2 varphi}) and the TMT-effective action,  eqs.(\ref{Seff varphi})-(\ref{K2 varphi}), in the spatially flat FLRW universe, the following Friedmann and scalar field equations are obtained:
\begin{equation}
\frac{\dot{a}^2}{{a}^2}=\frac{1}{3M_P^2}\Bigl[\Big(K_1-\frac{3}{2} \tilde{K}_2\frac{\dot{\varphi}^2}{2M_P^4}\Big)\cdot \frac{1}{2}\dot{\varphi}^2+
 U_{eff}^{(tree)}(\varphi)\Bigr],
\label{K-Frid prelim}
\end{equation}
\begin{equation}
\Bigl(K_1-3  \tilde{K}_2\frac{\dot{\varphi}^2}{2M_P^4}\Bigr)\ddot{\varphi}+\Bigl(K_1-\tilde{K}_2\frac{\dot{\varphi}^2}{2M_P^4}\Bigr)3
\frac{\dot{a}}{a}\dot{\varphi}+\Bigl(K_1'-\tilde{K}'_2\frac{\dot{\varphi}^2}{2M_P^4}\Bigr)\dot{\varphi}^2  +U_{eff}^{(tree)'}(\varphi)=0.
\label{K-FLRW varphi}
\end{equation}

\begin{figure}
\includegraphics[width=10.0cm,height=6cm]{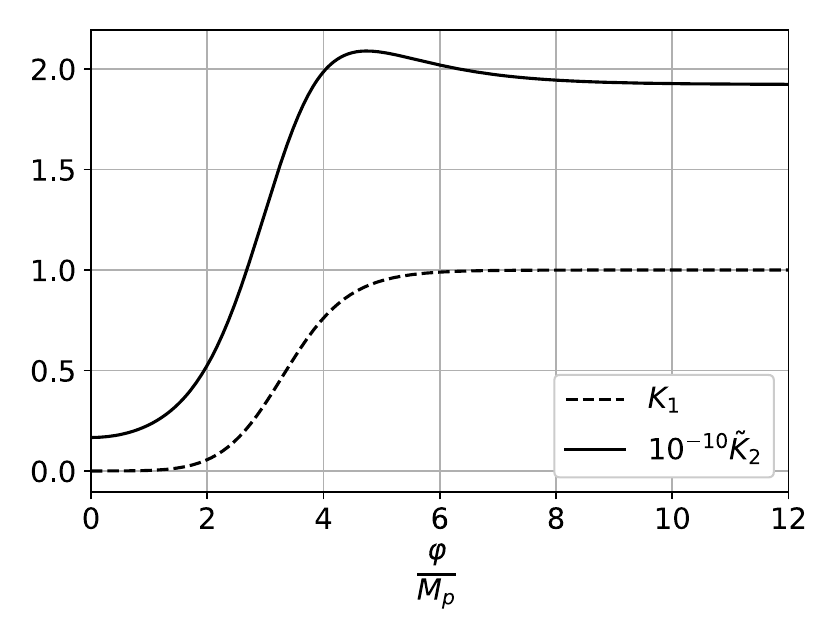}
%\end{capture}
\caption{Plots of $K_1(\varphi)$ and $\tilde{K}_2(\varphi)$ defined by eqs.(\ref{K1 varphi}) and (\ref{K2 varphi}), with the model parameters
$\lambda = 2.3\cdot 10^{-11}$, \, $\xi=\frac{1}{6}$, \,  $m=0.7GeV$, \, $b_p\approx 0.5$, \, $q^4=3\cdot 10^{-10}$ used in Fig.2. In addition 
the value  $b_k=1+\mathcal{O}(10^{-5})$ is used.}
\label{fig4}
\end{figure}

We will continue to explore the model by adding to the parameters used in figure \ref{fig2}, parameter $b_k=1+\mathcal{O}(10^{-5})$.
The graphs of
$K_1(\varphi)$ and $\tilde{K}_2(\varphi)$ are presented in figure \ref{fig4}. Except at the end, in this section we are only interested in the region of 
$\varphi$, where $\sqrt{\frac{8}{3}}\varphi\gg M_P$. In this region, the analytical form of $K_1(\varphi)$ and $\tilde{K}_2(\varphi)$ reduces to 
\begin{equation}
K_1(\varphi)\approx 1-185.5\cdot e^{-\sqrt{\frac{8}{3}}\frac{\varphi}{M_P}},
\label{K1 big varphi function}
\end{equation} 
\begin{equation}
\tilde{K}_2(\varphi)\approx 1.93\cdot 10^{10}\Big(1-185.5\cdot e^{-\sqrt{\frac{8}{3}}\frac{\varphi}{M_P}}\Bigr).
\label{K2 big varphi}
\end{equation} 
In the region $\varphi>6M_P$, where the TMT-effective potential $U_{eff}^{(tree)}(\varphi)$ is flat with sufficiently high accuracy, 
the functions $K_1$ and $\tilde{K}_2$ with a relative error of the order less than $10^{-2}$ can be used as the following constants
\begin{equation}
K_1(\varphi)\approx 1, \qquad \tilde{K}_2(\varphi)\approx 1.93\cdot 10^{10}
\label{K1 big varphi}
\end{equation} 
It is natural to assume that the initial value $\varphi_{in}$ is significantly greater than $6M_P$, that is, at the beginning of evolution, the approximations (\ref{K1 big varphi}) are valid with much greater accuracy than $10^{-2}$.

The structure of the cosmological equations (\ref{K-Frid prelim}) and (\ref{K-FLRW varphi}) differs significantly from the usual ones. First of all, in the place where the kinetic energy density should be in the energy density, we find an expression 
\begin{equation}
K_{eff}\stackrel{\mathrm{def}}{=}
\Big(K_1-\frac{3}{2} \tilde{K}_2\frac{\dot{\varphi}^2}{2M_P^4}\Big)\cdot \frac{1}{2}\dot{\varphi}^2,
\label{Kinetic eff}
\end{equation}
which, by analogy with the TMT-effective potential $U_{eff}^{(tree)}(\varphi)$, should be considered as the TMT-effective kinetic energy density.
What is very unusual is that $K_{eff}$ and the coefficient before $\ddot{\varphi}$ in eq.(\ref{K-FLRW varphi}) can be either positive or negative, depending on the values of $\dot{\varphi}^2$.

The classical evolution of the Universe implies that from the very beginning the total initial energy density must be positive 
\begin{equation}
\rho_{in}=K_{eff}^{(in)}+U_{eff}^{(tree)}(\varphi_{in})>0,
\label{rho in >0}
\end{equation}
from where we get  the upper bound on $\dot{\varphi}_{in}$ 
\begin{equation}
\frac{1}{2}\dot{\varphi}_{in}^2\lesssim  0.78\cdot U_{eff}^{(tree)}(\varphi_{in})\approx 0.8\cdot 10^{-10}M_P^4. 
\label{upper bound from rho in >0}
\end{equation}
We note that this constraint on  the permissible values of $\dot{\varphi}_{in}^2$ is somewhat stronger compared to the upper bound imposed by the condition $\zeta>0$, eq.(\ref{kin bigger grad 6 and 8 in}).

If in classical space-time, at the very beginning of cosmological evolution, the multiplier  $\bigl(K_1-3  \tilde{K}_2\frac{\dot{\varphi}_{in}^2}{2M_P^4}\bigr)$ in front of $\ddot{\varphi}$ in eq.(\ref{K-FLRW varphi}) is positive, then the classical  field $\varphi$ that drives this evolution has regular, "normal" dynamics.
But in general, this  multiplyer may be negative, that is, the evolution can start not only from the "normal" dynamics, but also from  stages
with "pathological" dynamics.  Another reason for the appearance of "pathological" dynamics may be the case when the TMT-effective kinetic energy density  $K_{eff}$ is negative.
It is convenient to introduce a notation
\begin{equation}
y=\frac{\dot{\varphi}^2}{2M_P^4}\lesssim y_{cr}=1.7\cdot 10^{-11},
\label{y}
\end{equation}
and $y_{cr}$ is the critical value of $y$ where the  multiplyer $\Bigl(K_1-3  \tilde{K}_2\frac{\dot{\varphi}^2}{2M_P^4}\Bigr)\equiv \Bigl(K_1-3  \tilde{K}_2y\Bigr)$ in front of $\ddot{\varphi}$ in eq.(\ref{K-FLRW varphi}) equals zero.
Considering the cases when $y\neq y_{cr}$ one can divide eq.(\ref{K-FLRW varphi}) by the factor standing in front of  $\ddot{\varphi}$. This reduces eq.(\ref{K-FLRW varphi}) to the following form
\begin{equation}
\ddot{\varphi}+f_1(y)3\frac{\dot{a}}{a}\dot{\varphi}+f_2(y)\frac{\dot{\varphi}^2}{2}  
+\frac{U_{eff}^{(tree)\prime}(\varphi)}{K_1-3  \tilde{K}_2\cdot y}=0,
\label{K-FLRW varphi after deviding}
\end{equation}
where the notations are introduced
\begin{equation}
 f_1(y)=\frac{K_1-\tilde{K}_2\cdot y}{K_1-3  \tilde{K}_2\cdot y}; \quad 
f_2(y)=\frac{K'_1-\tilde{K}'_2\cdot y}{K_1-3  \tilde{K}_2\cdot y}
\label{notations after deviding}
\end{equation}
The  $\varphi$ equation, presented in the form (\ref{K-FLRW varphi after deviding}), allows one to classify, depending on $y$,  possible stages of the evolution of the very early Universe according to their types of dynamics. 

\begin{itemize}

\item 

\underline{Type I  pathological dynamics} is realized if $K_{eff}^{(in)}<0$. Then  $K_1-3\tilde{K}_2\frac{\dot{\varphi}_{in}^2}{2M_P^4}<0$
as well. Together with the condition (\ref{upper bound from rho in >0}), ensuring  $\rho_{in}>0$, this leads to restrictions on the initial conditions under which the classical evolution of the Universe begins with the  first type of  pathological dynamics
\begin{equation}
0.35\cdot U_{eff}^{(tree)}(\varphi_{in})<\frac{1}{2}\dot{\varphi}_{in}^2<0.8\cdot U_{eff}^{(tree)}(\varphi_{in}). 
\label{1-st type pathol}
\end{equation}

\item 

If $K_{eff}^{(in)}>0$, but $K_1-3\tilde{K}_2\frac{\dot{\varphi}_{in}^2}{2M_P^4}<0$, then the classical evolution of the Universe begins with 
\underline{type II   pathological dynamics}, which is realized under the following  restrictions on the initial conditions
\begin{equation}
0.17\cdot U_{eff}^{(tree)}(\varphi_{in})\lesssim\frac{1}{2}\dot{\varphi}_{in}^2\lesssim  0.35\cdot U_{eff}^{(tree)}(\varphi_{in}). 
\label{2-nd type pathol}
\end{equation}

\item

The condition $K_1-3\tilde{K}_2\frac{\dot{\varphi}_{in}^2}{2M_P^4}>0$ means that the classical evolution of the Universe can start with \underline{normal dynamics} only if the following sufficiently strong upper limit on the initial value of $\dot{\varphi}^2$ is satisfied
\begin{equation}
\frac{1}{2}\dot{\varphi}_{in}^2\lesssim 0.17\cdot U_{eff}^{(tree)}(\varphi_{in})\approx 1.7\cdot 10^{-11}M_P^4, 
\label{normal}
\end{equation}
where at the last step we took into account that in the region $\varphi>6M_P$ the TMT- effective potential $U_{eff}^{(tree)}(\varphi)$ can be considered flat with a hight that is equal to  $U_{eff}^{(tree)}(\varphi_{in})\approx 10^{-10}M_P^4$ with sufficiently high accuracy.
Under the condtion (\ref{normal}), the TMT-effective kinetic energy density  $K_{eff}^{(in)}$ is positive and is  bounded from above by
\begin{equation}
0<K_{eff}^{(in)}\lesssim 8.6\cdot 10^{-2}\cdot U_{eff}^{(tree)}(\varphi_{in}) 
\label{cond for normal kinetic}
\end{equation}
with the upper bound being reached at  $\frac{\dot{\varphi}_{in}^2}{2M_P^4}\to 1.7\cdot 10^{-11}$.

\end{itemize}

\newcommand{\STAB}[1]{\begin{tabular}{@{}c@{}}#1\end{tabular}}

\begin{table}[!h]
  \caption{Types of dynamics at the very beginning of classical cosmological evolution}
  \centering
  \begin{tabular}{|c|l|r|r|r|r|}
  
     \hline
     \multirow{1}{*}{\STAB{\rotatebox[origin=c]{90}{dynamics \ }}}
     
     &  &  & &  & \\
     & \multicolumn{1}{c|}{range of} & \multicolumn{1}{c|}{sign of} & \multicolumn{1}{c|}{sign of} & \multicolumn{1}{c|}{sign of active force} & \multicolumn{1}{c|}{$w = \frac{p}{\rho}$} \\
     & \multicolumn{1}{c|}{$\tilde{y}=10^{10}y=10^{10}\frac{\dot{\phi}^2}{2M_P^4}$} & \multicolumn{1}{c|}{$K_{eff}$} & \multicolumn{1}{c|}{$f_1(y)$} & \multicolumn{1}{c|}{$\frac{U_{eff}^{(tree)'}}{K_1-3\tilde{K}_2y}$} & \multicolumn{1}{c|}{} \\ 
	 &  &  & &  & \\

     \hline
     \multirow{1}{*}{\STAB{\rotatebox[origin=c]{90}{pathological I\ }}} 
     
     &  &  & &  & \\
     &  &  & &  & \\
     & $0.52<\tilde{y}<0.8$ & $K_{eff}<0$ &  $f_1(y)>0$ &"wrong"  &   $w<-1$ \\
     &  &  & &  & \\
     & $0.35<\tilde{y}<0.52$ & $K_{eff}<0$ &  $f_1(y)<0$ &"wrong"  &   $w>-1$ \\
     &  &  & &  & \\
      
    \hline
    \multirow{1}{*}{\STAB{\rotatebox[origin=c]{90}{\ pathological II \ }}}
     
	 &  &  & &  & \\     
     &  &  & &  & \\
     & $0.17<\tilde{y}<0.35$ & $K_{eff}>0$ &  $f_1(y)<0$ &  "wrong"  &   $w>-1$ \\
     &  &  & &  & \\
     &  &  & &  & \\
     &  &  & &  & \\
     
     \hline
     \multirow{1}{*}{\STAB{\rotatebox[origin=c]{90}{normal \ }}} 
     
     &  &  & &  & \\ 
     & $\tilde{y}<0.17$ & $K_{eff}>0$ &  $f_1(y)>0$ &  regular  &   $w>-1$ \\
     &  &  & &  & \\
     &  &  & &  & \\ 
     \hline

  \end{tabular}
\label{table}
\end{table} 

\begin{figure}
\includegraphics[width=13.0cm,height=8cm]{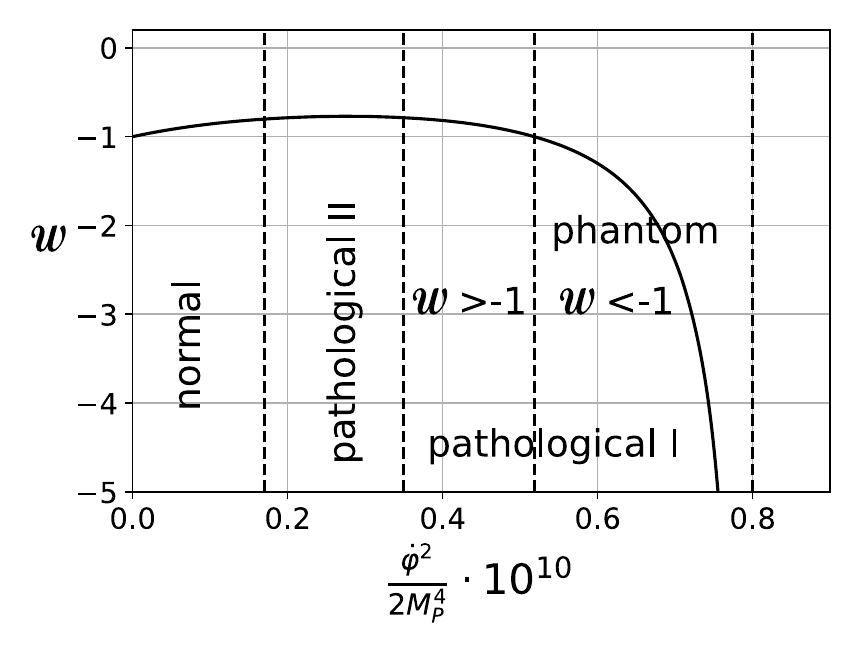}
%\end{capture}
\caption{Plot of the function (\ref{EOS}) describing the $\frac{\dot{\varphi}^2}{2M_P^4}$-dependence of the equation of state $w=\frac{p}{\rho}$ in the region $\varphi>6M_P$, where the TMT effective potential has a plateau $U_{eff}(\varphi)\approx 10^{-10}M_P^4$. The condition 
(\ref{upper bound from rho in >0}),  necessary for the total initial energy density to be positive, is manifested in the fact that the line  
$\tilde{y}=\frac{\dot{\varphi}^2}{2M_P^4}\cdot 10^{10}=0.8$ is a vertical asymptote for the function $w(y)$ and $w\to -\infty$ for $\tilde{y}\to 0.8^-$.
The results presented in eqs.(\ref{1-st type pathol})-(\ref{normal}), together with the shape of the graph $w(y)$, require dividing the interval $0<\tilde{y}<0.8$
 into four regions with different types of dynamics: 1)  region $0.52<\tilde{y}<0.8$ -  pathological type I with EOS $w<-1$, i.e. phantom dynamics;  
2)   region $0.35<\tilde{y}<0.52$ -  pathological type I  dynamics with EOS $w>-1$; 3)  region $0.17<\tilde{y}<0.35$ -  pathological type II  dynamics;
4)  region $0<\tilde{y}<0.17$ -  normal dynamics. }
\label{fig5}
\end{figure}

The results presented in Table \ref{table} for each type of dynamics (pathological type I, pathological type II and normal dynamics) show: 1) the sign of the TMT-effective kinetic energy density  $K_{eff}$; 2) the sign of $f_1(y)$ on which the sign of the friction force $f_1(y)3H\dot{\varphi}$ depends; 3) the sign of the active force in the $\varphi$ equation (\ref{K-FLRW varphi after deviding}); 4) the sign of $1+w$, where $w$ is the equation of state (EOS) $w=\frac{p}{\rho}$. As can be seen from the table, in fact, pathological dynamics of type I consists of two fundamentally different stages with different EOS. Using the (classical) TMT-effective energy-momentum tensor, eq.(\ref{Tmn 0 phi K1 K2 varphi}), and the expressions of $K_1$ and $\tilde{K}_2$ for $\varphi>6M_P$, eq.(\ref{K1 big varphi}), we find that
\begin{equation}
w=\frac{p}{\rho}=\frac{\tilde{y}-0.97\tilde{y}^2-1.07}{\tilde{y}-2.9\tilde{y}^2+1.07}, \qquad \text{where} \qquad \tilde{y}=y\cdot 10^{10}=\frac{\dot{\varphi}^2}{2M_P^4}\cdot 10^{10}.
\label{EOS}
\end{equation}
The graph of EOS is presented in figure \ref{fig5}, from which  it is evident that the evolution of the Universe can begin with $w<-1$. 
Thus, using the terminology adopted in the models of present-day  evolution of the Universe, we come to the conclusion 
that the classical evolution of the Universe can begin with the stage of phantom dynamics.

\begin{figure}
\includegraphics[width=13.0cm,height=8cm]{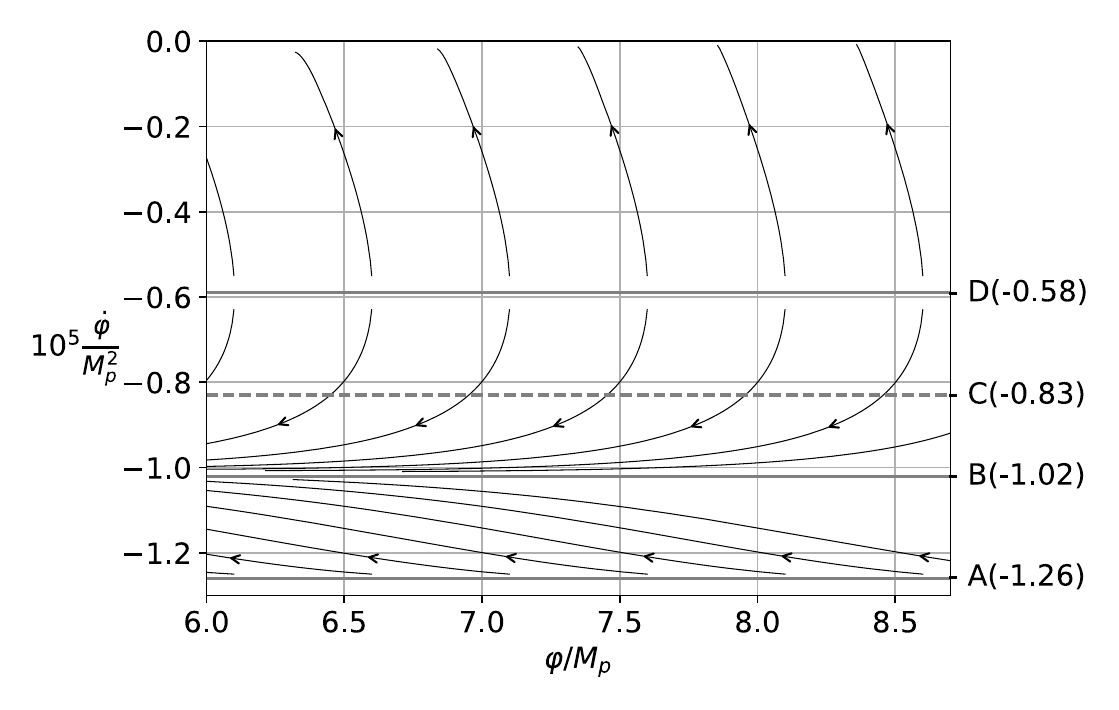}
%\end{capture}
\caption{Three disconnected sets of phase trajectories satisfying eq.(\ref{K-FLRW varphi after deviding}). The set of trajectories between lines A and B  corresponds to solutions of the pathological type I dynamics with EOS $w<-1$.  The set of trajectories between lines B and D  corresponds to solutions  with EOS $w>-1$, and line C separates  the dynamics with the pathological types I and II. The set of phase trajectories above line D corresponds to solutions with normal dynsmics. For illustration, for six chosen initial values of ${\phi}_{in}$, three  initial values of $\dot{\phi}_{in}$ are chosen respectively such that 1) close to line A (above A); 2)  close to line D (below D); 3) close to line D (above D).
Additional analysis of the dynamic properties is given in the main text.}
\label{fig6}
\end{figure}

\begin{figure}
\includegraphics[width=13.0cm,height=8cm]{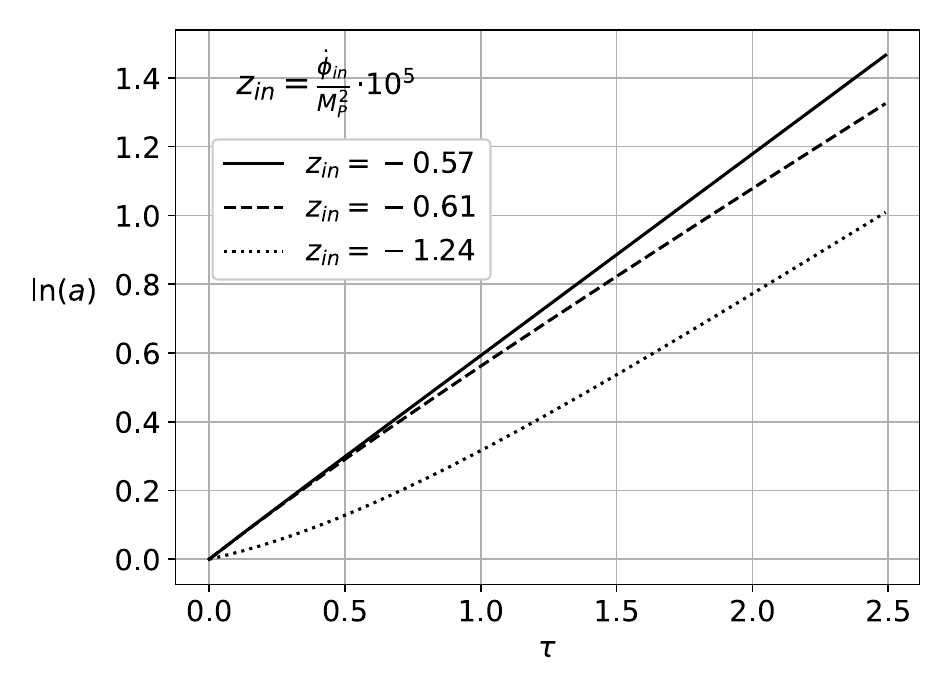}
%\end{capture}
\caption{Three graphs represent the dependences of $\ln a$  on  $\tau=10^{-5}M_Pt$ for solutions of eqs.(\ref{K-Frid prelim}), (\ref{K-FLRW varphi}) with three  initial values of $\dot{\phi}_{in}$  chosen  such that 1) close to line A in figure 6 (above A, $z_{in}\equiv \frac{\dot{\phi}_{in}}{M_P^2}\cdot 10^5=-1.24$); 2)  close to line D (below D, $z_{in}=-0.61$); 3) close to line D (above D,  $z_{in}=-0.57$), while the  initial value of ${\phi}_{in}$ is the same for three solutions.}
\label{fig7}
\end{figure}

After the presented classification of the types of dynamics depending on the initial values of $\dot{\phi}_{in}$, it is necessary to clarify the characteristic dynamic properties of each of them in the course of cosmological evolution. This can be conveniently done using the phase plane in figure \ref{fig6}, which shows the phase curves corresponding to the initial conditions close to the limiting ones for each type of dynamics. Accordingly, the phase plane is divided by horizontal lines into dynamically disconnected regions. Line A separates the region with $\rho>0$ from the region with $\rho<0$.  In the regions below line D, the sign of the active force in eq.(\ref{K-FLRW varphi after deviding}) is "wrong", which is the main reason for the existence of pathological dynamics.
Line B separates the regions with opposite sign of the friction term $f_1(y)3H\dot{\varphi}$ in eq.(\ref{K-FLRW varphi after deviding}). Phantom-like dynamics with EOS $w<-1$ can exist only in the region between lines A and B.  When the phase curves intersect the dotted line C, the TMT-effective kinetic energy $K_{eff}$ changes sign from $K_{eff}>0$ above C to $K_{eff}<0$ below C. In the region above line D, the active and frictional forces have regular signs, and  $K_{eff}>0$. Consequently,
the phase curves with initial conditions located above the line D correspond to normal dynamics and, as can be seen, the slow decrease of $\phi$ is accompanied by a rapid approach of $\dot{\phi}$ to zero, which is a characteristic property of the regime of the slow-roll inflation. It is interesting to note that line B is an attractor for the phase curves on both sides, i.e. not only for the phase curves in the region between A and B, but also for the phase curves in the region between D and B. 

Three cirves (dotted, dashed and solid) in figure \ref{fig7} show the cosmic time dependences of the scale factor $a(t)$  for three regions of the phase diagram with the same ${\phi}_{in}$ and three different values of $\dot{\phi}_{in}$: the first (dotted) with $\dot{\phi}_{in}$ slightly above line A , the second (dashed) with $\dot{\phi}_{in}$ slightly below line D , and the third (solid) with $\dot{\phi}_{in}$ slightly above line D. As can be seen, the graphs demonstrate predominantly exponential expansion. The only exception is the initial section of the first, dotted line, which means that for initial conditions close to line A, the initial stage of the cosmological expansion is non-exponential.

At the end of this section, at least brief comments  are appropriate.
The above analysis shows that pathological regimes of evolution are dynamically isolated from the regime with normal dynamics, which has been found to inevitably describe slow-roll inflation (see also sections \ref{normal dyn} and \ref{K slow roll}). It may also be noted that the initial pathological stage of the evolution of the Universe predicted by TMSM could be a model describing the unknown physics whose existence was proposed in the interpretation of the BGV theorem\cite{BGV}.

\subsection{Some important properties of normal dynamics}
\label{normal dyn}

 Considering the normal dynamics, i.e under the condition (\ref{normal}), let us turn to the $\varphi$ equation (\ref{K-FLRW varphi}) represented in the equivalent form  (\ref{K-FLRW varphi after deviding}), trying to compare it with the commonly used form of the Klein-Gordon equation in the FLRW Universe. 

Using eqs.(\ref{K1 big varphi function}), (\ref{K2 big varphi}) and (\ref{normal})  it is easy to see that $f_1(y)\geq 1$. Therefore, the presence of the factor $f_1(y)$ in eq.(\ref{K-FLRW varphi after deviding}) leads to an increase in  the usual friction force coefficient $3H$.
However, due to the factor $(K_1-3  \tilde{K}_2\cdot y)^{-1}\approx (1-1.93\cdot 10^{10}y)^{-1}$ in the last term of eq.(\ref{K-FLRW varphi after deviding}), an analogous effect occurs, leading to a similar strengthening of the usual active force $-U_{eff}^{(tree)\prime}(\varphi)$.

To estimate the third term in eq.(\ref{K-FLRW varphi after deviding}), we first use the relations (\ref{K1 big varphi function}) and  (\ref{K2 big varphi}) and find that
\begin{equation}
f_2(y)=\frac{K'_1-\tilde{K}'_2\cdot y}{K_1-3  \tilde{K}_2\cdot y}\approx 185.5\sqrt{\frac{8}{3}}\cdot \frac{f_1}{M_P}\cdot e^{-\sqrt{\frac{8}{3}}\frac{\varphi}{M_P}}.
\label{estim of 3-rd term}
\end{equation}
For  normal dynamics we can use the upper bound (\ref{normal}) on $\dot{\varphi}^2$, which together with  (\ref{estim of 3-rd term})
and the condition $\varphi>6M_P$ allow one to obtain an estimate for the third term in eq.(\ref{K-FLRW varphi after deviding}) in a form suitable for comparison with the second term in this equation
\begin{equation}
f_2\frac{\dot{\varphi}^2}{2}\lesssim 307.8 \frac{f_1}{M_P}\cdot e^{-\sqrt{\frac{8}{3}}\frac{\varphi}{M_P}}\sqrt{0.17\cdot U_{eff}^{(tree)}}
\frac{|\dot{\varphi}|}{\sqrt{2}}< 5\cdot 10^{-3}\cdot\frac{f_1}{M_P}\sqrt{U_{eff}^{(tree)}}\cdot |\dot{\varphi}|.
\label{estim of 3-rd term next}
\end{equation}
Since $K_{eff}>0$, it follows from eq.(\ref{K-Frid prelim}) that
\begin{equation}
\frac{\dot{a}}{a}=\frac{1}{\sqrt{3}M_P}\sqrt{K_{eff}+U_{eff}^{(tree)}}>\frac{1}{\sqrt{3}M_P}\sqrt{U_{eff}^{(tree)}},
\label{H ineq}
\end{equation}
Then for the contribution of the second term in eq.(\ref{K-FLRW varphi after deviding}) we obtain
\begin{equation} 
\Big|3f_1\frac{\dot{a}}{a}\dot{\varphi}\Big|>\sqrt{3}\frac{f_1}{M_P}\sqrt{U_{eff}^{(tree)}}|\dot{\varphi}|.
\label{estim of 2-nd term}
\end{equation}
Thus, the contribution of $f_2\frac{\dot{\varphi}^2}{2}$ in  eq.(\ref{K-FLRW varphi after deviding}) is at least $2.9\cdot 10^{-3}$ times smaller than $|3f_1\frac{\dot{a}}{a}\dot{\varphi}|$, and therefore it can be omitted in the further study of normal dynamics. In this approximation equation (\ref{K-FLRW varphi after deviding})  takes the form of  the Klein-Gordon equation in the FLRW Universe.

Applying the result (\ref{cond for normal kinetic}) to the Friedmann equation (\ref{K-Frid prelim}), we see that 
it can be viewed as a modification of the condition (\ref{Cond for begin infl}) required for inflation to begin
(it is assumed that $K_{eff}^{(in)}$ is greater than the initial TMT-effective gradient energy density). 
Thus,, we arrive at the following two fundamentally new conclusions:

\begin{itemize}

\item 
 \underline{classical cosmological evolution with  normal dynamics can begin only if} 
\newline
$\frac{1}{2}\dot{\varphi}_{in}^2< 0.17\cdot U_{eff}^{(tree)}(\varphi_{in})$;

\item

 \underline{classical cosmological evolution with normal dynamics}
\newline
 \underline{begins with the inflationary regime.}

Thus, {\em in the studied Two-Measure Standard Model of Higgs inflation, the problem of initial conditions for inflation does not arise.}

\end{itemize}

\subsection{About the slow-roll phase of inflation, 
\\
taking into account the K-essence type structure.}
\label{K slow roll}

 Now, using the results of the previous subsection, we are going to show that the slow-roll approximation is valid during inflation. Furthermore, we want to see what changes occur compared to what we had in section \ref{sub Higgs slow preliminary}, where the K-essence structure was completely ignored. Considering the shapes of $K_1(\varphi)$ and $K_2(\varphi)$, one
can expect that this  effect on inflation at $\varphi>6M_P$ is minor. But at the stage when
inflation is approaching its end, which happens at $\varphi<6M_P$ , where $K_1(\varphi)$ and $K_2(\varphi)$
noticeably decrease, this influence can be significant.

As we already know, the initial conditions (\ref{normal}) and (\ref{cond for normal kinetic}) not only ensure the onset of cosmological evolution with normal dynamics, but also the evolution starts with inflation. Here it is convenient to return  from eq.(\ref{K-FLRW varphi after deviding}) to its original form (\ref{K-FLRW varphi}).  We explore the region $\varphi>6M_P$, where, as we also know, the contribution of the third term in eq. (\ref{K-FLRW varphi}) is negligible compared to the second term.
From eq.(\ref{K-Frid prelim}) it follows that the first slow-roll condition requires that 
\begin{equation}
K_{eff}=K_1\cdot \frac{1}{2}\dot{\varphi}^2-\frac{3}{2}\cdot \tilde{K}_2\frac{\dot{\varphi}^4}{4M_P^4}\ll
 U_{eff}^{(tree)}(\varphi)
\label{first cond for slow A}
\end{equation} 
At the very beginning of inflation, due to the frictional force $\propto 3\frac{\dot{a}}{a}\dot{\varphi}$, $\dot{\varphi}^2$ decreases, and $\dot{\varphi}^4$ decreases even faster. Therefore, not only does 
$K_{eff}$ remain positive, but the term $\frac{3}{2}\cdot \tilde{K}_2\frac{\dot{\varphi}^4}{4M_P^4}$ quickly becomes subdominant compared to $K_1\cdot \frac{1}{2}\dot{\varphi}^2$. Hence,  the first slow-roll condition very quickly reduces to the standard one
\begin{equation}
K_1\cdot\frac{1}{2}\dot{\varphi}^2\approx \frac{1}{2}\dot{\varphi}^2\ll
 U_{eff}^{(tree)}(\varphi)\approx  10^{-10}M_P^4,
\label{first cond for slow}
\end{equation} 
which agrees with the condition  (\ref{normal}) under which normal dynamics, and therefore inflation, exists.

Thanks to the known shapes of $K_1(\varphi)$ and $\tilde{K}_2(\varphi)$ when $\varphi>6M_P$ and 
using (\ref{first cond for slow}) we  get
\begin{equation}
\tilde{K}_2\cdot \frac{\dot{\varphi}^2}{2M_P^4}\ll \frac{\tilde{K}_2}{K_1}
\frac{U_{eff}^{(tree)}(\varphi)}{M_P^4}\approx 2.
\label{K2 for second cond for slow}
\end{equation}  
\begin{figure}
\includegraphics[width=13.0cm,height=8cm]{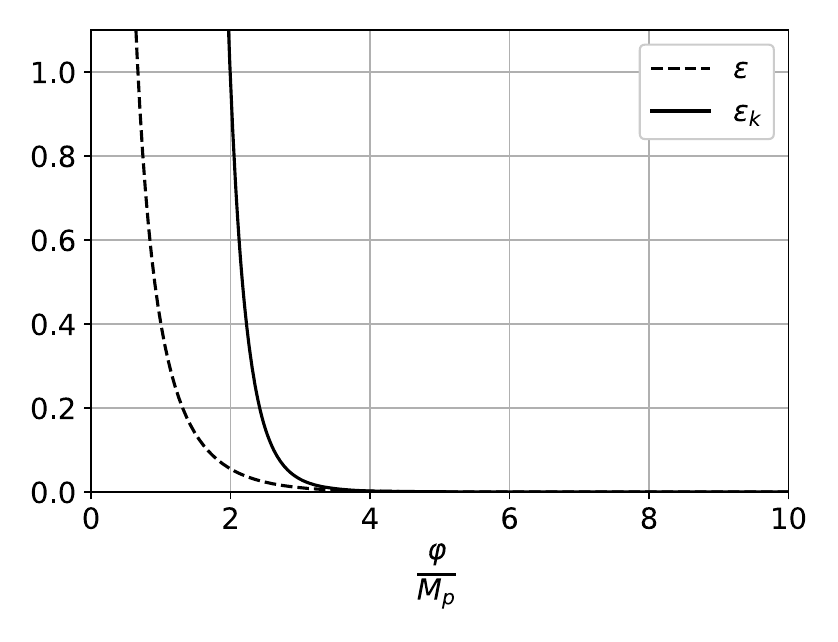}
%\end{capture}
\caption{Graphs of the functions $\epsilon(\varphi)$ and $\epsilon_k(\varphi)$ defined 
in eqs.(\ref{epsilon eta prelim}) and (\ref{epsilon K eta K}), respectively.}
\label{fig8}
\end{figure}
Therefore, in eqs. (\ref{K-Frid prelim}) and (\ref{K-FLRW varphi}) one can neglect the contributions  $\propto\tilde{K}_2(\varphi) \frac{\dot{\varphi}^2}{2M_P^4}$ compared to $K_1(\varphi)\approx 1$.  Hence,   the second slow-roll condition reduces to $|K_1(\varphi)\ddot{\varphi}|\ll |3\frac{\dot{a}}{a}K_1(\varphi)\dot{\varphi}|$, that for $\varphi>6M_P$
coincides with the usual one $|\ddot{\varphi}|\ll |3\frac{\dot{a}}{a}\dot{\varphi}|$.  Thus, the cosmological equations (\ref{K-Frid prelim}) and (\ref{K-FLRW varphi}) in the slow-roll approximation read
\begin{equation}
\frac{\dot{a}^2}{a^2}=\frac{1}{3M_P^2}U_{eff}^{(tree)}(\varphi),
\label{K-Frid slow}
\end{equation}
\begin{equation}
3\frac{\dot{a}}{a}K_1(\varphi)\dot{\varphi}=-U_{eff}^{(tree)\prime}(\varphi).
\label{K-FLRW varphi slow}
\end{equation} 
One can show\footnote{It is interesting that a somewhat similar equations appear in the Palatini-Higgs inflation considered in
the model with non-minimal derivative coupling \cite{derivative coupling}.}
that in this  slow-roll approximation the inflationary parameters
$\epsilon$ and $\eta$, eq.(\ref{epsilon eta prelim}), and the corresponding flatness conditions are replaced by
\begin{equation}
\epsilon_k=\frac{M_P^2}{2K_1}\Biggl(\frac{U_{eff}^{(tree)\prime}}{U_{eff}^{(tree)}}\Biggr)^2\ll 1, \qquad 
|\eta_k|=\frac{M_P^2}{U_{eff}^{(tree)}\sqrt{K_1}}\Biggl|\Biggl(\frac{U_{eff}^{(tree)\prime}}{\sqrt{K_1}}\Biggr)' \Biggr|\ll 1,
\label{epsilon K eta K}
\end{equation} 
where we used the subscript $k$ to indicate that the TMT-effective K-essence-type structure was used to derive these parameter expressions.
Since the deviation of $K_1(\varphi)$ from unity becomes significant only for $\varphi<6M_P$, the difference between $\epsilon$ and $\epsilon_k$  can only appear at the last stage of inflation. To illustrate the expected effect, figure \ref{fig8} shows graphs of the $\varphi$-dependencies of $\epsilon$ and $\epsilon_k$. As can be seen, $\epsilon_k$ approaches unity at higher $\varphi$. Thus, one of the expected effects of the K-essence type structure is an earlier end of inflation compared to the simplified  model of section \ref{Preliminary Higgs infl}.

\section{Prospects for further development of the Two-Measure electroweak SM (TMSM)}
\label{Further development}

\subsection{What is the TMSM}
\label{What is the TMSM}

To give a general idea of TMSM developed in ref.\cite{2-nd paper}, we need to start by describing the TMSM primordial action, which is chosen in the following form providing  the standard  $SU(2)\times U(1)$ gauge-invariant structure
\begin{equation}
S_{primordial}^{(TMSM)}=S_{gr}+S_{vac}+ S_{Higgs}+S_{gauge}+S_{ferm},
\label{S-TMSM}
\end{equation} 
where $S_{gr}$ and $S_{vac}$ are given by eqs. (\ref{S-gr}) and (\ref{S_vac}), and the contributions of the Higgs, gauge bosons and fermions, respectively, are determined as follows
\begin{equation}
S_{Higgs}=\int d^4x \left[(b_k\sqrt{-g}-\Upsilon) g^{\alpha\beta}\left( \mathit{D}_{\alpha}H\right)^{\dag} \mathit{D}_{\beta}H
 -(b_p\sqrt{-g}+\Upsilon)\lambda |H|^4-(b_p\sqrt{-g}-\Upsilon)m^2 |H|^2\right],
\label{S-H}
\end{equation}
\begin{eqnarray}
S_{gauge}=\int d^4x(b_k\sqrt{-g}-\Upsilon)\left[-\frac{1}{4} g^{\mu\alpha}g^{\nu\beta}\mathbf{F}_{\mu\nu}\mathbf{F}_{\alpha\beta}-\frac{1}{4} g^{\mu\alpha}g^{\nu\beta}B_{\mu\nu}B_{\alpha\beta}\right],
 \label{S gauge}
\end{eqnarray}
\begin{eqnarray}
S_{ferm}&=&\int d^4x\sum_{l=e,\mu,\tau} (b_l\sqrt{-g}-\Upsilon)\frac{i}{2} \bigl[\overline{L_l}\gamma^{\mu}\nabla_{\mu}L_l -(\nabla_{\mu}\overline{L_l})\gamma^{\mu}L_l+
\overline{l}_R\gamma^{\mu}\nabla_{\mu}l_R -(\nabla_{\mu}\overline{l}_R)\gamma^{\mu}l_R\bigr]
\nonumber
\\
&-&\int d^4x\sum_{l=e,\mu,\tau} (b_l\sqrt{-g}+\Upsilon)\cdot y^{(ch)}\left(\overline{L}_lH l_R +
\overline{l}_RH^{\dag}L_l\right)
\label{S l3}
\end{eqnarray}
The standard definition are used: 

For the operator  $\mathit{D}_{\mu}=\partial_{\mu}
-ig\mathbf{\hat{T} A}_{\mu}-i\frac{g'}{2}\hat{Y}B_{\mu}$ with the  primordial gauge coupling constants $g$ and $g'$.
 As usually, $\mathbf{\hat{T}}$ stands for the three generators of the $SU(2)$ group and $\hat{Y}$ is the generator of the U(1) group;

For the field strengths of the isovector $\mathbf{A}_{\mu}$ and isoscalar $B_{\mu}$.  $\mathbf{F}_{\mu\nu}=\partial_{\mu}\mathbf{A}_\nu -\partial_{\nu}\mathbf{A}_\mu +g\mathbf{A}_\mu\times \mathbf{A}_\nu$ and $B_{\mu\nu}=\partial_{\mu}B_\nu -\partial_{\nu}B_\mu$;

 $\gamma^{\mu}=V_{k}^{\mu}\gamma^{k}$, \, $\gamma^{k}$ are Dirac matrices;  $V_{k}^{\mu}$ is vierbein ($i, k$ are Lorentz indeces) In the fermion sector, for brevity, we limited ourselves to an explicit description of only three generations of charged leptons
\begin{equation}
L_l=\frac{1-\gamma_5}{2} \left(\begin{matrix} \nu^{(l)} \\ l \end{matrix}\right);
\qquad l_R=\frac{1+\gamma_5}{2} l ;  \qquad  l=e, \mu, \tau .
\label{GWS fermions 3}
\end{equation}
Three generations of up-quarks are added to (\ref{S l3}) in a similar manner.
To construct a generally coordinate invariant kinetic terms of the action for fermionic sector we have to use the  covariant operator
$\nabla_{\mu}=D_{\mu}+\frac{1}{2}\omega_{\mu}^{ik}\sigma_{ik}$,
 $\sigma_{ik}=\frac{1}{2}(\gamma_i\gamma_k -\gamma_k\gamma_i)$ and $\omega_{\mu}^{ik}$
 is the affine spin-connection
\begin{equation}
\omega_{\mu}^{ik}=\frac{1}{2}\Bigl[ V^{i\lambda}\bigl(\partial_{\mu} V^k_{\lambda}+\Gamma^{\beta}_{\mu\lambda} V^k_{\beta}\bigr)-
(i \leftrightarrow k)\Bigr]
\label{affine spin-conn}
\end{equation}
For up-quarks the action is constructed in a similar way.

 Utilizing the properties of the volume measure density $\Upsilon$, discussed in detail in section \ref{Adding matter fields}, additional factors associated with the existence of two types of volume elements are selected in a special way. Taking these properties into account in the previous sections of this paper  played a decisive role in the feasibility of Higgs inflation at small $\xi$.
 The parameters $b_k>0$, $b_p>0$,
$\lambda>0$ and $m^2>0$ are chosen to be positive.
 The density of the volume measure in (\ref{S gauge}) is chosen to be the same as in the kinetic term of $S_{Higgs}$. 

Turn now to  the Yukawa coupling terms in (\ref{S l3}) responsible for the fermion mass generation after SSB. The standard method is that for each generation of charged leptons, the Yukawa coupling constants $y_e$, $y_{\mu}$, $y_{\tau}$ are selected to ensure agreement with the experimental data.
As is well known, this aspect of SM contains a flaw indicating insufficient completeness of the theory:  in order to obtain the required values of the masses of charged leptons,  the corresponding Yukawa coupling constants must vary more than three orders of magnitude 
$y_e:y_{\mu}:y_{\tau}=m_e:m_{\mu}:m_{\tau}$. Moreover, if we try to obtain the mass of the Dirac neutrino using the same method as the masses of charged leptons, then this range of scatter of Yukawa constants must be increased by 6 orders of magnitude. In the quark sector, when trying to describe, for example, the masses of up-quarks through their Yukawa coupling to the Higgs field, the range of scatter of the corresponding Yukawa constants is 5 orders of magnitude. The situation with the choice of the value of the Yukawa coupling constant necessary to obtain the top quark mass is also  very strange. In the framework of the Glashow-Weinberg-Salam (GWS) theory where particle masses arise as a result of the SSB, the spread of Yukawa coupling constants required for the observed fermion mass hierarchy looks unnatural for a theory that claims to correctly describe nature.
In this regard, the SM extension implemented in TMSM provides an unexpected opportunity to circumvent this hierarchy problem.
In ref. \cite{2-nd paper} it is shown that in order to construct a realistic TMSM with three generations of leptons, it is possible
 to suppose that {\bf Yukawa coupling constant $y^{(ch)}$ is chosen to be universal for all charged leptons}.
 Similar construction for up-quarks is also realized in ref. \cite{2-nd paper}. Then the problem of the mass hierarchy is solved simply 
by using the freedom to choose the coefficients in the linear combination of the densities of the volume 
measures $\sqrt{-g}$ and $\Upsilon$ when constructing the volume elements. Similar to selecting signs 
 in the primordial action of the bosonic sector, eqs.(\ref{S-H 12}) and (\ref{S gauge}), for charged leptons 
we choose  volume elements $(b_l\sqrt{-g}-\Upsilon)d^4x$ for kinetic terms  and $(b_l\sqrt{-g}+\Upsilon)d^4x$
 for the Yukawa coupling terms  in the primordial action. The basis for the success of the idea that makes it possible to obtain 
the masses of charged leptons with one universal  Yukawa coupling constant in the primordial action lies in the possibility
of such a choice of the parameters $b_l$ ($l=e, \mu, \tau$ ) 
that the deviations of their values from unity are small.

{\em The main idea of the cosmological realization of the TMSM }studied in this paper and developed further in the work
  \cite{2-nd paper} is based on the fact that, unlike all other local quantum fields of the SM, the Higgs field has a component in the form of a condensate that fills the entire space of the Universe and is described by a classical scalar field. Cosmological averaging of this field provides us with  
 the homogeneous scalar field $\phi(t)$ which drives the evolution of the FLRW Universe and, in particular, the Higgs inflation  studied in this paper.  {\em Such a definition of the field $\phi(t)$ makes it meaningless to include $\phi(t)$ in the quantization procedure in the SM context along with the local quantum fields of the SM. Instead, the field $\phi(t)$ together with the function $\zeta$ and the curvature describe the classical cosmological background on which the electroweak SM exists as a local quantum field theory.}
\underline{If we consider the  GWS theory as} \underline{a pattern, then at each specific stage of the evolution of the cosmological background, the}\newline \underline{particle field theory is realized as a  cosmologically modified copy of the electroweak SM}.

\subsection{A concise summary of the main results of the TMSM in ref. \cite{2-nd paper} and a qualitative description of how it works}
\label{concise summary}

In  paper \cite{2-nd paper} the main idea formulated above is studied in detail for two stages of the evolution of the cosmological background: for slow-roll inflation and for the stage of approaching a vacuum.
Mainly due to the presence of $\zeta(\phi(t))$ in all equations of motion, all TMSM coupling constants turn out to be a kind of running  (classical) TMT-effective parameters.  During the evolution of 
the cosmological background, changing these parameters yields new results that are important for both particle physics and inflation:

  1) The classical running TMT-effective Higgs self-coupling increases from $\lambda\sim 10^{-11}$ at the stage of slow-roll inflation
 (which ensures consistency with the Planck's CMB data at $\xi=\frac{1}{6}$)
 to $\lambda_{SM}\sim 0.1$ at the stage close to the vacuum; 

2) The mass term in the TMT-effective Higgs potential changes sign from positive to negative,
which provides SSB in the standard way of GWS theory; 

3) The classical running constants of the gauge and Yukawa couplings change by several orders of magnitude;

4) The GWS theory is reproduced in such a way that the observed  hierarchy of fermion masses is obtained quite naturally. 

5) It is clearly shown that the theory, considered on the cosmological background at the the slow-roll inflation regime, is renormalizable.
  Taking into account quantum corrections 
 in the one-loop approximation preserves the slow-roll inflation regime and does not violate the vacuum stability.

6) At the classical level, the constraint (\ref{constraint using M and zetav}) is a self-consistency condition for the equations of motion derived from the {\em classical} primordial TMT action (\ref{S-gr-H-VB}). That is why, to take into account the quantum corrections to the constraint in ref. \cite{2-nd paper}, the {\em quantum} effective primordial TMT action in the 1-loop approximation was first found. Then, the self-consistency of the equations of motion derived from this action gives us a constraint that takes into account the quantum corrections. It turns out that the quantum corrections to the constraint are small, and their influence on cosmological evolution is negligible.

In the following part of this section we give a qualitative explanation of how the listed results can be achieved if
 the TMSM is described as a set of cosmologically modified copies of the GWS model, with each copy existing as a local quantum field theory defined on the classical cosmological background at the appropriate stage of its evolution.
The TMSM model, formulated in this way, dictates that we should start with the TMSM-primordial action (\ref{S-TMSM})-(\ref{S l3}) with the following two modifications: 
 a) unlike the standard TMT procedure, it is necessary to perform the  transition to the Einstein frame in this TMSM-primordial action; 
b) substituting for $\phi$, $\zeta$, $R$ the expressions $\phi_{(back)}(t)$, $\zeta_{(back)}$, $R_{(back)}$, which describe a certain stage of the evolution of the cosmological background, we obtain an action that describes a cosmologically modified copy of the GWS model corresponding to this background. For more details, see section 5.1  in ref. \cite{2-nd paper}. To briefly show how this is done, it suffices to restrict ourselves here to the $SU(2)$ Higgs doublet  $H$, neglecting for the time being the explicit inclusion of gauge bosons. Having carried out the above modifications, we arrive at the following  
\underline{tree-level} {\em action for the TMSM quantum Higgs field sector, considered on a cosmological background:}
\begin{eqnarray}
S^{(Higgs)}_{(on \, back)}=\int\sqrt{-\tilde{g}}d^4{\mathsf x}&&\Biggl[\frac{b_k-\zeta_{back}}{(1+\zeta_{back})\Omega_{back}}\tilde{g}^{\alpha\beta}H_{,\alpha}^{\dag}H_{,\beta}-\frac{\xi}{\Omega_{back}}R_{back}|H|^2
\nonumber
\\
&& -\frac{b_p+\zeta_{back}}{(1+\zeta_{back})^2\Omega_{back}^2}\lambda |H|^4-\frac{b_p-\zeta_{back}}{(1+\zeta_{back})^2\Omega_{back}^2}m^2 |H|^2\Biggr],
 \label{L H back general}
\end{eqnarray}
where $\Omega_{back}=1+\xi\frac{\phi_{(back)}^2}{M_P^2}$.
Bearing in mind that the functional quantization procedure will be applied to the action (\ref{L H back general}), it is necessary to redefine the Higgs field in such a way as to absorb the factor 
$\frac{b_k-\zeta_{back}}{(1+\zeta_{back})\Omega_{back}}$ in front of the kinetic term in the Lagrangian, bringing it to canonical form.
A similar structure of dependence on $\zeta_{back}$ appears for gauge bosons and fermion fields, which implies the need to redefine them in order to bring their kinetic terms to canonical form.

{\bf I. On the cosmological background near vacuum}, where $\zeta_{back}=\zeta_v$, the Lagrangian of the bosonic sector (see eqs.(\ref{S-H}), (\ref{S gauge}))
reduces to the following 
\begin{eqnarray}
L_{(on \, back \,near\,vac)}^{(bos)}&=&
 \frac{1}{2}(b_k-\zeta_v)\eta^{\alpha\beta}\left(\mathit{D}_{\alpha}H\right)^{\dag} \mathit{D}_{\beta}H-\frac{1+b_p}{4}\lambda |H|^4+\frac{1-b_p}{4}m^2 |H|^2\Bigr]
\nonumber
\\
&-&\frac{1}{4}(b_k-\zeta_v)\eta^{\mu\alpha}\eta^{\nu\beta}\left(\mathbf{F}_{\mu\nu}
\mathbf{F}_{\alpha\beta}+B_{\mu\nu}B_{\alpha\beta}\right),
\label{L bos back near vac}
\end{eqnarray}
where the choice of the zero value of CC was taken into account, and $\eta_{\alpha\beta}$ is the metric in Minkowski space.
As a result of the redefinition of the fields
\begin{equation}
\sqrt{\frac{b_k-\zeta_v}{2}}\cdot H=\mathcal{H}, \quad  \sqrt{b_k-\zeta_v}\cdot \mathbf{A}_\mu=\mathcal{A}_\mu, \quad \sqrt{b_k-\zeta_v}\cdot B_\mu=\mathcal{B}_\mu
\label{H in LPP near vac}
\end{equation}
and the primordial gauge coupling constants $g$, $g'$
\begin{equation}
\frac{g}{\sqrt{b_k-\zeta_v}}=\mathtt{g}, \quad \frac{g'}{\sqrt{b_k-\zeta_v}}=\mathtt{g}',
\label{gauge param near vac}
\end{equation}
the Lagrangian (\ref{L bos back near vac}) takes the canonical form for the bosonic sector of the electroweak GWS theory with the Higgs potential $\frac{1}{4}\lambda_{SM}(\mathcal{H}^{\dag}\mathcal{H})^2-\frac{1}{2}m_{SM}^2\mathcal{H}^{\dag}\mathcal{H}$,
where 
\begin{equation}
\lambda_{SM}=\frac{1+b_p}{(b_k-\zeta_v)^2}\lambda, \qquad m_{SM}^2=\frac{1-b_p}{2(b_k-\zeta_v)}m^2;
\label{lambda and m2 in LPP near vac}
\end{equation}
 Then the VEV with
$v=m_{SM}/\sqrt{\lambda_{SM}}\approx 246 \, GeV$
 and the Higgs boson mass $m_h\approx 125 \, GeV$, as in the GWS theory,
are obtained if\footnote{The relation between $v\approx 246 \, GeV$ and the value of $\sigma$, eq.(\ref{v2 m mv}), where the TMT-effective potential $U_{eff}(\phi)$ has a minimum (see also eq.(\ref{Ueff in all})),    is given by the formula
\begin{equation}
\sigma=\sqrt{\frac{2}{b_k-\zeta_v}}v\approx 8.7\cdot 10^4\, GeV.
\label{sigma to v relation}
\end{equation}
} \label{foot 16}
\begin{equation}
b_k-\zeta_v = 1.6\cdot 10^{-5} \quad \text{and} \quad m\approx 0.7 \, GeV, \quad \text{that is} \quad \frac{m^2}{M_P^2}\approx 8\cdot 10^{-38}.
\label{bk}
\end{equation}
 This is one of the results  of "the classical running TMT-effective Higgs self-coupling", mentioned in point 1) above. The choice of the parameter $b_p\approx \frac{1}{2}\zeta_v$ in  action (\ref{S-H 12}) provides the effect described  in point 2) above; interestingly, the estimate $b_p- \frac{1}{2}\zeta_v\sim 10^{-8}$ provides sufficient duration  of the slow-roll inflation. 
\newline
Generated by the Higgs phenomenon masses of $W$ and  $Z$-bosons, obtained after redefinitions (\ref{H in LPP near vac}) and (\ref{gauge param near vac}) in the  Lagrangian (\ref{L bos back near vac}), have the standard expressions
\begin{equation}
M_{W}=\frac{\mathtt{g}}{2}\cdot v, \qquad  
M_{Z}=\frac{\sqrt{\mathtt{g}^2+\mathtt{g}^{\prime 2}}}{2}\cdot v, \quad \text{as well as}\quad M_{A}=0 \quad \text{for photon}.
\label{MW and MZ eff}
\end{equation} 
As usual in the electroweak SM, there are relations between the values of the gauge coupling constants near vacuum, $\mathtt{g}$ and $\mathtt{g}'$, the Weinberg angle $\theta_W$ and the electric charge $e=\sqrt{4\pi\alpha}$ : \, 
$e=\mathtt{g}\sin\theta_W$, \,  $e=\mathtt{g}'\cos\theta_W$.

In ref. \cite{2-nd paper} it was shown that performing similar procedures with fermions on a cosmological background near a vacuum completely reproduces the  fermion sector of the GWS theory  with the following expressions for the TMT-effective Yukawa coupling constants  and masses of charged leptons and up-quarks:
\begin{equation}
Y_l= \frac{b_l+\zeta_v}{(b_l-\zeta_v)\sqrt{b_k-\zeta_v}}y^{(ch)},  \quad m_l=Y_l\frac{v}{\sqrt{2}};  \quad l=e, \mu, \tau, 
\label{ml and fl appr vac}
\end{equation}
\begin{equation}
Y_q=\frac{b_q+\zeta_v}{(b_q-\zeta_v)\sqrt{b_k-\zeta_v}}y^{(up)}, \quad m_q=Y_q\frac{v}{\sqrt{2}}; 
   \quad q=u,  c,  t . 
\label{mq appr vac}
\end{equation}
With the choice of the primordial universal Yukawa coupling parameters $y^{(ch)}\approx 10^{-10}$ for charged leptons and $y^{(up)}\approx 10^{-9}$ for up-quarks,
the measured fermion masses \cite{Part Data} are obtained with the following choice of the primordial parameters $b_l$ and $b_q$:
\begin{equation}
b_e\approx \zeta_v+1.7\cdot 10^{-2}; \quad b_{\mu}\approx \zeta_v+ 8\cdot10^{-5}; \quad b_{\tau}\approx \zeta_v+4.9\cdot 10^{-6},
\label{b_l values}
\end{equation}
\begin{equation}
b_u\approx \zeta_v+3.7\cdot 10^{-2}; \quad b_c\approx \zeta_v+6.7\cdot 10^{-5}; \quad b_t\approx \zeta_v+5\cdot 10^{-7}.
\label{b_q values}
\end{equation}
Thus, instead of six Yukawa coupling constants that differ significantly in order of magnitude, which is the case in the standard formulation, in TMSM
we can limit ourselves to one Yukawa coupling constant for all three generations of leptons and, similarly, one Yukawa coupling constant for all three generations of quarks. And the necessary hierarchy of fermion masses is achieved thanks to the six model parameters  $b_l$, $l=e,\mu,\tau$, and $b_q$, $q=u,c,t$. {\em The fact that all these six parameters are close to $\zeta_v$ allows us to consider the described model of fermion mass generation as more natural than in the standard model.}
It should be added that when obtaining these results it was  taken into account that the parameter $b_k$ is also very close to $\zeta_v$.

Exactly as in the usual formulation of the GWS theory, agreement with the experimental data obtained using the effective four-fermion Lagrangian  is achieved with the  standard VEV $v\approx 246\, GeV$, from which, using the redifinitions of the gauge coupling constants in eqs.(\ref{gauge param near vac}) and the  value of $(b_k-\zeta_v)$ obtained 
above, eq.(\ref{bk}), we find  the values of the primordial model parameters $g$ and $g'$ in the primordial action (\ref{S gauge})
\begin{equation}
g\approx 2.6\cdot 10^{-3}, \qquad g'\approx 1.4\cdot 10^{-3}.
\label{g and gprime}
\end{equation} 

{\bf II. On the cosmological background at the stage of  the slow-roll inflation}, which is realized at $\phi \gtrsim 14.2M_P$ (which corresponds to $\varphi\gtrsim 6M_P$),  the background scalar  $\zeta_{back}$ is bounded from above by the value $\zeta_{back}\lesssim 5\cdot 10^{-3}$.
Therefore, up to a very small relative correction, one can neglect  $\zeta_{back}$ in the combinations $(1+\zeta_{back})$, $(b_k-\zeta_{back})$,  $(b_p\pm\zeta_{back})$, $(b_l\pm\zeta_{back})$ and $(b_q\pm\zeta_{back})$. 
Proceeding in this approximation when constructing the action for quantum matter fields of TMSM on the background at the stage of slow-roll inflation, we can simply set $\zeta_{back}=0$.
Then, for example, the Lagrangian (\ref{L H back general}) reduces to the following  
\begin{equation}
 L(H)|_{( \phi_{back}>14.2M_P)}=\frac{M_P^2}{\xi\phi_{back}^2}\tilde{g}^{\alpha\beta}H_{,\alpha}^{\dag}H_{,\beta}
 -\frac{M_P^4}{2\xi^2\phi_{back}^4}\lambda |H|^4-\frac{M_P^2}{2\xi\phi_{back}^2}m^2 |H|^2
-\frac{M_P^2}{\phi_{back}^2}R_{back}|H|^2,
 \label{L bos back slow roll}
\end{equation}
where the background scalar curvature $R_{back}= -\frac{\lambda}{2\xi^2} M_P^2$. 
Since the background is in a slow-roll  inflationary stage, the background field $\phi_{back}$ can be viewed as a slow dynamical variable compared to $H$, which is quantized at Planck scale and is therefore a fast variable.
Then, using redefinition  
\begin{equation}
\tilde{H}({\mathsf x})=\frac{M_P}{\sqrt{\xi}\phi_{back}(t)}H({\mathsf x}),
\label{redif H in slow}
\end{equation}  
we can neglect the term $\propto\dot{\phi_{back}}/\phi_{back}$, and, up to relative corrections of order $\lesssim 3\cdot 10^{-2}$,  the Lagrangian (\ref{L bos back slow roll}) is reduced to the form
\begin{equation}
L(\tilde{H} )|_{(slow \, roll \, infl)} =\tilde{g}^{\alpha\beta}\tilde{H}_{,\alpha}^{\dag}\tilde{H}_{,\beta} -V(\tilde{H})
\nonumber
 \label{L bos after redef}
\end{equation}
\begin{equation}
V(\tilde{H})= \frac{1}{2}\lambda |\tilde{H}|^4+\frac{1}{2}m^2 |\tilde{H}|^2+\xi R_{back}|\tilde{H}|^2. 
\label{V tilde H}
\end{equation}
Using the  estimate for $m^2$ in eq.(\ref{bk}),  we see that the second term in $V(\tilde{H})$ is negligible compared to the last term. 
The potential $V(\tilde{H})$ has a minimum at
\begin{equation}
|\tilde{H}_0|^2=  \frac{1}{2}\tilde{v}^2=- \frac{\xi}{\lambda}R_b=\frac{1}{2\xi}M_P^2 
\label{tilde H2min}
\end{equation}
with
\begin{equation}
 \tilde{v}=\frac{1}{\sqrt{\xi}}M_P=\sqrt{6}M_P
\label{tilde v}
\end{equation}
if $\xi=\frac{1}{6}$.
This means that, {\em up to relative corrections of order $\lesssim 3\cdot 10^{-2}$, during the entire slow-roll inflation stage, the potential $V(\tilde{H})$} of the Higgs field $\tilde{H}({\mathsf x})$  {\em has a minimum independent of the slowly rolling  background field $\phi_b$}. 
Note that by analogy with the GWS theory, near the minimum of the potential $V(\tilde{H})$ in the matrix of the $SU(2)$  gauge transformations it is necessary to use $\tilde{v}=\sqrt{6}M_P$ instead of $v$.
\newline
Continuing in the described above approximation, i.e. setting $\zeta_{back}=0$, one can see that the TMT-effective gauge coupling parameters at the slow-roll inflation stage are equal to their primordial values  given by eq.(\ref{g and gprime}), while
the TMT effective Yukawa coupling parameters are almost
 the same for all leptons,
$Y^{(ch)}_{eff}|_{(slow\, roll\, infl)}\approx y^{(ch)}\approx 10^{-10}$,
and similarly for all up-quarks
$Y^{(up)}_{eff}|_{(slow\, roll\, infl)}\approx y^{(up)}\approx 10^{-9}$.
\newline
{\em The quantization of the TMSM on the background formed at the slow-roll inflation} is also studied in detail in ref. \cite{2-nd paper} and the 1-loop effective potential (corresponding to this background) is calculated. In contrast to the generally accepted theory, due to the smallness of the primordial model parameters of the gauge coupling $g$ and $g'$, the radiative corrections $\propto g^4\sim g'^4$ have a weak effect on the effective quartic Higgs self-interaction.
On the other hand, the negative fermion contribution to the 1-loop effective potential at the slow-roll inflation stage is proportional to $(Y^{(ch)}_{eff}|_{(slow\, roll\, infl)})^4\sim 10^{-40}$ and 
$(Y^{(up)}_{eff}|_{(slow\, roll\, infl)})^4\sim 10^{-36}$. Due to the extreme smallness of these parameters, the fermion contribution turns out to be many orders of magnitude smaller than the contribution of the gauge fields. Thus, the minimum of the studied one-loop effective potential is absolute, and the vacuum is stable. 
A detailed analysis in \cite{2-nd paper}  shows that the back reaction of one-loop corrections to the shape of the TMT-effective potential  at $\varphi\gtrsim 8.1M_P$, with relative corrections $\lesssim 3\cdot 10^{-2}$, is expressed in the fact that instead of the tree-level potential (\ref{1-st plateau 1}) we get
\begin{equation}
\tilde{U}(\varphi)\equiv U_{eff}^{(1-loop\, corrected)}(\varphi)|_{\varphi\gtrsim 8.1M_P}= \frac{\lambda M_P^4}{8\xi^2}
\Bigl(1+1.4\cdot10^{-2}\cdot\frac{\varphi}{M_P}\Bigr) \Bigl(1-8e^{-\sqrt{\frac{2}{3}}\frac{\varphi}{M_P}} \Bigr).
\label{varphi >6MP 1-loop}
\end{equation}
It is easy to see by direct calculation that the flatness conditions imposed on the TMT-effective potential are satisfied with a large margin
\begin{equation}
\epsilon|_{\varphi\gtrsim 8.1M_P}=\frac{M_P^2}{2}\Biggl(\frac{\tilde{U}'}{\tilde{U}}\Biggr)^2\approx\frac{10^{-4}}{\Bigl(1+1.4\cdot10^{-2}\cdot\frac{\varphi}{M_P}\Bigr)^2}\ll 1, 
\label{epsilon}
\end{equation} 
\begin{equation}
|\eta|_{\varphi\gtrsim 8.1M_P}=M_P^2\Bigl|\frac{\tilde{U}''}{\tilde{U}} \Bigr|\approx \frac{16}{3}\cdot e^{-\sqrt{\frac{2}{3}}\frac{\varphi}{M_P}} \ll 1.
\label{eta}
\end{equation} 
Therefore, we can claim that, as in the classical model studied in this paper, taking into account quantum corrections in the 1-loop approximation does not change the fact that inflation quickly passes into the slow-roll regime.

\subsection{Preheating. Key findings from the preliminary study}
\label{Preheating}

Understanding the preheating mechanisms in the TMSM cosmological scenario is extremely important and will be studied in detail in the future. However, here we can present a preliminary approach to addressing the preheating problem by considering one possible preheating channel:  the production of fermions interacting with a coherently oscillating background scalar field shortly after the end of inflation. When such a scalar field is the inflaton, this channel of post-inflationary preheating has been thoroughly studied in the literature \cite{GrKof1}, \cite{GrKof2}, \cite{GiudiceRiottoTkachev}, \cite{Garcia-Bellido}. In the model we study, the role of the inflaton is played by the only non-zero component of the cosmologically averaged Higgs field. At the end of this section, we will discuss some fundamentally new aspects of this scenario that naturally arise within the cosmological realization of the TMSM.

First of all, it is necessary to understand how, despite the smallness of the primordial model parameters  $y^{(ch)}\approx 10^{-10}$, $y^{(up)}\approx 10^{-9}$, the preheating can be effective. To do this, let us consider more closely the behavior of the function $\zeta(\phi)$, determined by constraint (\ref{constraint using M and zetav})  in terms of $\phi$ (or by (\ref{zeta via varphi}) in terms of $\varphi$). According to the results presented in figure \ref{fig8}, inflation terminates at $\varphi\approx 2M_P$. Let's consider what happens for $\phi \lesssim 0.1M_P$, where we have taken into account that $\varphi =\phi$ with high accuracy for $\varphi \lesssim 0.1M_P$.
Assuming that the contribution of $X_{\phi}/M_P^4$ is negligible (see footnote 10), the behavior of $\zeta(\phi)$ in the entire interval from $\varphi\gg M_P$ to $\varphi= \sigma=\phi$ is given by curve 3 in figure 2. However, this graph does not provide sufficient information about the behavior of 
	$\zeta(\phi)$ after inflation. With the values of the model parameters used in figure 2, for $\phi \lesssim 0.1M_P$, constraint  (\ref{constraint using M and zetav}) reduces to the form
\begin{equation}
\zeta|_{\phi \lesssim 0.1M_P}\approx\frac{2q^4\zeta_v(1+\zeta_v)}{2q^4(1+\zeta_v)+\frac{\lambda}{4}\frac{\phi^4}{M_P^4}}
\approx1-4.8\cdot 10^{-3}\frac{\phi^4}{M_P^4},
\label{constraint for phi<0.1}
\end{equation}
where, as in figure 2, to obtain numerical estimates we used the choice $V_1=V_2$, that is, $\zeta_v=1$. 
Therefore, neglecting relative corrections of the order $ \lesssim 5\cdot 10^{-7}$,  {\em one can use the value of $\zeta$ 
in the vacuum, $\zeta_v=1$,  as the background scalar $\zeta_{back}$ during cosmological evolution  when $|\phi |\lesssim 0.1M_P$}. 
Then with the same accuracy  we will use $b_k-\zeta= b_k-1$,  $b_l-\zeta= b_l-1$ ($l=e, \mu, \tau$) and $b_q-\zeta= b_q-1$ ($q=u, c, t$).

Using cosmological equations (\ref{K-Frid prelim}) and (\ref{K-FLRW varphi}) following from the TMT-effective action, eqs.(\ref{Seff}), (\ref{Leff}),  the TMT-effective potential $U_{eff}^{(tree)}(\phi)$, eq.(\ref{V eff}), and definitions of the  functions $K_1(\phi)$ and $K_2(\phi)$, eqs.(\ref{K1 phi}) and (\ref{K2 phi}),   we find that for $|\phi |\lesssim 0.1M_P$ the cosmological dynamics is described by equations\footnote{Following the remark in footnote 4, we choose the correction 
$\delta$ to the integration constant, eq.(\ref{M 2sqrt}),   so that the value of the TMT-effective potential  in vacuum, eq.(\ref{U vac}), is equal to zero .}
\begin{equation}
\ddot{\phi}+3\frac{\dot{a}}{a}\dot{\phi}+\frac{3}{4(b_k-1)}\lambda\phi^3-\frac{1}{4(b_k-1)}m^2\phi=0,
\label{K-FLRW varphi reh}
\end{equation}
\begin{equation}
\frac{\dot{a}^2}{{a}^2}=\frac{1}{3M_P^2}\Bigl[\frac{b_k-1}{2}\cdot
\frac{\dot{\phi}^2}{2}+\frac{3\lambda}{32}\bigl(\phi^2-\sigma^2\bigr)^2\Bigr],
\label{K-Frid prelim reh}
\end{equation}
where $\sigma^2=\frac{m^2}{3\lambda}$ (see eqs.(\ref{v2}), (\ref{lambda and m2 SM}) and (\ref{v2 m mv}) for the values $\zeta_v=1$ and $b_p=1/2$). The appearance of the non-canonical kinetic term is a residual effect of the K-essence structure of the TMT-effective action, which, as we investigated in section \ref{impact of K-ess}, plays an important role at the very initial stage of the evolution of the Universe.
The redefinition $\boldsymbol{\phi}(t)=\sqrt{\frac{b_k-1}{2}}\phi(t)$ brings these equations to the canonical form\footnote{With such a redefinition, for example, the following equality is valid:  $\frac{\lambda_{SM}}{4}\boldsymbol{\phi}^4=\frac{\lambda}{4}\phi^4$.}
\begin{equation}
{\boldsymbol{\ddot\phi}}+3\frac{\dot{a}}{a}{\boldsymbol{\dot\phi}}+\lambda_{SM}\boldsymbol{\phi}^3
-m_{SM}^2\boldsymbol{\phi}=0,
\label{K-FLRW varphi reh}
\end{equation}
\begin{equation}
\frac{\dot{a}^2}{{a}^2}=\frac{1}{3M_P^2}\Bigl[
\frac{\boldsymbol{\dot{\phi}}^2}{2}+\frac{\lambda_{SM}}{4}\bigl(\boldsymbol{\phi}^2-v^2\bigr)^2\Bigr],
\label{K-Frid prelim reh}
\end{equation}
where the SM parameters
$\lambda_{SM}$ and $m_{SM}$ are defined by eq.(\ref{lambda and m2 in LPP near vac}); see also footnote 16. Note that, unlike the similar redefinition of the quantum Higgs field in eq.(\ref{H in LPP near vac}), we are now working with a classical scalar field.
 Thus, unexpectedly small deviations of $\zeta$ and $b_k-\zeta$ from their values in vacuum already at $\phi\leq 0.1 M_P$ lead to a remarkable result:  after the end of inflation, but {\em long before reaching the vacuum state}, the TMT-effective self-coupling parameter of the Higgs field $\phi$ runs from the model parameter $\lambda=2.3\cdot 10^{-11}$  to a value practically coinciding with the self-coupling constant 
$\lambda_{SM}$ of the Higgs field $\boldsymbol{\phi}$ in the GWS theory. Below we will see that similar effect occurs with the TMT-effective Yukawa coupling parameters. 

We are interested in oscillatory solutions for $\boldsymbol{\phi}(t)$, that begin with the value  
$\boldsymbol{\phi}=\sqrt{\frac{b_k-1}{2}}\phi\approx 2.8\cdot 10^{-2}\phi\sim 7\cdot 10^{15}GeV$, that is 
$\boldsymbol{\phi}^2\approx 8\cdot 10^{26}v^2$. Preheating models with the SM Higgs potential $\frac{1}{4}\lambda_{SM}\bigl(\boldsymbol{\phi}^2-v^2\bigr)^2$ are studied  in papers \cite{Freese}, \cite{Vis},  where   $v^2$ can be safely neglected compared to $\boldsymbol{\phi}^2$ during preheating;  for a detailed discussion of this point see the earlier paper \cite{GKLS}. Then the Higgs potential is adequately modeled by a pure quartic term  $\frac{\lambda_{SM}}{4}\boldsymbol{\phi}^4$. Oscillatory solutions in this model   are well studied in refs. \cite{GKLS} and \cite{Brandenberger}, where using different approaches a very similar solutions were found. In the approach described in \cite{GKLS}, it is shown that after averaging over several oscillations,  
eqs.(\ref{K-FLRW varphi reh}) and (\ref{K-Frid prelim reh}) rewritten with the potential $\frac{\lambda_{SM}}{4}\boldsymbol{\phi}^4$ and expressed in terms of conformal time $\eta=\int\frac{dt}{a(t)}$ and field $\boldsymbol{\varphi}=a\boldsymbol{\phi}$, take the form
\begin{equation}
\boldsymbol{\varphi}^{\prime\prime}+\lambda_{SM}\boldsymbol{\varphi}^3=0,
\label{K-FLRW varphi reh av conf}
\end{equation}
\begin{equation}
a^{\prime 2}=\frac{1}{3M_P^2}\Bigl(
\frac{\boldsymbol{\varphi\prime}^2}{2}+\frac{\lambda_{SM}}{4}\boldsymbol{\varphi}^4\Bigr)\equiv\frac{1}{3M_P^2,}\rho_{\boldsymbol{\varphi}}
\label{K-Frid prelim reh av conf}
\end{equation}
where $^\prime$ stands for the derivative with respect to $\eta$. The energy density $\rho_{\boldsymbol{\varphi}}$ of the oscillating field $\boldsymbol{\varphi}(\eta)$ is  expressed in terms of the amplitude of the oscillations $\tilde{\boldsymbol{\varphi}}$: 
$\rho_{\boldsymbol{\varphi}}=\frac{\lambda_{SM}}{4}\tilde{\boldsymbol{\varphi}}^4$.  From eq.(\ref{K-Frid prelim reh av conf}) it follows that
\begin{equation}
a(\eta)=\sqrt{\frac{\lambda_{SM}}{12}}\frac{\tilde{\boldsymbol{\varphi}}^2}{M_P}\eta, \qquad t=\sqrt{\frac{\lambda_{SM}}{48}}\frac{\tilde{\boldsymbol{\varphi}}^2}{M_P}\eta^2.
\label{a(eta) t(eta2)}
\end{equation}
Then, using a dimensionless conformal time 
\begin{equation}
\tau=\sqrt{\lambda_{SM}}\tilde{\boldsymbol{\varphi}}\eta=(48\lambda_{SM}M_P^2)^{1/4}\sqrt{t}
\label{tau}
\end{equation}
in ref. \cite{GKLS} is shown that solution of eq.(\ref{K-FLRW varphi reh av conf}) can be represented as
$\boldsymbol{\varphi}=a\boldsymbol{\phi}=\tilde{\boldsymbol{\varphi}}f(\tau)$, where the function $f(\tau)$ has an amplitude equal to unity and obeys  the canonical equation for the elliptic function. The latter has the solution in terms of an elliptic cosine
\begin{equation}
f(\tau)=cn\bigl(\tau,\frac{1}{\sqrt{2}}\bigr).
\label{f cn}
\end{equation}
 The period of the oscillations of $f(\tau)$  (in units of $\tau$) is $T=7.416$. The  first term $\propto \cos\frac{2\pi\tau}{T}$ in the harmonic expansion of $f(\tau)$ has the amplitude 0.955. The potential energy density $\frac{\lambda_{SM}}{4}\boldsymbol{\varphi}^4$ of the background field $\boldsymbol{\varphi}$ averaged over a period of oscillation is equal to $\frac{1}{3}\rho_{\boldsymbol{\varphi}}$. 

The oscillatory solutions described above were used in the paper \cite{GrKof1} within the framework of a model where, in addition to the potential $\frac{1}{4}\lambda\phi^4$ of the  inflaton background field, its Yukawa coupling with the quantum field of fermions is taken into account. The authors of  \cite{GrKof1}  found that in such a model, the particle production  occurs in the regime of the parametric excitation, which can provide an effective preheating mechanism. Below it will be shown that the cosmological realization of TMSM  includes a stage that naturally, i.e. without any additional assumptions, contains all the components of the preheating model \cite{GrKof1}.

To describe fermions as quantum fields on the cosmological background formed by the coherently oscillating  scalar field $\phi(t)$ at 
 $|\phi|\lesssim 0.1M_P$ we begin with representing the  fermionic primordial TMT-action (\ref{S l3}) in the Einstein frame. Taking into account the results formulated after eq.(\ref{constraint for phi<0.1}), the corresponding transformations are reduced to the following:
$g_{\mu\nu}=\frac{1}{1+\zeta_{back}}g_{\mu\nu}=\frac{1}{2}\tilde{g}_{\mu\nu}$, \, $V^{\mu}_k=\sqrt{2}\tilde{V}^{\mu}_k$, \,
$V_{\mu}^k=\frac{1}{\sqrt{2}}\tilde{V}_{\mu}^k$, \,
$\omega_{\mu}^{ik} \to \tilde{\omega}_{\mu}^{ik}=\frac{1}{2}\Bigl[ \tilde{V}^{i\lambda}\Bigl(\frac{\partial\tilde{V}^{k}_{\lambda}}{\partial{\mathsf x}^{\mu}}+\{^{\beta}_{\mu\lambda}\}\tilde{V}^{k}_{\beta}\Bigr)-(i \leftrightarrow k)\Bigr]$. Omitting gauge bosons, 
it is also necessary to make the replacement $\nabla_{\mu}\to\tilde{\nabla}_{\mu}=\partial_{\mu}+\frac{1}{2}\tilde{\omega}_{\mu}^{ik}\sigma_{ik}$. The resulting fermionic action on the described cosmological background takes the form
\begin{equation}
S^{(ferm)}_{(on \, back)}=\int \Bigl[L^{(lept)}_{(on \, back)}+L^{(quarks)}_{(on \, back)}\Bigr] \frac{1}{4}\sqrt{-\tilde{g}}d^4x, 
\label{Sferm LPP on back}
\end{equation}
with the following Lagrangian for charged leptons \footnote{The notation $\tilde{\gamma}^{\mu}=\tilde{V}_{k}^{\mu}\gamma^{k}$ is used.}
\begin{equation}
L^{(lept)}_{(on \, back)}=
\sum_{l=e,\mu,\tau}\Bigl[ (b_l-1)\sqrt{2}i \Bigl(\overline{l}\tilde{\gamma}^{\mu}\tilde{\nabla}_{\mu}l -(\tilde{\nabla}_{\mu}\overline{l})\tilde{\gamma}^{\mu}l\Bigr)-(b_l+1) \frac{y^{(ch)}}{\sqrt{2}}\overline{l}l\phi \Bigr],
\label{L lept on back}
\end{equation}
where $l=l_L+l_R$ and the definition (\ref{H cosm av}) was used. The Lagrangian  
$L^{(quarks)}_{(on \, back)}$ for up-quarks has a similar structure, and for brevity we omit it. 
It is convenient to continue with the notations $\psi_f$, $b_f$ ($f=e,\mu,\tau$  for charged leptons; $f=u,c.t$  for up-quarks) and
\begin{equation}
Y_f=\frac{b_f+1}{(b_f-1)\sqrt{b_k-1}}y, \,  \text{where} \, y=y^{(ch)}\approx 10^{-10} \, \text{ for}\, e,\mu,\tau;\, y=y^{(up)}\approx 10^{-9} \, \text{for}\, u,c.t.
\label{Y f}
\end{equation}
Then, after using the canonically normalized scalar field  $\boldsymbol{\phi}=\sqrt{\frac{b_k-1}{2}}\phi$, the Dirac equations for the fermion fields $\psi_f$ 
take the form
\begin{equation}
\bigl[i\tilde{\gamma}^{\mu}\tilde{\nabla}_{\mu} -Y_f\cdot\boldsymbol{\phi}(t)\bigr]\psi_f=0.
\label{psi eq with t}
\end{equation}
In the FRW space-time $\tilde{\gamma}^0=\gamma^0$, $\tilde{\gamma}^i=\frac{1}{a(t)}\gamma^i$, where the $\gamma^{\alpha}$'s are Dirac matrices in the Minkowski space-time, and $\frac{1}{2}\gamma^{\mu}\tilde{\omega}_{\mu}^{\alpha\beta}\sigma_{\alpha\beta}=\frac{3}{2}(\frac{\dot{a}}{a})\gamma^0$. Transforming the fields $\boldsymbol{\varphi}=a\boldsymbol{\phi}$ and 
$\boldsymbol{\psi}_f=a^{3/2}\psi_f$ in combination with the use of  conformal dimensionless time $\tau$, eq.(\ref{tau}), and the  above definition $\boldsymbol{\varphi}=\tilde{\boldsymbol{\varphi}}f(\tau)$ allows us to reduce eq.(\ref{psi eq with t}) to the following
\begin{equation}
\Bigl[i\sqrt{\lambda_{SM}}\tilde{\boldsymbol{\varphi}}\gamma^0\frac{\partial}{\partial\tau} +
\vec{\gamma}\vec{\nabla}-Y_f\cdot\boldsymbol{\varphi}\Bigr]\boldsymbol{\psi}_f=0.
\label{psi eq with tau}
\end{equation}
In ref.\cite{GrKof1}, to find the eigenfunctions of quantum fluctuations, the Dirac equation is reduced to a second-order equation for the auxiliary field $X^{(f)}(\tau,\vec{x})$. In the model we are studying, the corresponding ansatz has the form
$\boldsymbol{\psi}_f=\Bigl(i\sqrt{\lambda_{SM}}\tilde{\boldsymbol{\varphi}}\gamma^0\frac{\partial}{\partial\tau} +
\vec{\gamma}\vec{\nabla}+Y_f\cdot\boldsymbol{\varphi}\Bigr)X^{(f)}$. 
Next step\footnote{See also ref.\cite{GrKof2}} is the decomposition of the auxiliary spinor 
$X^{(f)}(\tau,\vec{x})$ into eigenmodes of the form $X^{(f)}_{\mathbf{k}}(\tau)R_{\pm}e^{i\mathbf{k}\mathbf{x}}$, 
where $R_{\pm}(\mathbf{k})$ are eigenvectors of the helicity operator $\mathbf{k}\mathbf{\Sigma}$: $\mathbf{k}\mathbf{\Sigma}R_{\pm}(\mathbf{k})=\pm 1$ and $\gamma^0R_{\pm}(\mathbf{k})=+1$. The temporal part of the eigenmode obeys the equation
\begin{equation}
\frac{d^2X^{(f)}_{\mathbf{k}}}{d\tau^2}+\Bigl(\kappa^2+q^{(f)}f^2-i\sqrt{q^{(f)}}\frac{df}{d\tau}\Bigr)X^{(f)}_{\mathbf{k}}=0,
\label{mode X eq}
\end{equation}
where the comoving momentum $\mathbf{k}$ enters the equation in the combination 
$\frac{\mathbf{k}^2}{\lambda_{SM}\tilde{\boldsymbol{\varphi}}^2}=\kappa^2$. The background oscillations enter in the form 
$f(\tau)=cn\bigl(\tau,\frac{1}{\sqrt{2}}\bigr)$, eq.(\ref{f cn}),  having unit amplitude. 
For a fermion $f$, the interaction force with periodic oscillations $f^2(\tau)=\bigl(cn\bigl(\tau,\frac{1}{\sqrt{2}}\bigr)\bigr)^2$ is determined by the dimensionless resonance parameter $q^{(f)}$, the expression of which in the model we are studying is reduced to the following combination of observable quantities of the GWS  theory
\begin{equation}
q^{(f)}=\frac{Y_f^2}{\lambda_{SM}}=\frac{1}{\lambda_{SM}}\Bigl(\frac{\sqrt{2}m_f}{v} \Bigr)^2,
\label{qf}
\end{equation}
where $m_f$ is the mass of the fermion $f$ and $v\approx 246GeV$. 

In \cite{GrKof1}, eq.(\ref{mode X eq}) is solved numerically and the $\tau$-dependence of the comoving occupation number of particles $f$ in a given spin state is found with the formula
\begin{equation}
n^{(f)}_{\mathbf{k}}(\tau)=\frac{1}{2}-\frac{1}{\sqrt{\kappa^2+q^{(f)}\cdot f^2(\tau)}}\Bigl[\kappa^2 Im \bigl( X^{(f)}_{\mathbf{k}}X^{(f)*}_{\mathbf{k}}\bigr)
+\frac{1}{2}\sqrt{q^{(f)}}\cdot f(\tau) \Bigr]
\label{nk}
\end{equation}
The energy density of created fermions $f$ is $\rho^{(f)}=
\frac{1}{2\pi^3}\int n^{(f)}_{\mathbf{k}}\bigl(\kappa^2+q^{(f)}\cdot f^2\bigr)^{1/2}d^3k$.
The results of numerical calculations presented in \cite{GrKof1} show that for all values of the the resonance parameter $q$ (i.e. $q^{(f)}$ in the model under consideration) in the range $10^{-4}\leq q\leq 100$ the fermion occupation number oscillates with a high frequency (period $<T/2$). The smoothed occupation number obtained by averaging $\bar{n}^{(f)}_{\mathbf{k}}(\tau)=\frac{1}{T}\int_{\tau}^{\tau+T}n^{(f)}_{\mathbf{k}}(\tau)d\tau$ can be represented in a form
$\bar{n}^{(f)}_{\mathbf{k}}(\tau)=F^{(f)}_{\mathbf{k}}\sin^2\nu_{\mathbf{k}}\tau$. The envelope function $F^{(f)}_{\mathbf{k}}$ and the period $\frac{\pi}{\nu_{\mathbf{k}}}$ depend on $q^{(f)}$. A semi-analytical approach, originally developed in ref.\cite{MostepGrib} , allows  to obtain an expression for the envelope function $F^{(f)}_{\mathbf{k}}$ of maximum occupation number via solutions of 
eq.(\ref{mode X eq}). It turns out\cite{GrKof1} that the fermionic modes get fully excited, reaching their time-averaged value $F^{(f)}_{\mathbf{k}}/2$ in just tens of oscillations\footnote{It should be noted that the small number of oscillations allows us to consider that we are using an acceptable approximation when we neglect  $v^2=m_{SM}^2/\lambda_{SM}$ compared to $\boldsymbol{\varphi}^2$ in eqs (\ref{K-FLRW varphi reh}) and (\ref{K-Frid prelim reh}).
 For comparison, in the perturbative approximation \cite{KLS1}, the timing for exitation via decay of the inflaton to pair of fermions is about $10^{14}$ inflaton oscillations.} of the coherently oscillating background scalar field. Following the authors of \cite{GrKof1}, one can speak of  the effect of parametric excitation of fermions, where $q^{(f)}$ plays the role of the resonance parameter.
For three generations of charged leptons and three generations of up-quarks considered in the context of TMSM, we arrive at the following estimates: $q^{(e)}\sim 3\cdot 10^{-6}$, \,  $q^{(\mu)}\sim 6\cdot 10^{-4}$, \,  $q^{(\tau)}\sim  10^{-2}$, \,
 $q^{(u)}\sim 10^{-9}$, \,  $q^{(c)}\sim 5\cdot 10^{-4}$, \,  $q^{(t)}\sim 10$.

The efficiency of the preheating  is determined by the fraction of the background scalar field energy 
$\rho_{\boldsymbol{\phi}}=\frac{\lambda_{SM}}{4}\tilde{\boldsymbol{\varphi}}^4$ transferred to the created fermions $\rho_f$.
The greater $q$, the greater the efficiency. In the case of the broad resonance excitation, $q\gg 1$,
in \cite{GrKof1} the following estimate was obtained:
$\rho_f/\rho_{\boldsymbol{\phi}}\sim 0.1\cdot Y_f^2\cdot q^{1/4}$.  Of all the standard model fermions, this estimate can only be applied to the t-quark, and for the preheating efficiency of the t-quarks, we obtain an estimate  $\rho_t/\rho_{\boldsymbol{\phi}}\sim 0.18$. The study in paper \cite{GiudiceRiottoTkachev} confirms the expected result that the preheating  efficiency  of very heavy fermions with masses of order $10^{17}-10^{18} GeV$ is high.
For fermion and boson preheating after hybrid inflation see ref.\cite{Garcia-Bellido}.

Apparently, there is a need to discuss some aspects of preheating in the model we are studying that are fundamentally new in comparison with preheating models known in the literature.
The hypothesis of the presence in the Lagrangian of the terms $\sim \phi^2\chi^2$ and $\sim \bar{\psi}\psi\phi$, describing the interaction of the inflaton field $\phi$ with the scalar $\chi$ and the fermion $\psi$ particle fields, is inevitable for all known preheating models,  while experimental data on the values of the corresponding coupling constants are absent.
 In this regard, in the cosmological implementation of  the TMSM the situation is fundamentally new. For example, the Yukawa interactions in the fermionic sector of the TMSM, eq.(\ref{S l3}),  considered on the background produced by the cosmologically averaged Higgs field during its oscillations after inflation, make the additional postulation of an interaction of the type $\sim \bar{\psi}\psi\phi$ completely superfluous. Moreover, it turns out that soon after the end of inflation the TMT-effective   Yukawa coupling parameters practically coincide with the Yukawa coupling constants of the GWS theory. It is easy to see that a similar effect is observed for the coupling parameters of gauge bosons.
In contrast to the preheating in conventional Higgs inflation models, where finding the coupling parameters at high energies implies  renormalizability (which poses a serious problem for the theory), in the cosmological implementation of the TMSM, the TMT-effective coupling parameters behave like classical running coupling constants (see section \ref{concise summary}).

\section{Discussion}
\label{Discussion}
In the discussion of the basic ideas and main results of the paper, it is worth concentrating on the role played by the combination of the presence of two volume measures  in the TMSM with the flatness of the TMT-effective potential. 
To understand {\em how Higgs inflation becomes possible at small $\xi$}, we should start by recalling the special properties of the model that allow us to overcome the apparent contradiction between physics at inflationary-scale energies and the energies of accelerator experiments. The choice of a small value of the non-minimal coupling constant $\xi$ can be consistent with the CMB data only if the primordial model parameter $\lambda$ of the Higgs field selfcoupling is very small: as was shown,  $\lambda=2.3\cdot 10^{-11}$ if  $\xi=1/6$. But such a value of $\lambda$ is completely unsuitable for the 
implementation of the GWS theory. The resolution of this contradiction in the studied TMSM turns out to be possible only due to two volume measures  in the primordial action. The main consequence of this is that $\lambda$ becomes a kind of running classical coupling constant, the value of which at energies near the Higgs field vacuum turns out to be approximately $10^{10}$ times larger than during the inflation stage. The very possibility of such a huge difference is the result of the fact that the value of the coefficient $b_k$ in the volume element $(b_k\sqrt{-g}-\Upsilon)d^4x$ is chosen extremely close to $\zeta_v$: 
$b_k=\zeta_v+\mathcal{O}(10^{-5})$. 

It would be natural to assume that the dynamics of the fields obtained from the primordial action of the TMSM must ensure that  all total volume measures
$(b_i\sqrt{-g}\pm \Upsilon) d^4x=(b_i\pm \zeta)\sqrt{-g} d^4x$ are positive throughout the evolution of the Universe. Direct verification made in this paper and in paper \cite{2-nd paper} shows that this requirement is met. The only exception is the total volume measure $(b_p\sqrt{-g}-\Upsilon) d^4x$ with the Higgs field mass term in  action (\ref{S-H 12}) (and, in the more general formulation,  eq.(\ref{S-H})). When $\zeta$ increases during cosmological evolution from $\zeta\approx 0$ at the inflationary stage to  
$\zeta=\zeta_v=\sqrt{\frac{V_1}{V_2}}>\frac{1}{2}$ near vacuum, at some point in cosmic time $\zeta$ intersects the value  $b_p$ and the total volume measure 
$(b_p\sqrt{-g}-\Upsilon) d^4x=(b_p-\zeta)\sqrt{-g} d^4x$ becomes negative. This pre-geometric effect (see the last paragraph in section \ref{Why Two-Measure theory})  is of particular interest because it manifests itself as the appearance of a negative squared mass of the Higgs field.

The choice of the dimensionless parameters $b_k$ in the bosonic sector, $b_l$, $l=e, \mu, \tau$, and $b_q$, $q=u, c, t$, in the fermionic sector, deserves a separate discussion. It is tempting to explain the closeness of all these parameters to $\zeta_v$ by a violation of some initial balance between $\sqrt{-g}$ and $\Upsilon$. However, as long as the mechanism responsible for the breaking of this type of symmetry remains completely unknown, we are forced to regard the choice of these parameters  simply as an adjustment to achieve agreement with the measured quantities. 

Now let us return to the problem mentioned in the introductory section \ref{sec:intro} that 
 {\em the initial conditions for inflation are unnatural}. Such claims are widespread in the literature
 and are usually based on statistical estimates that the initial conditions
 (\ref{Cond for begin infl}) required for inflation to begin are exponentially rare. 
This argument became even more convincing when it turned out that models with plateau-like
 potentials are the most preferable in light of the Planck's CMB results. As the authors of 
the paper \cite{Stein}  noted, for a flat potential of height $\sim 10^{-10}M_P^4$ 
one would expect that initial values of the kinetic and gradient energy densities exceeding 
$\sim 10^{-10}M_P^4$, up to the Planck scale, are the most probable; and conversely, 
values of these energies satisfying the conditions (\ref{Cond for begin infl}) and necessary
 for the onset of inflation have a much lower probability. However, as shown in this paper, 
{\em this argument turns out to be completely inapplicable in the context of TMSM}. 
First of all, given the K-essence type structure, 
we know (see eq.(\ref{upper bound from rho in >0})) that for the initial energy density $\rho_{in}$ to be positive,
 the initial value $\frac{1}{2}\dot{\varphi}_{in}^2$ must satisfy the condition 
$\frac{1}{2}\dot{\varphi}_{in}^2\lesssim 0.78\cdot U_{eff}^{(tree)}(\varphi_{(in)})$.
However, a more fundamental argument can be advanced,
unrelated to the K-essence structure. In this regard, we recall that the choice made in this paper
 for the solution of the cosmological equations applicable to the description of our Universe
 imposes the requirement that the scalar function $\zeta$ be strictly positive. Violating this 
constraint would, firstly, contradict the conditions for the applicability of the equations derived 
from the principle of least action, and, secondly, would imply a change in the orientation of the 
space-time manifold of our Universe during its evolution. In section \ref{sub initial}, by 
analyzing the constraint (\ref{constr for initial}) determining the values of $\zeta$, we found that 
in order to satisfy the requirement $\zeta>0$ at the beginning of the classical evolution of the Universe,
 the initial values  $\dot{\varphi}_{in}$ and $(\nabla\varphi)_{in}$ must satisfy the upper bounds 
(\ref{kin bigger grad 6 and 8 in}) and (\ref{grad  bigger  kin 6 and 8 in}), with which conditions 
(\ref{Cond for begin infl})  are satisfied. This allowed us to conclude 
that there is a wide range of model parameters for which the conditions (\ref{Cond for begin infl})
 for the onset of inflation are satisfied.

The absence of the problem of initial conditions for Higgs inflation in the developed model is part of a more general result, namely, that {\em  inflation can be preceded by other, rather exotic initial stages of classical cosmological evolution}. This result is of fundamental theoretical importance, and its origin deserves a qualitative discussion. Since at $\varphi>6M_P$ the  TMT-effective potential  has a plateau of almost constant height, the possibility of the onset of cosmological evolution turns out to be independent of the initial value $\varphi_{in}>6M_P$. Therefore, from the pair of initial conditions $(\varphi_{in}, \dot{\varphi}_{in})$ the possibility of the onset of evolution turns out to depend only on $\dot{\varphi}_{in}$.
On the other hand,  the TMT primordial action (\ref{S-gr-H-VB}) leads to a TMT-effective action (\ref{Seff varphi})-(\ref{K2 varphi}) having a K-essence-type structure: the  TMT-effective kinetic energy density $K_{eff}=\Big(K_1-\frac{3}{2} \tilde{K}_2\frac{\dot{\varphi}^2}{2M_P^4}\Big)\cdot \frac{1}{2}\dot{\varphi}^2$, eq.(\ref{Kinetic eff}),  is a quadratic form in $\dot{\varphi}^2$. This structure implies that the condition of positive initial total energy density, $\rho_{in}>0$, allows the  TMT-effective kinetic energy density $K_{eff}$ to be negative at the initial $\dot{\varphi}_{in}$.
 This makes it possible that the evolution of the Universe starts with an equation of state $w=\frac{p}{\rho}<-1$, i.e. proceeds according to the dynamics known in dark energy models as phantom. Other pathological initial states with $w>-1$ are also possible. However, as it was shown in
 section \ref{normal dyn}, if the evolution starts with normal dynamics, then from the very beginning it proceeds in the inflationary regime.

Finally, it's worth noting that the description of fermion preheating in the TMSM model, presented in section \ref{Preheating}, yields quantitative results consistent with those obtained in conventional  models. Taken together with the other results obtained in this paper, this can be seen as further argument in favor of the TMSM over conventional models.

\appendix

\section{Justification of the TMT approach developed in the paper}
\label{Justification}

\subsection{Why Two-Measure theory}
\label{Why Two-Measure theory}

To justify the structure of the primordial  action  in the Two-Measure Theory\footnote{For more systematic and complete presentation of the TMT see recent review \cite{Universe}} presented in section \ref{sec:TMSM}, a clearly formulated new idea  is required: \underline{all possible degrees of freedom}
 contained in the primordial action should be considered as independent of each other, and all relationships between them appear as a result of applying the principle of least action. For brevity, we will call this approach to constructing a theory "the principle of maximum dynamic independence (PMDI)". 
The key to this formulation is the word "all" in the expression "all possible degrees of freedom", and we will now discuss what is meant by this.

Let us start with the {\em Palatini formulation of gravity}, in which the metric tensor and the (affine) connection are treated as independent variables. As is well known, in the absence of a non-minimal coupling of matter with curvature, the theory of gravity obtained by variation of the original action coincides with Einstein's general  relativity (GR). 
However, in the presence of, for example, a non-minimal coupling of a scalar field with a curvature scalar, the solution of the equations obtained by variation with respect to the affine connection leads to connection coefficients for which the covariant derivative of the metric tensor is non-zero.  Violation of the metricity condition means that the space-time is non-Riemannian, and to formulate the theory in a Riemannian space-time, it is necessary to perform the corresponding Weyl transformation of the metric tensor. Typically, such a transformation is performed in the original action in such a way as to eliminate the non-minimal coupling of the scalar field with the curvature scalar, and the set of variables used in this case is called the Einstein frame. 

Considering the formulation of Palatini's gravity from the point of view of the applied mathematical apparatus, we inevitably come to the conclusion that we are dealing with a 4-dimensional differentiable manifold at the stages of equipping it with  an affine connection and a metric structure (see, for example, \cite{Hawking}). 
{\em The construction of the TMT can be considered as the next step} in the implementation of the PMDI.  Indeed, in an oriented 4-dimensional differentiable manifold, even before equipping it with an affine connection and a metric structure, one can define  a volume measure as a 4-form, which can be represented as in eq.(\ref{Phi}), $dV_{\Upsilon}=\Upsilon d^4x=4!d\varphi_1\wedge d\varphi_2\wedge d\varphi_3\wedge d\varphi_4$. After the metric structure is defined on the manifold, if the latter is orientable, $dV_g=\sqrt{-g}d^4x$ can also be chosen as a 4-form to describe the volume measure. The possibility of such a choice is based on the fact that under general coordinate transformations with a positive Jacobian (i.e. without changing the orientation of the 4-manifold), $\sqrt{-g}$ and $\Upsilon$ are transformed according to the same rule. 
The standard approach (see section 2.8 in ref.\cite{Hawking}) is to use only the volume element $dV_g$, by default assuming that the volume element $dV_{\Upsilon}$ is redundant in this situation.  However, a closer look at the problem, when following PMDI, functions $\varphi_a$ are added to the set of degrees of freedom in action, shows that:

1. Only in empty space does the solution of the equations of the gravity model with two measures lead to the result $\Upsilon=const.\cdot \sqrt{-g}$, i.e. there is no fundamental difference between the volume measures $dV_{\Upsilon}$ and $dV_g$. However, even in empty space the gravity model with two measures can acquire new interesting properties compared to the gravity model with only one measure $dV_g$.

2. When the original action contains terms describing matter fields, and the volume measure $dV_{\Upsilon}$ is included in the action in addition to $dV_g$, then the scalar $\zeta\equiv\Upsilon /\sqrt{-g}$ turns out to be a function of the matter fields, which radically changes the dynamics of the fields.

The validity of the  assertion 2 was shown in section \ref{sec:TMSM}, where the effect of two volume measures on the field dynamics was studied in detail. It remains to demonstrate the validity of the   assertion 1, namely, that using both volume measures in empty space does indeed lead to Einstein's gravity model, in which after solving the equations it is found that there is no fundamental difference between the volume measures $dV_g$ and $dV_{\Upsilon}$. This will be done in subsection \ref{Empty space}.
When constructing models with two volume measures, we assume that the total volume measures have the form of linear combinations 
$(\alpha\sqrt{-g}+\beta\Upsilon)d^4x$, in which the constant dimensionless real coefficients $\alpha$, $\beta$ can be different depending in which term of the Lagrangian density they are present with in the primordial action. Bearing in mind that gravity in GR has a geometric nature, we will use the term 
\underline{pre-geometry} for effects caused by the structure of the total volume measures used in the primordial action. In subsection \ref{Adding matter fields} we will discuss some additional very important properties of the pre-geometry.

\subsection{Empty space}
\label{Empty space}

In examining the TMT model in empty space, we first consider the purely gravitational part of the primordial action, which in general can be chosen as
$\int (\alpha\sqrt{-g}+\beta\Upsilon)\left( -\frac{M_P^2}{2}\right)g^{\mu\nu}R_{\mu\nu}(\Gamma)d^4x$, where  $\alpha$ and $\beta$ are arbitrary  parameters. The coefficient $\alpha$ can be extracted as a common factor from the parentheses and absorbed by redefining the original parameter $M_P$ (we keep the original notation for $M_P$). After this, by redefining the  functions $\varphi_a$ by  $(\frac{\beta}{\alpha})^{1/4}\varphi_a \to \varphi_a$, we reduce the purely gravitational part of the primordial action to the form
$\int (\sqrt{-g}+\Upsilon)\left( -\frac{M_P^2}{2}\right)g^{\mu\nu}R_{\mu\nu}(\Gamma)d^4x$, which will be used in what follows.
So, let us consider the TMT model in the empty space with the following primordial action 
\begin{equation}
S_{empty}=\int\Biggl[(\sqrt{-g}+\Upsilon)\left( -\frac{M_P^2}{2}\right)g^{\mu\nu}R_{\mu\nu}(\Gamma)-\sqrt{-g}V_1-
\frac{\Upsilon^2}{\sqrt{-g}}V_2\Biggr]d^4x,
\label{S-empty}
\end{equation} 
The notation in (\ref{S-empty}) implying the use of Palatini's formulation and the explanation of the presence of two vacuum-like terms were given in section \ref{action}.
Here it is assumed that  $V_1<0$ and $V_2<0$.
The action takes into account the effect of pre-geometry and includes the following independent variables: the affine connection $\Gamma^{\lambda}_{\mu\nu}$, the metric tensor $g_{\mu\nu}$ and four scalar functions  $\varphi_a$, (a=1...4), with the help of which the measure density  $\Upsilon$ is constructed.
Varying the action with respect to the functions $\varphi_a$, similar to what was done in section \ref{sec:Equations of motion} (see eqs (\ref{varphiB}), (\ref{B}), (\ref{var varphi})), under condition $\Upsilon(x)\neq 0$ we obtain that the equality
\begin{equation}
-\frac{M_P^2}{2}g^{\mu\nu}R_{\mu\nu}(\Gamma)={\mathcal M}+2\zeta V_2
\label{var varphi empty}
\end{equation} 
must be satisfied, where $\zeta=\frac{\Upsilon}{\sqrt{-g}}$ and  ${\mathcal M}$ is a constant of
integration. Variation with respect to $g^{\mu\nu}$ yields  the equation
\begin{equation}
(1+\zeta)\left(-\frac{M_P^2}{2}\right)R_{\mu\nu}(\Gamma)
- \frac{1}{2}g_{\mu\nu}\left(-\frac{\tilde{M}_P^2}{2}\right)g^{\alpha\beta}R_{\alpha\beta}(\Gamma)
+\frac{1}{2}g_{\mu\nu}(V_1-\zeta^2V_2)=0.
\label{Grav.eq empty}
\end{equation}
The condition for compatibility of eqs.(\ref{var varphi empty}) and  (\ref{Grav.eq empty}) is a constraint in the form of the following expression for a constant scalar $\zeta$
 \begin{equation}
\zeta=\frac{{\mathcal M}-2 V_1}{{\mathcal M}-2V_2},
\label{zeta empty redef}
\end{equation}
As noted above in assertion 1, the latter means that the solution in empty space reveals that the volume measures $dV_{\Upsilon}$ and $dV_g$ differ only by a constant factor. 
From the constancy of $\zeta$ it follows that the variation of $\Gamma^{\lambda}_{\mu\nu}$ leads to an equation whose solution is the equality
$\Gamma^{\lambda}_{\mu\nu}=\{^{\lambda}_{\mu\nu}\}$,
where $\{^{\lambda}_{\mu\nu}\}$  are the Christoffel's connection
coefficients of the metric $g_{\mu\nu}$. This means that spacetime is Riemannian, there is no need for a Weyl transformation and
$R_{\mu\nu}(\Gamma)=R_{\mu\nu}(g)$ is the Ricci tensor of the spacetime with the metric $g_{\mu\nu}$. Substituting the value of $\tilde{\zeta}$ into eq.(\ref{Grav.eq empty}), we obtain Einstein's equations in empty space
\begin{equation}
R_{\mu\nu}(g)=
- \frac{1}{2M_P^2}\cdot\frac{{\mathcal M}^2-4V_1 V_2}{{\mathcal M}-2 V_2}g_{\mu\nu}.
\label{Grav.final eq empty redef}
\end{equation}
An important novelty of the solution of the model in empty space in TMT compared to the convensional theory is that by choosing the constant of integration
${\mathcal M}$, we can implement a solution with any TMT-effective  cosmological constant $\Lambda$. In particular, by choosing
 ${\mathcal M}=2\sqrt{V_1V_2}$, we obtain a solution with $\Lambda =0$ and 
\begin{equation}
\zeta|_{\Lambda =0}=\sqrt{\frac{V_1}{V_2}}.
\label{zeta empty lambda 0}
\end{equation}
Note that in this case, the value of $\zeta|_{\Lambda =0}$ coincides, up to a very small correction, with the value  $\zeta_v$ in the Higgs field vacuum found in section \ref{sec:Higgs vacuum}.

\subsection{Adding matter fields}
\label{Adding matter fields}

Continuing the justification for the choice of the primordial  action (\ref{action}), we take the action in empty space (\ref{S-empty}) as the gravitational term to which the matter terms must be added. As matter we consider the electroweak SM, and to describe it in the context of TMT we must take into account that in addition to rewriting the SM action in a general coordinate-invariant form, it is necessary to choose the volume measures  with which the SM Lagrangian terms enter the primordial action. But before doing this, we must return to the mathematical  issue discussed in section \ref{Why Two-Measure theory} about the existence of two volume measures $dV_g$ and $dV_{\Upsilon}$ and take into account important differences between them. The density $\sqrt{-g}$ of the volume measure
$dV_g$  is positive-definite and the volume measure $dV_g$ is defined  if the manifold is oriented, \cite{Hawking}. But the density $\Upsilon$ of the volume measure $dV_{\Upsilon}$ is sign-indefinite and, moreover, the sign of the 4-form
$dV_{\Upsilon}=
4!d\varphi_1\wedge d\varphi_2\wedge d\varphi_3\wedge d\varphi_4$ determines the sign of the orientation of the manifold (for more details see, for example, \cite{Lee}). Therefore, using linear combinations  
$(\alpha_i\sqrt{-g}+\beta_i\Upsilon)d^4x$ ('i' lists the terms in the action) as a total  volume measures in constructing the  terms of the primordial action, {\em we leave the sign of the space-time orientation undefined}. And only as a result of choosing a solution to the equations of motion, the sign of $\Upsilon =\zeta\sqrt{-g}$ is fixed, and, therefore, the sign of the orientation. In essence, we are discovering {\em the TMT-effect of  spontaneous restoration of the sign of the space-time orientation}. To understand how this effect occurs, let us consder the simplest example described by eqs. (\ref{zeta empty redef}) and (\ref{Grav.final eq empty redef}) (for more complicated case see eqs.(\ref{U vac})-(\ref{zeta v}) in section \ref{sec:Higgs vacuum}). To obtain a solution with $\Lambda=0$, we choose the integration constant ${\mathcal M}=2\sqrt{V_1V_2}$ and as a result found that $\zeta=\sqrt{\frac{V_1}{V_2}}>0$ and, therefore, $\Upsilon>0$. However, we could have chosen ${\mathcal M}=-2\sqrt{V_1V_2}$, which also ensures $\Lambda=0$. But with such a choice\footnote{In the case studied in section \ref{sec:Higgs vacuum}, instead of the choice (\ref{M 2sqrt}) for the integration constant  ${\mathcal M}$ we could have chosen ${\mathcal M}=-2\sqrt{V_1V_2}(1+\delta)$. But with such a choice, the constraint (\ref{zeta v before choice M}) leads to
$\zeta_v=-\sqrt{\frac{V_1}{V_2}}+ \mathcal{O}\left(\frac{\lambda \sigma^4}{|V_2|}\right)$.}, the constraint (\ref{zeta empty redef}) leads to
 $\zeta=-\sqrt{\frac{V_1}{V_2}}<0$ and, therefore, $\Upsilon<0$. Therefore, changing the sign of the integration constant ${\mathcal M}$, we obtain solutions with opposite signs of the space-time orientation.

The conclusion that follows from the described TMT effect of spontaneous restoration of space-time orientation is so fundamental that it allows us to formulate in the most concentrated form the main difference between TMT and conventional field theories. In conventional field theories, it is assumed by default that the sign of space-time orientation can be chosen arbitrarily. But in TMT, as we have seen, one of the two possible signs of orientation of the space-time manifold of our Universe is fixed by the choice of the solution of cosmological equations. And since the sign of orientation coincides with the sign of the scalar function  $\zeta$, which plays a decisive role in the dynamics of fields, 
{\em the physical results predicted by TMT turn out to be different for different signs of  the space-time orientation}.

Since the primordial action does not imply the existence of a definite space-time orientation, and also because of the the sign-indefinitness of $\Upsilon$, when constructing the primordial action terms using linear combinations of $\sqrt{-g}$ and $\Upsilon$, it should be assumed that the coefficients in front of 
$\Upsilon$ in the total volume measures can be positive  as well as negative. That is, assuming  $\alpha_i>0$,  $\beta_i>0$, the total volume measures can have the form 
$(\alpha_i\sqrt{-g}\pm\beta_i\Upsilon)d^4x$. This conclusion is a direct consequence of the fact that the structure of the primordial action takes into account pre-geometry in an essential way, and this is crucial for achieving the main results of this paper. and  in paper \cite{2-nd paper}. Below we give a qualitative description of how this works. But before that, one more remark, more of a technical nature, needs to be made.
 Let $\tilde{L}_i$ denote the i-th term of the SM Lagrangian that enters the  primordial action with volume measure 
$(\alpha_i\sqrt{-g}\pm\beta_i\Upsilon)d^4x$. Since the functions  $\varphi_a$ have already been redefined to obtain the gravitational term (\ref{S-empty}) of the action, to eliminate the coefficient $\beta_i$, we can factor it out of the brackets and absorb it by redefining either the SM parameters or the SM fields in $\tilde{L}_i$. As a result, the i-th term of
 the Lagrangian density takes the form $(b_i\sqrt{-g}\pm \Upsilon)L_i d^4x$, 
where $b_i=\frac{\alpha_i}{\beta_i}$ and $L_i$ is obtained from  the mentioned redefinitions in $\tilde{L}_i$. 
This is the form in which all the terms of the Two-Measure Standard Model are written in this paper and in paper \cite{2-nd paper}.

We can now explain how the mathematical features of TMT described above are related to the possibility of obtaining the results of this paper and ref.\cite{2-nd paper}. The main role in this is played by the choice of signs in front of $\Upsilon$ in the volume measures $(b_i\sqrt{-g}\pm \Upsilon) d^4x$ \,: with a minus sign in the kinetic terms of all SM fields and with a plus sign in all other terms. In the framework of this paper, where gauge and fermion fields are completely excluded from consideration, the corresponding primordial action for the Higgs field is given by eq.(\ref{S-H 12}).
 Due to this structure of the primordial action, during the evolution of the Universe,   the TMT-effective selfcoupling constant of the Higgs field changes from
 the parameter value $\lambda\sim 10^{-11}$ in the inflationary epoch to the value  $\lambda_{\sigma}$, eq.(\ref{lambda and m2 SM}), related to the GWS theory. This occurs due to the change of $\zeta$ from a value close to zero during the inflationary epoch to the value  $\zeta_v=\sqrt{\frac{V_1}{V_2}}$ near vacuum. In the latter case, this leads to the appearance in the denominator of the expression (\ref{lambda and m2 SM}) for $\lambda_{\sigma}$ of a very small number $b_k-\zeta_v$, which compensates for the smallness of the parameter $\lambda$.

The only exception in the choice of signs in the volume measures is the minus sign in front of $\Upsilon$ in the volume measure of the Higgs field mass term, that allows to provide a mechanism for the change of the Higgs mass squared term during cosmological evolution from positive to negaive (see also the last paragraph in sec.\ref{more about the TMT})

\section{Kinetic energy and $\zeta$ during inflation in the slow-roll regime}
\label{append 2}

In this appendix we will analyze the function $\zeta(\varphi, X_{\varphi})$ on a flat segment of the TMT-effective potential $U_{eff}(\varphi)$, where
$U_{eff}(\varphi)\approx 10^{-10}M_P^4$.
 The latter is realized at $\varphi\gtrsim 6M_P$, where the constraint that provides us with the function  $\zeta(\varphi, X_{\varphi})$ is given by 
eq.(\ref{constr for initial}), and for convenience we reproduce it here once again:
\begin{equation}
\zeta(\varphi,X_{\varphi})=\frac{32q^4e^{-4\sqrt{\xi}\frac{\varphi}{M_P}}-\frac{\lambda}{8\xi^2}\delta_p
+\frac{X_{\varphi}}{M_P^4}}
{32q^4e^{-4\sqrt{\xi}\frac{\varphi}{M_P}}+\frac{\lambda}{8\xi^2}-\frac{X_{\varphi}}{M_P^4}},
\label{constr for initial 1}
\end{equation}
Let's remember that  $\lambda\approx 2.3\cdot 10^{-11}$, \, $\xi=\frac{1}{6}$, \, $q^4=3\cdot 10^{-10}$, \, $b_p=\frac{1}{2}(1+\delta_p)$, \, $\delta_p\approx 10^{-8}$,
that is $\frac{\lambda}{8\xi^2}\approx 10^{-10}$ and $32q^4e^{-4\sqrt{\xi}\frac{\varphi}{M_P}} \lesssim 5.3\cdot 10^{-13}$ for $\varphi\gtrsim 6M_P$. In the context of the FLRW Universe, we are limited to $X_{\varphi}$ as kinetic energy, that is, $X_{\varphi}=\frac{1}{2}\dot{\varphi}^2$. 
In sections \ref{pathol and normal} and \ref{normal dyn} it  was shown that
the classical cosmological evolution starts with a normal dynamics only if the initial value $\dot{\varphi}_{in}$ satisfies the condition 
$\frac{1}{2}\dot{\varphi}_{in}^2< 0.17\cdot U_{eff}^{(tree)}(\varphi_{in})$. Moreover,  under this condition, cosmological evolution begins with inflation.

Taking into account that for $\varphi>6M_p$ with very high accuracy $K_1(\varphi)=1$, the cosmological eqs (\ref{K-Frid slow}) and (\ref{K-FLRW varphi slow}), describing inflation in the slow-roll approximation, are reduced to
\begin{equation}
\frac{\dot{a}^2}{a^2}=\frac{1}{3M_P^2}U_{eff}^{(tree)}(\varphi), \qquad 3\frac{\dot{a}}{a}\dot{\varphi}=-U_{eff}^{(tree)\prime}(\varphi).
\label{K=1 cosm eq slow}
\end{equation}
 From these equations it follows that in the slow-roll approximation the following equality is satisfied
\begin{equation}
X_{\varphi}=\frac{1}{2}\Bigl(\frac{d\varphi}{dt}\Bigr)^2=\frac{M_P^2}{6U_{eff}}(U_{eff}^{(tree)'})^2.
\label{phi dot 2}
\end{equation}
From this we find that for $\varphi > 6M_P$  the contribution of $X_{\varphi}$ to the constraint (\ref{constr for initial 1}) is bounded from above
according to the following estimate 
\begin{equation}
\frac{X_{\varphi}}{M_P^4}\approx 128\lambda \cdot e^{-2\sqrt{\frac{8}{3}}\frac{\varphi}{M_P}} <9\cdot 10^{-18}.
\label{estimate for Xvarphi}
\end{equation}
Then, using the parameter values mentioned after eq.(\ref{constr for initial 1}), we see that 
the first term in the numerator of eq.(\ref{constr for initial 1}) is leading and  is about $10^{5}$ times larger than the second and third terms.
As is known from the graph of $\zeta(\varphi)$ with the same set of parameters,  curve 3 in figure \ref{fig2}, $\zeta(\varphi)$ crosses zero at 
$\varphi \approx 14M_P$.
We conclude that in the interval of $6M_P<\varphi < 14M_P$, where $U_{eff}^{(tree)}(\varphi)$ is flat  and the slow-roll approximation is valid,  the scalar 
$\zeta$ can vary within the range
\begin{equation}
0<\zeta\approx\frac{32q^4}{\lambda/8\xi^2}\cdot e^{-\sqrt{\frac{8}{3}}\frac{\varphi}{M_P}} \lesssim 5\cdot 10^{-3}.
\label{zeta on flat segment}
\end{equation}

\end{document}